\documentclass{aa}  
\usepackage{natbib}
\usepackage{braket}
\usepackage{graphicx}
\usepackage{booktabs}
\usepackage{setspace}
\usepackage{comment}
\usepackage[colorlinks=true,linkcolor=blue,citecolor=blue, urlcolor=blue]{hyperref}
\usepackage{txfonts}
\newcommand\HI{H {\small{I}}}

\usepackage{color}
\usepackage{breqn}
\usepackage{amsmath}
\usepackage[normalem]{ulem} % For \sout{}

\newcommand{\soutPC}{\bgroup\markoverwith{\textcolor{cyan}{\rule[0.5ex]{2pt}{1pt}}}\ULon}

\newcommand{\soutFS}{\bgroup\markoverwith{\textcolor{red}{\rule[0.5ex]{2pt}{1pt}}}\ULon}
\begin{document}

   \title{Dust and gas modelling in radiative transfer simulations of disc-dominated galaxies with RADMC-3D}

   \titlerunning{Modelling gas and dust in RT simulations of galaxies}

   \author{Francesco Sinigaglia      \inst{1,2}\fnmsep\thanks{\email{francesco.sinigaglia@unige.ch}}
          \and
          Miroslava Dessauges-Zavadsky\inst{1}
          \and
          Lucio Mayer\inst{2}
          \and
          \\
          Pedro R. Capelo\inst{2}
          \and 
          Valentina Tamburello\inst{2}
          }

   \authorrunning{F. Sinigaglia et al.}

   \institute{Département d’Astronomie, Université de Genève, Chemin Pegasi 51, CH-1290 Versoix, Switzerland
        \and Department of Astrophysics, University of Zurich, Winterthurerstrasse 190, CH-8057 Z\"urich, Switzerland
             }

   \date{Received \today; accepted XYZ}

% \abstract{}{}{}{}{} 
% 5 {} token are mandatory
 
  \abstract
   {Bridging theory and observations is a key task in modern astrophysics to understand the formation and evolution of galaxies. With the advent of state-of-the-art observational facilities, an accurate modelling of galaxy observables through radiative transfer simulations coupled to hydrodynamic simulations of galaxy formation must be performed.}
   {We present a novel pipeline, dubbed \texttt{RTGen}, based on the Monte Carlo radiative transfer code \texttt{RADMC-3D} , and explore the impact of the physical assumptions and modelling of dust and gas phases on the resulting galaxy observables. In particular, we thoroughly address the impact of the dust abundance, composition, and grain size, as well as implement approximate models for the atomic-to-molecular transition and study the resulting emission from molecular gas.}
   {We apply Monte Carlo radiative transfer a posteriori to determine the dust temperature in six different hydrodynamic simulations of isolated galaxies. Afterwards, we apply ray tracing to compute the spectral energy distribution, as well as continuum images and spectral line profiles.}
   {We find our pipeline to predict accurate spectral energy distribution distributions of the studied galaxies, as well as continuum and CO luminosity images, in good qualitative agreement with literature results from both observations and theoretical studies. In particular, we find the dust modelling to have an important impact on the convergence of the resulting predicted galaxy observables, and that an adequate modelling of dust grains composition and size is required.}
   {We conclude that our novel framework is ready to perform high-accuracy studies of the observables of the interstellar medium, reaching few tens percent convergence under the studied baseline configuration. This will enable robust studies of galaxy formation, and in particular of the nature of massive clumps in high-redshift galaxies, through the generation of reliable and accurate mock images mimicking observations from state-of-the-art facilities such as JWST and ALMA.}


   \maketitle 

%-------------------------------------------------------------------
%-------------------------------------------------------------------
%-------------------------------------------------------------------

\section{Introduction}

%Main points:
%\begin{itemize}
%    \item galaxy formation studied through simulations
%    \item simulations can be used to produce images to be compared to real ones 
%    \item running RT is too expensive, we need it a posteriori
%    \item outline
%\end{itemize}

The advent of modern astronomical ground-based and space telescopes probing a large portion of the electromagnetic spectrum has triggered an unprecedented recent advancement in the understanding of the processes shaping galaxy formation and evolution. In particular, the James Webb Space Telescope \citep[JWST;][]{Gardner2006}, the Atacama Large Millimeter/submillimeter Array \citep[e.g.][]{Wootten2009}, as well as optical telescopes such as the Very Large Telescope (VLT\footnote{\url{https://www.eso.org/public/teles-instr/paranal-observatory/vlt/}}), the twin Keck\footnote{\url{https://www.keckobservatory.org/our-story/telescopes/}} telescopes, and the Subaru telescope\footnote{\url{https://subarutelescope.org/en/about/}}, amongst others, have delivered unprecedented high-resolution photometric and spectroscopic observations up to high redshift, with the current highest-redshift spectroscopically confirmed galaxy being at $z\sim 14.32$ \citep[][]{Robertson2024}.

Nonetheless, developing a solid theoretical understanding of the empirical measurements is not a trivial task due to the high degree of complexity and nonlinearity of the involved physical processes. In this sense, numerical simulations represent the ideal tool to investigate galaxy formation and evolution from a theoretical perspective. Over the last decades, a huge effort has been performed to refine hydrodynamic simulations to reproduce the observed properties with great accuracy \citep[see, e.g.][for detailed reviews]{Vogelsberger2020,Crain2023}. Despite the general success in this field, simulations have to cope with the complex problem of dealing with a wide range of different scales and a huge dynamical range. In fact, achieving both high resolution and large volume is computationally prohibitive even for modern world-class supercomputing facilities. In particular, the resolution is typically not large enough to simulate all the involved physical processes down to parsec scales. As such, a vast range of physical phenomena are modelled in a subgrid fashion, by adopting prescriptions which try to mimic the effect of such processes despite the lack of resolution to properly resolve them. 

In this context, adequately simulating the phenomena involving the interaction between radiation and matter and the transport of energy and momentum happening thereof would imply solving the equations of radiative transfer at all the steps of the simulations, which would make the computation even more costly. In fact, while there exist some cases in which the radiative transfer is solved on the fly in hydrodynamic simulations \citep[e.g.][]{Gnedin2001,Gnedin2009,Gnedin2010,Lupi2018,Rosdahl2018,Kannan2022}, in the vast majority of the cases the computation is performed a posteriori. In particular, a posteriori radiative transfer simulations are based on the assumption that radiation propagates at much faster speed than that inferred from typical dynamical time scales involved in the motion of matter, and hence, one can simulate the radiative transfer following the propagation of photons at frozen time, neglecting the potential micro-motion of matter meanwhile. In this sense, a number of radiative transfer codes have been presented in the literature, achieving great accuracy and becoming a standard in the generation of mock images and spectra to perform galaxy formation and evolution studies, such as \texttt{TRAPHIC} \citep{Pawlik2008}, \texttt{RADMC-3D} \citep{Dullemond2012}, \texttt{DESPOTIC} \citep{Krumholz2014}, \texttt{SKIRT} \citep[][]{Camps2015,Baes2015,Camps2020}, \texttt{CLOUDY} \citep{Ferland1998,Ferland2017}, \texttt{ART$^2$} \citep{Li2020}, \texttt{Powderday} \citep{Narayanan2021}, amongst others. %These techniques have been thoroughly explored in the literature, achieving great accuracy and becoming a standard in the generation of mock images and spectra to perform galaxy formation and evolution studies. 
%In particular, they have been employed to study 

%In order to be able to accurately simulate galaxy images in a given band of the electromagnetic spectrum without relying on approximated prescription, solving the radiative transfer question become crucial. The generation of mock images and spectra has nowadays become a standard tool to properly compare theory with observations. 

In this paper, we present a novel pipeline -- dubbed \texttt{RTGen} --  to perform end-to-end a posteriori radiative transfer of galaxy simulations, including both dust continuum and spectral line transfer computations. In particular, we rely on the \texttt{RADMC-3D} code \citep{Dullemond2012}, a highly-flexible software performing Monte Carlo simulations to determine the dust temperature and ray-tracing photons by solving the radiative transfer equation afterwards. In particular, we show first how to obtain continuum images at arbitrary wavelengths as well as the full spectral energy distribution (SED). Afterwards, we present our modelling of the atomic-to-molecular transition to split the atomic and molecular phases of the gas. Eventually, we combine the previous techniques to predict images of CO fine-structure transitions. To address the variations between different simulations and investigated models, we compare both the predicted SEDs and images, as well as distributions of different quantities when relevant.

The paper is organized as follows. Sect.~\ref{sec:refsim} presents the suite of simulations used as simulated sample in this work. Sect.~\ref{sec:radmc3d} summarizes the modelling of the various physical components that we carry out. Sect.~\ref{sec:results} presents the analysis of the main results, and Sect.~\ref{sec:time_evolution} discusses the time evolution of the predicted observables. We conclude in Sect.~\ref{sec:conclusions}.

\begin{figure*}
    \centering  \includegraphics[width=\textwidth]{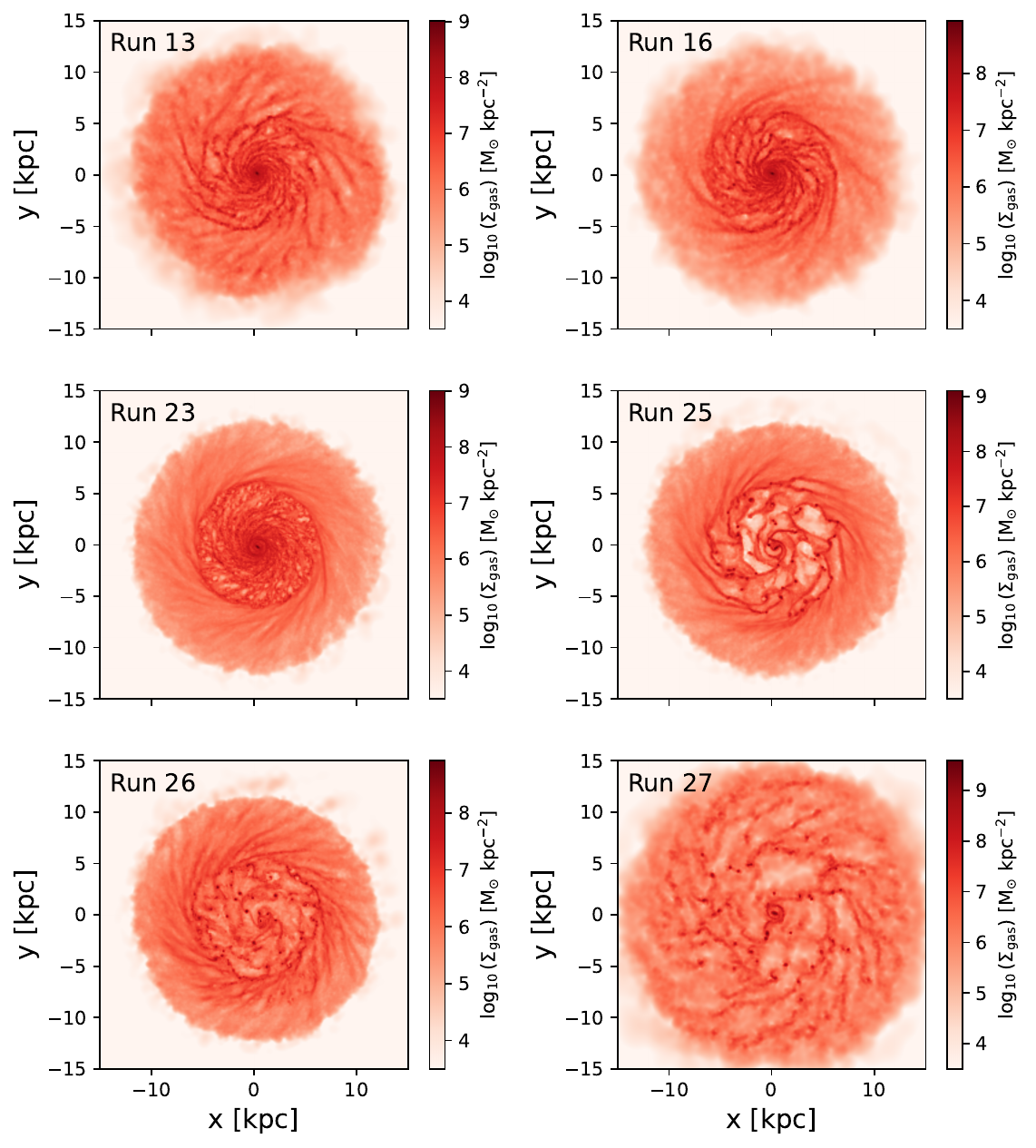}
    \caption{Gas mass surface density maps, viewed face-on, for the six different runs analyzed in this paper, after $\Delta t=0.2 \, {\rm Gyr}$ from the end of the relaxation phase. The run each projection refers to is reported inside the corresponding panel.
    }
    \label{fig:gas_density}
\end{figure*}

\begin{figure*}
    \centering  \includegraphics[width=\textwidth]{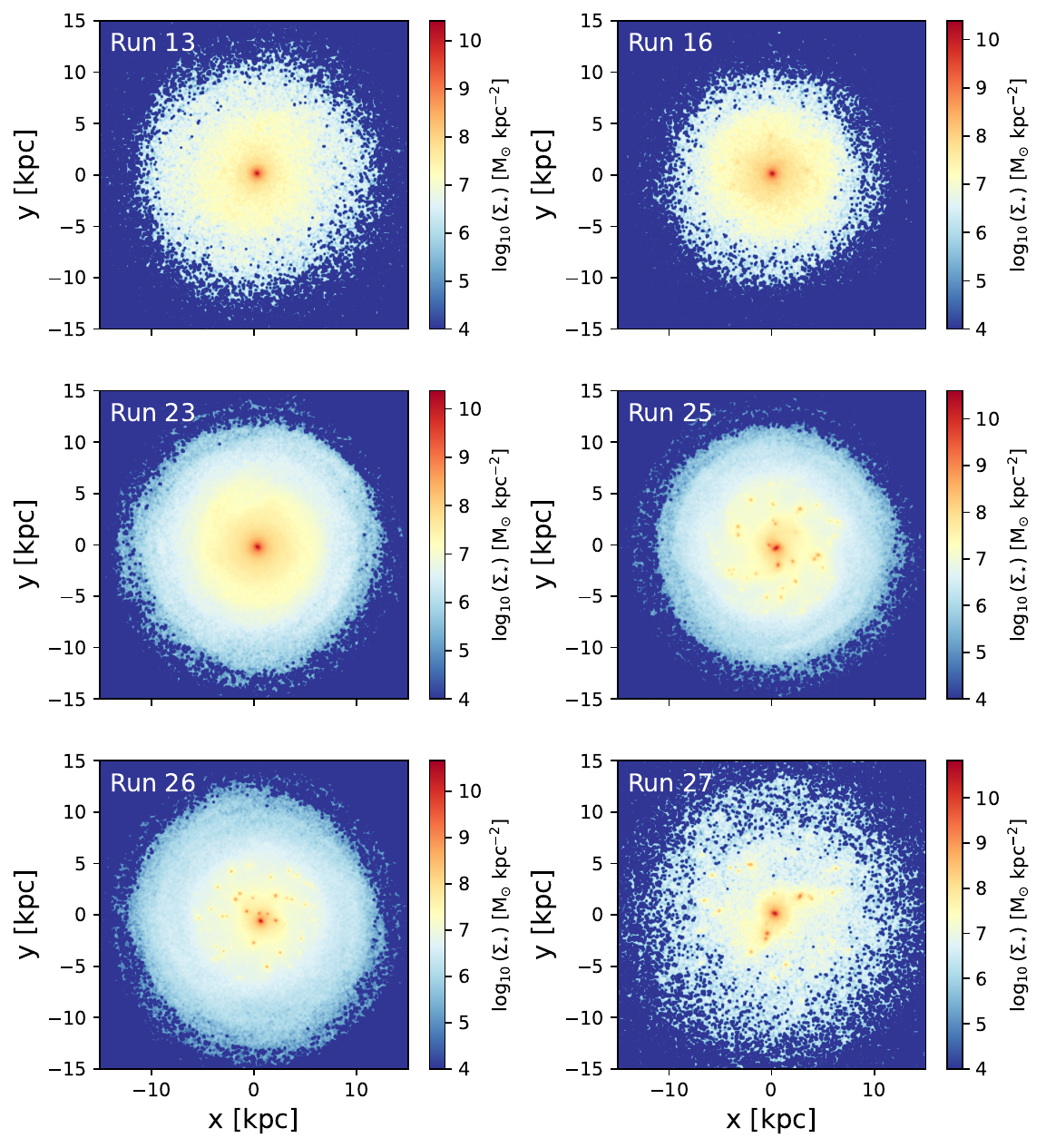}
    \caption{Stellar mass surface density maps, viewed face-on, for the six different runs analyzed in this paper, after $\Delta t=0.2 \, {\rm Gyr}$ from the end of the relaxation phase. The run each projection refers to is reported inside the corresponding panel.
    }
    \label{fig:stellar_density}
\end{figure*}

%-------------------------------------------------------------------
%-------------------------------------------------------------------
%-------------------------------------------------------------------

\section{Hydrodynamic simulations of isolated galaxies}\label{sec:refsim}

In this section, we briefly summarize the simulations adopted throughout this work. We refer the reader to \citet{Tamburello2015}, for a detailed description of the setup and the general results.

The suite of simulations presented in \citet{Tamburello2015} comprises $28$ simulations spanning $11$ different models of galaxy structural parameters. The simulations were run using \texttt{GASOLINE2} \citep{Wadsley2017}, a more recent version of the original $N$-body + smoothed particle hydrodynamics (SPH) code \texttt{GASOLINE} \citep[][]{Wadsley2004}. \texttt{GASOLINE2} adopts a binary tree structure to compute pair-wise gravitational interactions and smooths the gravitational force using a spline kernel function. The equations of hydrodynamics are solved relying on the SPH Lagrangian formulation, and include radiative and Compton cooling, as well as subgrid models for star formation, supernova feedback, metal enrichment from the explosion of supernovae and stellar winds. Radiative cooling is implemented for a mixture of hydrogen, helium, and metals, where collisional ionization equilibrium (CIE) is assumed for the latter (but not for hydrogen and helium) and modelled according to the tabulated data from \citet{Shen2010,Shen2013}, who assumed CIE using the \texttt{CLOUDY} code \citep{Ferland1998} and set a cooling temperature floor $T_{\rm min}=10~{\rm K}$. In the feedback model adopted for this suite of simulations -- known as the `blastwave model' \citep[][]{Stinson2006} -- cooling is switched off inside the radius of the blastwave produced by  the explosion of supernovae type II for a time interval $\Delta t\sim 10$--30~Myr, but it is not disabled in supernovae type I explosions.     

Star formation is triggered in gas particles which (i) have density $\rho > 10 \,m_{\rm H}\, {\rm cm}^{-3}$, where $m_{\rm H}$ is the hydrogen mass, (ii) have temperature $T<3\times 10^4 \, {\rm K}$, and (iii) are part of  converging flow. The star formation rate (SFR) in star-forming particles is assigned according to a Schmidt law and newly-formed stars are sampled from a \citet{Kroupa2001} initial mass function. 

Such subgrid prescriptions for star formation and feedback were proven to be capable of reproducing the properties of galaxies across a large dynamical range and a variety of morphological types. As stressed in \citet{Tamburello2015}, this indicates that the adopted models -- while purely phenomenological -- encapsulate the salient features of the energetics of the processes shaping galaxy formation, and hence, directly impacting the stability of forming discs. This aspect is crucial to reliably study the processes responsible for the formation and phenomenology of giant clumps -- massive star-forming regions -- as was done in \citet{Tamburello2015,Tamburello2017} and will be done in a companion paper. 

The initial conditions of the simulations were set up as in \citet{Mayer2001b,Mayer2002}, following the original method by \citet{Hernquist1993}. In particular, the models are constituted of a dark matter halo following a \citeauthor{NFW1997} (\citeyear{NFW1997}; NFW) radial density profile and embedding a disc with both stellar and gas components, and explore a variety of structural parameters. While these simulations consist of high-resolution non-cosmological simulations of isolated galaxies, the initial conditions were set up in such a way to mimic the features of cosmological simulations. In particular, \citet{Tamburello2015} chose a sample of $12$ massive isolated galaxies at $3.8<z<5.2$ from the \texttt{ARGO} simulation \citep{Feldmann2015,Fiacconi2015}, in an epoch when they are still in the outskirts of the main host and not experiencing mergers. 

All simulations were run adiabatically for $\Delta t=1 \, {\rm Gyr}$, to allow the galaxies to relax and achieve stability. After the relaxation phase, all discs are stable, with the \citet{Toomre1964} parameter $Q> 1$ in all the simulations and all over the simulated volume. Afterwards, the simulations were run for $\Delta t=2 \, {\rm Gyr}$ more, switching on radiative cooling and other subgrid physics models (see below).
As stated by \citet{Tamburello2015}, assuming that galaxies are initialized at $z\sim 3$, after the relaxation phase they should represent discs at $z\sim 2$.

The $11$ different studied models are summarized in Table 1 of \citet{Tamburello2015}. In particular, they consist in adopting different sets of values of the following parameters:

\begin{itemize}

    \item velocity at the virial radius of the halo $V_{\rm vir}$: set to be in the range $100<V_{\rm vir}<180 \, {\rm km \, s^{-1}}$, to match the asymptotic velocity of the most massive galaxies in ARGO \citep{Fiacconi2015};
    
    \item NFW halo concentration $c$: set to be in the range $6<c<15$ \citep[e.g.][]{Bullock2001,Ludlow2014}, with $c=6$ constituting the most representative case for the studied galaxies;
    
    \item gas fraction $f_{\rm gas}=M_{\rm gas}/(M_{\rm gas}+M_{\rm star})$: set to either $f_{\rm gas}=0.3$ -- an intermediate value in the selected \texttt{ARGO} galaxies \citep{Fiacconi2015} -- or $f_{\rm gas}=0.5$, as favoured by observations of galaxies at cosmic noon \citep[e.g.][]{Dessauges2015,Tacconi2020};
    
    \item inclusion of feedback (FB);
    
    \item inclusion of metal cooling (MC);
    
    \item inclusion of metal thermal diffusion (MTD);
    
    \item adoption of Geometric Density SPH (GDSPH);
    
    \item number of particles;
    
    \item softening length $\epsilon$: set to either $\epsilon=50 \, {\rm pc}$ or $\epsilon=100 \, {\rm pc}$.
    
\end{itemize}

Out of the $28$ simulations, $22$ were run at low resolution, using $n_{\rm p,dm}=1.2\times 10^6$ dark matter particles, $n_{\rm p,gas}=10^5$ gas particles and $n_{\rm p,s}=10^5$ star particles, while the remaining $6$ ones were run at higher resolution employing $n_{\rm p,dm}=2\times 10^6$ and $n_{\rm p,gas}=n_{\rm p,s}=10^6$.

In this work, we consider $6$ simulations, with different $V_{\rm vir}$, $c$, $f_{\rm gas}$, cooling, number of particles, and $\epsilon$. However, all the considered runs include feedback, and none of them adopt MTD nor GDSPH. In particular, the features of the chosen simulations are summarized in Table \ref{tab:sims}, a simplified version of Table 2 of \citet{Tamburello2015} restricted to just the simulations used herein.

\begin{table*}
    \centering
    \scriptsize
    \begin{tabular}{ccccccccccccc}
    \toprule
    %Run & $V_{\rm vir}$ [km s$^{-1}$] & $c$ & $f_{\rm gas}$ & MC & Res. & $\epsilon$ [pc] & Clumpy \\
    %\midrule
    %$13$ & $150$ & $6$ & $0.3$ & yes & low & $100$ & no\\
    %$16$ & $150$ & $10$ & $0.3$ & no & low & $100$ & no\\
    %$23$ & $150$ & $10$ & $0.3$ & yes & high & $100$ & no\\
    %$25$ & $150$ & $10$ & $0.5$ & no & high %& $100$ & yes\\
    %$26$ & $150$ & $10$ & $0.5$ & yes & high & $50$ & yes\\
    %$27$ & $180$ & $6$ & $0.5$ & no & low & $100$ & yes\\
    Run & $V_{\rm vir}$ [km s$^{-1}$] & $c$ & $f_{\rm gas}$ & MC & Res. & $\epsilon$ [pc] & $M_\star\times 10^{10}$ [M$_\odot$] & $M_{\rm h}\times 10^{10}$ [M$_\odot$] & $M_{\rm gas}\times 10^{10}$ [M$_\odot$] & SFR$\times 10^{-2}$ [M$_\odot$ yr$^{-1}$] & $r_{{\rm e},\star}$ [kpc] & $r_{{\rm e,gas}}$ [kpc]\\
    \midrule
    %$13$ & $150$ & $6$ & $0.3$ & yes & low & $100$ & $3.62\times 10^{10}$ & $139.67\times 10^{10}$ & $0.90\times 10^{10}$ & $4.9\times 10^{-2}$ & $2.73$ & $4.77$\\
    %$16$ & $150$ & $10$ & $0.3$ & no & low & $100$ & $3.68$ & $125.96$ & $0.84$ & $6.3$ & $2.18$ & $3.87$\\
    %$23$ & $150$ & $10$ & $0.3$ & yes & high & $100$ & $3.71$ & $125.96$ & $0.81$ & $6.2$ & $2.04$ & $4.01$\\
    %$25$ & $150$ & $10$ & $0.5$ & no & high & $100$ & $3.57$ & $125.96$ & $0.95$ & $23.1$ & $2.01$ & $5.44$\\
    %$26$ & $150$ & $10$ & $0.5$ & yes & high & $50$ & $3.76$ & $125.96$ & $0.76$ & $5.6$ & $1.98$ & $5.86$\\
    %$27$ & $180$ & $6$ & $0.5$ & no & low & $100$ & $5.98$ & $241.35$ & $1.83$ & $50.2$ & $2.50$ & $6.88$\\
    $13$ & $150$ & $6$ & $0.3$ & yes & low & $100$ & $3.6$ & $139.7$ & $0.9$ & $4.9$ & $2.7$ & $4.8$\\
    $16$ & $150$ & $10$ & $0.3$ & no & low & $100$ & $3.7$ & $125.9$ & $0.8$ & $6.3$ & $2.2$ & $3.9$\\
    $23$ & $150$ & $10$ & $0.3$ & yes & high & $100$ & $3.7$ & $125.9$ & $0.8$ & $6.2$ & $2.0$ & $4.0$\\
    $25$ & $150$ & $10$ & $0.5$ & no & high & $100$ & $3.6$ & $125.9$ & $0.9$ & $23.1$ & $2.0$ & $5.4$\\
    $26$ & $150$ & $10$ & $0.5$ & yes & high & $50$ & $3.8$ & $125.9$ & $0.8$ & $5.6$ & $1.9$ & $5.9$\\
    $27$ & $180$ & $6$ & $0.5$ & no & low & $100$ & $5.9$ & $241.4$ & $1.9$ & $50.2$ & $2.5$ & $6.9$\\
    \toprule
    \end{tabular}
    \caption{Summary of the parameters adopted in the simulations used in this work, extracted from the full suite of simulations from \citet{Tamburello2015}. The first column reports the number of the run (following the original numbering by \citealt{Tamburello2015}), the second column displays the velocity at virial radius ($V_{\rm vir}$), the third column shows the concentration ($c$), the fourth column the gas fraction ($f_{\rm gas})$ at the initial conditions, the fifth column whether metal cooling (MC) is included or not, the sixth column whether low or high resolution is used (i.e. how many particles are used), the seventh column reports the chosen gravitational softening length ($\epsilon$), the eighth, ninth, and tenth columns report the stellar mass ($M_\star$), the halo mass ($M_{\rm h})$, and the gas mass ($M_{\rm gas})$ 
    after $\Delta t=2$ Gyr from the end of the relaxation phase, respectively, the eleventh column shows the SFR after $\Delta t=2$ Gyr from the end of the relaxation phase, and the twelfth and thirteenth columns list the stellar half-mass radius ($r_{e,\star}$) and gas half-mass radius ($r_{e,{\rm gas}}$) after $\Delta t=2$ Gyr from the end of the relaxation phase, respectively. The halo virial mass of run 13 is different from the other runs with $V_{\rm vir}=150~{\rm km~s}^{-1}$ because the initial conditions are assumed at different redshift from the \texttt{ARGO} simulations, as explained in Sect.~\ref{sec:refsim}.}
    \label{tab:sims}
\end{table*}

Figures~\ref{fig:gas_density} and \ref{fig:stellar_density} show, respectively, gas mass and stellar mass surface density maps of the studied simulations, viewed face-on, after $\Delta t=0.2 \, {\rm Gyr}$ from the end of the relaxation phase. As can be easily noticed, the discs display different morphologies and clumpiness properties, depending on the physics models and the structural parameters adopted in each simulation. In particular, runs~13, 16, and 23 feature a smooth regular disc, whereas runs~25, 26, and 27 display the presence of clumps, as a result of the higher gas fraction used in the simulation, which favours fragmentation \citep[][]{Tamburello2015}.

%-------------------------------------------------------------------
%-------------------------------------------------------------------
%-------------------------------------------------------------------

\section{Radiative transfer with RADMC-3D}\label{sec:radmc3d}

\texttt{RADMC-3D}\footnote{\url{https://github.com/dullemond/radmc3d-2.0}} \citep{Dullemond2012} is a software package designed to perform radiative transfer simulations in astrophysical problems defined on an Eulerian grid of arbitrary geometry. The code performs first a Monte Carlo run to determine the dust temperature, and subsequently ray-traces the photons of specified wavelength along the line of sight. In particular, both continuum and atomic/molecular spectral line radiative transfer are implemented in \texttt{RADMC-3D}. We review the basic working principles underlying \texttt{RADMC-3D}, including a detailed description of Monte Carlo radiative transfer, in Appendix~A. We refer the interested reader to  \citet{Steinacker2013,Noebauer2019}, for a more complete review of Monte Carlo radiative transfer techniques.

Importantly, \texttt{RADMC-3D} does not include any built-in model for the physical system to be simulated. Hence, the practitioner should supply to the code all the inputs of the computation to be performed, and in particular (we will discuss all the inputs more in detail in the remainder of the paper):

\begin{itemize}

    \item the gas mass density;
    
    \item the stellar mass density, or the position of single stars, depending on the complexity of the problem and the number of total stars, as well as the effective temperature and the radius of the stars;
    
    \item the dust mass density.
    
\end{itemize}

In particular -- as discussed in more detail in Sect.~\ref{sec:radmc3d_input} -- the input fields must be supplied in a grid format. This implies that a suitable interpolation of particles onto a mesh must be performed in the case of SPH simulations.

For these reasons, an additional amount of work is required on the side of the user. However, this reflects into a great flexibility, as the user is not bound to any predefined model and can vastly explore the associated parameter space. In this paper -- the first of a series -- we focus on the study of the different assumptions and modeling of the resulting observable features of galaxies. In particular, we perform a variety of experiments changing the underlying physical assumptions and quantifying the differences by computing the SED, as well as images of the continuum at fixed wavelengths and of selected spectral lines.

In what follows, we present the modelling of the different physical components of the interstellar medium (ISM), needed as input of \texttt{RADMC-3D}.

\subsection{Dust modelling}\label{sec:dust_modelling}

Modelling the dust in a numerical simulation is a non-trivial task, which involves tracking the detailed evolution of the formation, growth, and destruction of dust grains \citep[see, e.g.][and references therein]{Aoyama2017,Dave2019,Esmerian2022,Dubois2024}.

In the simulations we use in this work, there is no on-the-fly dust modelling, and hence, we need to model it a posteriori. In what follows, we present our modelling of the dust and discuss the different implemented models and options. While this is certainly an additional degree of freedom and hence a source of potential uncertainty, we will show that it is very instructive to study a few representative cases to understand how sensitive the predicted observables from radiative transfer simulations are to the underlying model for the dust. In particular, we will present the models for the dust abundance, composition, and grain size.

We stress %\MDZ{stress (maybe better here)}
from now that our framework does not include  polycyclic aromatic hydrocarbon emission from stochastically-heated small dust grains \citep[see, e.g.][]{Camps2015}.

\subsubsection{Dust abundance}

As anticipated, the dust abundance -- namely the number of grains and their mass -- in a given spatial region of the simulation is the result of a complex balance between dust grain formation, growth and destruction. Here, since these processes are not modelled inside our simulations, we make use of literature results and perform our modelling a posteriori. 
In particular, we rely on the results by \citet{Li2019}, who found the relation between the dust-to-gas ratio (DGR) and the metallicity ($Z$) from the \texttt{SIMBA} simulation to be just weakly dependent on redshift from $z=0$ to $z=6$, and in reasonable agreement with observations \citep[e.g.][and references therein]{RemyRuyer2014,Zanella2018,DeVis2019,Peroux2020,Popping2022,Valentino2024}. In particular, we adopt the following scaling relation \citep{Li2019}:
\begin{equation} \label{eq:dgr}
    \log_{10}({\rm DGR}) = 2.445 \log_{10}(Z/Z_\odot) - 2.029 \, ,
\end{equation}
with a scatter $\sigma_{\rm DGR}\sim 0.31$ dex. While this relation was originally fitted by relating the mean global DGR and metallicity for a large and diverse sample of galaxies, we assume here that it holds cell by cell, in a spatially resolved fashion. This approach allows us to capture at least the dependence of the DGR on the gas metallicity. It has been shown that the DGR strongly varies as a function of gas properties beyond metallicity, such as density and temperature \citep[see, e.g.][]{Dubois2024}. However, in our simulations the cold phase of the ISM is not resolved, therefore it is not meaningful to incorporate these dependencies in our model. In this way, we first predict a DGR value for each cell given the metallicity therein, and subsequently add a random component to mimic the scatter, randomly-sampled from a log-normal distribution with $\sigma=\sigma_{\rm DGR}$. Assuming the validity of such a scaling relation at the cell level allows the model to have an additional degree of freedom, with respect to assuming a fixed DGR throughout the galaxy. %However, we will also test the assumption of a fixed DGR, by averaging the resulting DGR obtained via Eq. \ref{eq:dgr} for consistency.  
We also remind the reader that these tests are performed mostly as proofs of concept, but that the flexibility of our framework would in principle allow us to arbitrarily vary the assumed model and explore different scenarios. We restrict our study here to the two cases mentioned above. We also notice that the adoption of MTD in the simulations may have a significant impact on the spatial distribution of the metallicity, and consequently affect the cell-wise value of the DGR. We leave a more thorough investigation of these aspects related to the modelling of the DGR for future publications.

\subsubsection{Dust composition}\label{sec:dust_composition_theory}

The interstellar dust is formed from elements expelled by stars undergoing the Asymptotic Giant Branch phase \citep[e.g.][]{Marigo2001, Ferrarotti2006, Schneider2014, Goldman2022, DellAgli2023} and from supernovae explosions \citep[e.g.][]{Todini2001,Nozawa2007,Nozawa2011,Sarangi2015,Gall2018}. The two main constituents of the interstellar dust are silicate and carbonaceous grains -- oxygen-rich and carbon-rich respectively -- whose relative abundance depends on the chemical enrichment of the ISM from stars. In particular, adopting the C/O ratio as a proxy for the dust composition, stars with C/O$>1$ will tend to form carbonaceous grains, while stars with C/O$<1$ will preferentially form silicate grains \citep[see, e.g.][]{Draine2003}. To investigate the impact of the relative abundance between the two dust species, we perform herein the radiative transfer simulation under the following assumptions:

\begin{itemize}

    \item C/O $=1$: $50\%$ silicates, $50\%$ carbonaceous;
    
    \item C/O $=0.8$: $\sim 55\%$ silicates, $45\%$ carbonaceous;
    
    \item C/O $=1.2$: $\sim 45\%$ silicates, $55\%$ carbonaceous;
    
    \item C/O $=0$: $100\%$ silicates;
    
    \item C/O $\rightarrow\infty$: $100\%$ carbonaceous;
    
\end{itemize}

While the first three cases are found to be within a plausible C/O ratio, the last two cases are clearly unphysical and represent extreme conditions. Nonetheless, it is still instructive to investigate the impact of a radically different dust composition on the results of radiative transfer simulations. 

\texttt{RADMC-3D} allows the user to specify as many dust types as desired, each one with an associated weight representing the mass fraction. Therefore, it is straightforward to implement an arbitrary number of dust species in the computation. However, one should take care of computing a proper opacity for each of the studied dust types. We review the theory behind the computation of opacity curves in Sect.~\ref{sec:opac_comp}.

\subsubsection{Dust grain size}\label{sec:dust_size_theory}

The interstellar dust grain size typically ranges from $a\sim 0.0005\, \mu {\rm m}$ to $a\sim 0.250 \, \mu {\rm m}$ and its distribution has been found to follow the power law $n(a)\sim a^{-3.5}$ \citep[e.g.][]{Mathis1977,DraineLee1984,ODonnel1997,Hirashita2013}. 
As such, we can implement this scaling law in our modelling in order to pursue a higher degree of realism. We notice that other dust grain size distributions have been implemented and tested in the literature, such as the log-normal grain size distribution \citep{Hirashita2015} and a modified power-law model \citep{Jones2017,Ysard2024}. Nonetheless, we do not test such models in this work.

To implement the $n(a)\sim a^{-3.5}$ scaling law in our framework, we can proceed in two different ways:

\begin{itemize}

    \item compute an average opacity table taking into account the grain size distribution through the use of weights. In this way, we will obtain only one opacity table and perform the radiative transfer computation with just one effective dust species. In the general case where we admit a silicate-carbonaceous dust composition, this translates into two different opacity tables and dust species -- one for silicates and the other for carbonaceous grains;
    
    \item model explicitly the grain size distribution. In this case, we first subdivide the $n(a)$ distribution into $m$ bins within the range specified above. Then, we compute an opacity table for each grain size bin. Finally, we treat the $m$ bins as independent dust species. Following the reasoning above, for a silicate-carbonaceous composition, this implies treating $2\times m$ dust species. In this work, we assume $m = 20$ bins, evenly spaced in log scale, when performing this calculation.
    
\end{itemize}

As control examples, we also test the following assumptions that dust grains have a fixed size:  $a=0.001~\mu{\rm m}$, $a=0.01~ \mu{\rm m}$, and $a=0.1~ \mu{\rm m}$.

It is clear that -- as anticipated -- going from a fixed grain size to an effective grain size taking into account the grain size distribution, to an explicit modelling of the size distribution achieves a progressively higher degree of realism. However, while the transition from the first to the second implies just a slightly more expensive computation of the opacity table, going to the full modelling of the grain size distribution is significantly more expensive in terms of computing resources and requires $m$ times the memory required by the two other cases in order to store the dust grids. In this paper we always adopt limited mesh sizes (typically $n_{\rm cell}=256^3$), and hence, restrict the used memory to a treatable amount. As such, we can afford performing the full grain size modelling in a few cases, despite its longer computing time. However, we stress that in larger-scale problems this may start to become memory-demanding and in some cases achieve prohibitive resource requirements.

As a final note, we point out that the explicit modelling of the grain size distribution needs to be taken into account only in cases -- such as the present one -- in which also dust emission is considered, while for dust absorption and scattering the effective scaling corresponds exactly to the explicit one \citep[see, e.g.][]{Steinacker2013}.

\subsection{H$_2$/HI splitting}\label{sec:h2hi_modelling}

Modelling the transition between the atomic and the molecular phase of the interstellar gas is a tricky problem in simulations with too low resolution to resolve the multi-phase interstellar medium, as well as without on-the-fly radiative transfer computation and a chemistry network \citep[see, however,][]{Thompson2014,Dave2016,Polzin2024}. To overcome this issue, the common approach that has been adopted in the literature is to postprocess hydrodynamic simulations employing a subgrid model \citep[see][for a recent compilation of models and their application to cosmological simulations]{Diemer2018}. Such models should ideally try to account for the ensemble of complex processes determining the phase of the gas and which are unresolved or only partially resolved in the simulation, such as the shielding of molecular gas by other gas molecules, by atomic gas and by dust \citep[e.g.][]{SpitzerZweibel1974,Sternberg1988,Krumholz2013,Sternberg2014,Bialy2017,Capelo2018}, the non-trivial distribution of dust and in particular the spatial departure from the gas distribution \citep[e.g.][]{Gnedin2008,Bekki2015,Hopkins2016}, and the line overlapping effect\footnote{The line overlapping takes place when spectral lines of shorter wavelength and of higher order overlap with the spectral lines of longer wavelength and of lower order.} \citep[e.g.][]{Draine1996,GnedinDraine2014}, amongst others.

Several approximate solutions to quickly compute the molecular fraction have been proposed in the literature, which can be grouped into three classes \citep{Diemer2018}:

\begin{itemize}

    \item observed correlation between the molecular fraction $f_{\rm mol}$\footnote{We define the molecular fraction as $f_{\rm mol}=\rho_{\rm H2}/\rho_{\rm gas,tot}$, where $\rho_{\rm H2}$ and $\rho_{\rm gas,tot}$ are the molecular and the total gas density, respectively.} and the midplane pressure of galaxies, which can be estimated by using the surface density and velocity dispersion \citep[e.g.][]{Blitz2004,Blitz2006,Leroy2008,Robertson2008};
    
    \item prescriptions for $f_{\rm mol}$ depending on density, metallicity, and the ultraviolet (UV) radiation field, calibrated on high-resolution simulations solving for the chemistry \citep[e.g.][]{GnedinKravtsov2011,GnedinDraine2014,Lupi2018,Capelo2018};
    
    \item analytical equilibrium models of molecular clouds accounting for H$_2$ creation and destruction \citep[e.g.][]{Sternberg1988,Krumholz2013,Sternberg2014,Bialy2017}.
    
\end{itemize}

In this work, we adopt the second class of models. In fact, the first class of models has been shown to hold in good approximation only when projected onto the sky plane, but not at the volumetric level \citep{Diemer2018}. Also, such models do not include the UV field, hence there is no added value in running the full radiative transfer calculation to compute them. In contrast, both the second and third class of models explicitly rely on the UV radiation field. As such, since the main point of this computation is to support the usefulness of computing the UV field from radiative transfer, both classes are suited to our purpose. We choose here to test the second, and leave a more thorough investigation and comparison of the outcome of the two different classes for future work.  

As anticipated, a fundamental ingredient to compute the aforementioned models is the UV radiation field. While it is possible to make some assumptions and obtain a quick proxy of the radiation field from young stars \citep[see][]{Diemer2018}, we can here compute the radiation field self-consistently within our framework with a dedicated radiative transfer Monte Carlo run. In particular, following \citet{Diemer2018} we choose to compute the radiation field at $\lambda=1000 \, \AA$, and to adopt the normalization of the observed radiation field in the solar neighborhood from \citet{Draine1978}: $F=3.43 \times 10^{-8}$ photons s$^{-1}$ cm$^{-2}$ Hz$^{-1}$.

We also notice that no UV background is included here, as we are dealing with non-cosmological simulations. In future works, we will investigate the impact of UV background on radiative transfer simulations.

In particular, we consider the subgrid model by %\citet{GnedinDraine2014}, which builds upon 
\citet{GnedinKravtsov2011}. %and take into account the possible line overlap in the radiative transfer. 
This model is calibrated over tens of high-resolution simulations of isolated disc galaxies and solving explicitly the chemical evolution of the gas.

The %\citet{GnedinKravtsov2011} 
model parametrizes the molecular fraction as:
\begin{equation}
    f_{\rm mol} = \left(1+\frac{\Sigma_c}{\Sigma_{\rm HI+H_2}} \right) \, ,
\end{equation}
where $\Sigma_c$ is a critical density which depends on the radiation field $U$ and on the DGR $D$:
\begin{equation}
    \Sigma_c=2\times 10^7 \frac{{\rm M_\odot}}{{\rm kpc^3}}\, \left( \frac{\left[\ln(1+gD^{3/7}(U/15)^{4/7})\right ]^{4/7}}{D\sqrt{1+UD^2}} \right) \, ,
\end{equation}
where
\begin{equation}
    g=\frac{1+\alpha s + s^2}{1+s} \, , \quad \alpha = 5\frac{U/2}{1+(U/2)^2}\, ,
\end{equation}
and 
\begin{equation}
    s = \frac{0.04}{D + 1.5\times 10^{-3} \ln(1+(3~U)^{1.7})} \, .
\end{equation}

It is worth mentioning that a similar approach to ours -- combining the UV field from radiative transfer simulations with approximate models for the H$_2$/\HI{} transition -- has been studied by \citet{Gebek2023} and compared to the results from the prescriptions from \citet{Diemer2018}. The authors find that the two approaches yield very similar results in terms of \HI{} abundance and distribution because most of the atomic hydrogen resides in low-density gas, which is not efficiently converted into H$_2$. Nonetheless, this does not necessarily hold for the molecular component, for which we hence expect our approach based on the explicit UV field from full three-dimensional radiative transfer simulations to be more relevant than for \HI{}.

\subsection{Converting H$_2$ into molecular density}

To be able to properly simulate the molecular lines within our pipeline, we need to obtain an estimate of the number density of the target molecule. To this end, because the simulations adopted in this work do not solve for the chemistry, we need to resort to an effective model based on what is available from the simulation. The solution we choose here is the model-agnostic assumption that the target molecule number density is a function of the H$_2$ number density, which was previously obtained within the pipeline as described in Sect.~\ref{sec:h2hi_modelling}. 

In this paper, we focus on the study of radiative transfer of lines emitted by the CO molecule. In this sense, we notice that while the CO-to-H$_2$ conversion factor connecting CO luminosity to H$_2$ mass \citep[$\alpha_{\rm CO}$, see, e.g.][]{Bolatto2013} has been extensively studied in the literature, the inverse conversion from H$_2$ mass to CO mass (or alternatively, densities) is less well-constrained from simulations \citep[see, e.g.][]{Khatri2024} and it is not accessible from observations. Also, we notice that one cannot obtain the sought H$_2$-to-CO conversion by inverting the CO-to-H$_2$ conversion, because one would only obtain the CO luminosity and would not be able to infer the CO mass/density from it. In fact, such a relation is not bijective since the two quantities are connected by the radiative transfer computation and one can only pass from mass to luminosity and not vice versa.

To circumvent this issue, we decide here to adopt the following forward-modelling strategy. To marginalize over the uncertainty on the modelling of the H$_2$-to-CO relative abundance, we start by assuming that the H$_2$-to-CO conversion factor is a constant `effective' factor, which we call $\beta_{\rm CO}$ by analogy with the inverse conversion factor $\alpha_{\rm CO}$. After choosing an initial guess informed by the results from the literature \citep[namely $\beta\sim 10^{-8}-10^{-4}$,][]{Khatri2024}, we run the radiative transfer computation and obtain the CO(1-0) luminosity. Afterwards, we use the local $\alpha_{\rm CO}=4.3 \, {\rm M}_{\odot} \, {\rm pc}^{-2} \, ({\rm K \, km \, s}^{-1}  )^{-1} $ from Milky Way observations \citep[see][and references therein]{Bolatto2013} to convert the resulting CO luminosity into H$_2$ mass. At this stage, we compare the predicted H$_2$ mass in this way and the `true' H$_2$ mass obtained from the modelling of the molecular-to-atomic transition as in Sect.~\ref{sec:h2hi_modelling} and fit the $\beta_{\rm CO}$ conversion factor until the former H$_2$ mass converges to the latter within an arbitrary tolerance. We stop the optimization once the convergence reaches a $\lesssim 1\%$ level.

We notice that the conversion from H$_2$ mass to CO mass as just a fixed ratio $\beta_{\rm CO}$ is probably a simplistic assumption. However, we also assume a constant $\alpha_{\rm CO}$, and considering a more detailed model would arise in having strong degeneracies between such two parameters. Therefore, since we are here interested in showing a proof of concept, we regard this level of modelling as sufficient for now. We will explore more complex models in future works, where we plan to introduce the dependence of $\alpha_{\rm CO}$ on metallicity and where we will work with cosmological simulations with self-consistent dust treatment. In particular, there is strong observational evidence that the $\alpha_{\rm CO}$ factor varies significantly between different galaxies and even within the same galaxy \citep[see, e.g.][]{Teng2023,DenBrok2023}.

\subsection{Stellar modelling}\label{sec:star_model}

In our framework, the dust temperature is determined via Monte Carlo simulations by propagating photon packages placed initially according to the stellar density distribution and an input stellar spectrum. To adequately model the spatial distribution of photons across the grid, we then need to provide a grid of stellar densities, as well as a corresponding global spectrum. To produce the stellar density grid, because star particles are collisionless in the SPH formulation, we interpolate them onto a regular cubic mesh (the same used for all the inputs of \texttt{RADMC-3D}, see Sect.~\ref{sec:radmc3d_input}) via a Cloud-In-Cell (CIC) scheme. In addition, we explicitly verified that using the SPH kernel to interpolate the stellar particles induces negligible differences compared to the results obtained with the CIC kernel. Because the simulation resolution is not high enough to resolve single stars, to generate the input spectrum we treat star particles as stellar populations and model the spectrum of each one by means of the \texttt{STARBURST99} stellar population spectral synthesis code \citep{Leitherer1999}, following the approach adopted by \citet{Liang2019} to model stellar spectra in simulated galaxies at cosmic noon similar to the ones investigated herein. We run \texttt{STARBURST99} assuming the \texttt{PADOVA} stellar evolution tracks \citep[see, e.g.][and references therein]{Bressan1993,Marigo2001,Bressan2012} and a \citet{Kroupa2001} initial mass function in the stellar mass range $[0.1,100]~{\rm M}_\odot$, consistent with the setup used for the hydrodynamic simulations. Afterwards, given the age (computed as the difference between formation time and time of the snapshot under analysis), the metallicity, and the mass of each star particle, we interpolate these quantities onto the same spatial grid as the one introduced in Sect.~\ref{sec:radmc3d_input}. Then, we generate a synthetic spectrum per each cell, interpolating within a grid of values of metallicity and age provided by the \texttt{STARBURST99} simulations. Eventually, we obtain a single global spectrum for the whole grid by summing all the spectra throughout the simulation grid. 

We notice that for the baryon particle mass resolution $\gtrsim 10^5~{\rm M}_\odot$, stellar populations that are enshrouded in their dusty clouds may need to be treated by means of specific stellar population synthesis accounting for dusty
%\MDZ{dusty, instead of `dust'} 
star-forming regions \citep[see, e.g.][]{Jonsson2010,Ma2019}, such as e.g. \texttt{MAPPINGS-III} \citep{Groves2008} or \texttt{TODDLERS} \citep{Kapoor2023}. Nonetheless, the simulations used in this work have baryon mass particle resolution which is equal or lower than the mentioned threshold, 
%\MDZ{I am confused here: Are you sure the correct phrasing is not the opposite: the simulations used in this work have baryon mass particle resolution which is equal or LOWER than the mentioned threshold?}, 
therefore it is safe to use \texttt{STARBURST99} herein. We, however, caution the reader that a different approach implementing the aforementioned libraries may be needed when postprocessing simulations with worse resolutions.

\subsection{Interpolation of particles onto grids for \texttt{RADMC-3D}}\label{sec:radmc3d_input}

In order to be able to perform simulations through \texttt{RADMC-3D}, the user must supply the input files in mesh format. To this end, we interpolate the gas particles of the simulations onto a $(256,256,256)$ cell grid spanning a volume $V=(30 \, {\rm kpc})^3$ (corresponding to a physical cell resolution of $l\sim 117$ pc) using the three-dimensional \citet{Wendland1995} $C^2$ kernel. In this way, we obtain the following mesh fields: gas density, gas velocity along the three coordinate directions, and gas metallicity. 

As anticipated in Sect.~\ref{sec:star_model}, the stellar density grid is obtained by means of a CIC interpolation scheme onto the same $(256,256,256)$ mesh used for the interpolation of the other fields mentioned above (gas density, velocity, and metallicity).

\begin{figure}
    \centering
    \includegraphics[width=\columnwidth]{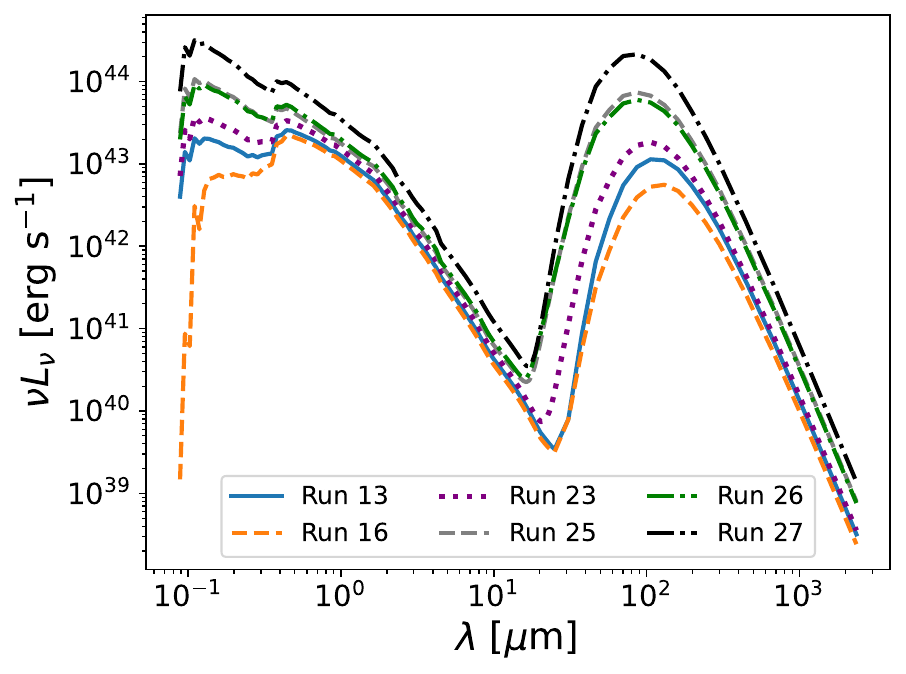}
    \caption{SED prediction from the baseline analysis for the six runs analyzed in this work.}
    \label{fig:sed_allsims}
\end{figure}

%\begin{figure}
%    \centering
    %\includegraphics[width=\columnwidth]{dust_temperature.pdf}
    %\caption{Dust temperature probability density distribution function for the silicate (blue solid) and carbonaceous (orange dashed) dust species in run~13 and for the analogous species in run~25 (green dotted-dashed and purple dotted, respectively).}
    %\label{fig:dust_temperature}
%\end{figure}

%-------------------------------------------------------------------
%-------------------------------------------------------------------
%-------------------------------------------------------------------

\section{Analysis, results and discussion}\label{sec:results}

In this section, we present the results obtained in each phase of the analysis and perform a comparative analysis of the different simulations investigated in this work. Throughout the work, we only consider the face-on projection of the six studied galaxies to be sure to perform a fair comparison, and leave the assessment of the impact of inclination for future research. When presenting our results, we designate as `baseline analysis' the radiative transfer run with the following fiducial setup: time snapshot at $\Delta t=0.2 ~{\rm Gyr}$, with DGR following the scaling relation reported in Eq. \ref{eq:dgr}, dust composition modelled as an effective mixture of silicate and carbonaceous grains with C/O $=0.8$, and grain size distribution following an effective scaling $n(a)\sim a^{-3.5}$.

When relevant, we compare the results of our simulations specifically for run~13 and run~25, because they represent examples of a smooth and a clumpy disc, respectively.

We test the impact of the simulation's resolution (number of SPH particles) in Appendix~\ref{sec:test_resolution}. In fact, while the resolution is not a parameter that enters in our model and it rather represents an intrinsic property of the used simulation, it is still instructive to be aware of its impact on the predicted observables.

\subsection{Dust continuum radiative transfer}

The first step in our procedure consists in running the Monte Carlo radiative transfer simulations to determine the dust temperature throughout the simulation box. Then, once the cell-wise dust temperature has been computed, we obtain the SED and mock images at arbitrary wavelength via ray tracing\footnote{Ray tracing consists in the procedure of following the spatial trajectory of photons while solving the radiative transfer equation.}. 

Figure~\ref{fig:sed_allsims} shows the resulting SEDs throughout the whole probed wavelength range for the six simulations studied in this paper. Here and in the SED plots which will be presented in the remainder of the paper, the model for the SED is computed through radiative transfer at $n_{\lambda}=150$ different wavelengths, evenly spaced in log scale, throughout the probed spectral interval, so as to ensure a proper sampling of the SED. The shape of the SED looks similar across different simulations, and is in reasonable agreement with the SEDs of galaxies at cosmic noon\footnote{We, however, warn the reader that the galaxies with lower gas fraction in our sample are not necessarily granted to be representative of cosmic noon galaxies.} measured from observations \citep[e.g.][]{Ouchi2008}, as well as from simulations \citep[e.g.][]{Liang2019}. This holds especially true if one takes into account the uncertainty in our modelling, given that the hydrodynamic simulations we consider herein do not include a built-in dust treatment nor a self-consistent cosmological evolution. To assess the uncertainty of the DGR on the predicted SED, we compute the variance from $50$ simulations varying the parameters of Eq. \ref{eq:dgr} within the uncertainties reported by \citet{Li2019} and adopting a proper covariance matrix. The resulting uncertainty is typically of the order of one/few percent, and hence, negligible compared to other sources of uncertainty and to the deviations induced by the different investigated models for the dust composition and grain size. We hereafter do not display these uncertainties in the SED plots for visual clarity.

\begin{figure}
    \centering
    \includegraphics[width=\columnwidth]{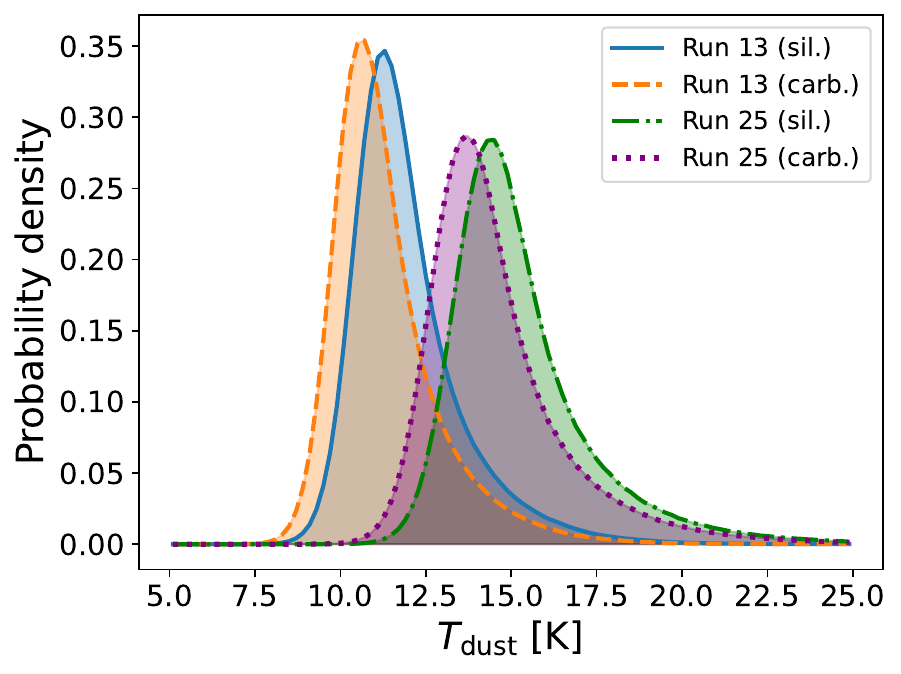}
    \caption{Dust temperature probability density distribution function for the silicate (blue solid) and carbonaceous (orange dashed) dust species in run~13 and for the analogous species in run~25 (green dotted-dashed and purple dotted, respectively).}
    \label{fig:dust_temperature}
\end{figure}

Furthermore, we perform diagnostics of the resulting dust temperature and analyze the resulting distribution. We adopt in this case the fiducial baseline dust composition. We plot the distribution of the in-cell (i.e. not weighted by mass) dust temperatures for run~13 and run~25 in Figure~\ref{fig:dust_temperature}, distinguishing between the silicate (blue solid for run~13, green dotted-dashed for run~25) and carbonaceous grains (orange dashed for run~13, purple dotted for run~25). The resulting dust temperatures display consistent distributions in all cases, within a range 1--60~K\footnote{We limit the histogram to $25$~K for visual purposes.}, and the majority of the distribution lies in a range 10--20~K. Carbonaceous grains are found to be slightly cooler than silicate grains by a few Kelvin degrees than silicate grains. In addition, to perform a consistent comparison with metrics for the dust temperature derived from observations, we compute also the `mass-weighted' dust temperature, $T_{\rm mw}$, and the `peak temperature', $T_{\rm peak}$. Specifically, the former is computed as the cell-wise average temperature weighted by dust mass, whereas the latter corresponds to the peak temperature obtained by fitting a black body template to the infrared (IR) SED \citep{Casey2012,Casey2014}. We obtain $T_{\rm mw}\sim 22.8~{\rm K}$ and $T_{\rm peak}\sim 24.1~{\rm K}$ for run~13, and $T_{\rm mw}\sim 27.4~{\rm K}$ and $T_{\rm peak}\sim 31.5~{\rm K}$ for run~25. These values are in reasonable agreement with the detailed study of the dust temperature based on simulations of galaxies at $z=2$ \citep[e.g.][]{Liang2019}, as well as a plethora of observational results %\MDZ{Proposed change: 
of $20~{\rm K}\lesssim T_{\rm peak}\lesssim 40~{\rm K}$ \citep[see, e.g.][]{Magnelli2014,Simpson2017,Thomson2017,Zavala2018,Schreiber2018}.

It is worth mentioning that the dust from run~25 is, on average, warmer than that from run~13. This fact can be explained by noticing that the UV part of the SED of run~25 (gray dashed line in Figure~\ref{fig:sed_allsims}) is significantly brighter than its counterpart from run~13 (blue solid line in Figure~\ref{fig:sed_allsims}), due to a larger abundance of young stellar populations emitting energetic radiation. This is in turn due to the fact that the higher gas fraction in run~25 favours disc fragmentation and star formation. Furthermore, from the fit of the modified black body model to the IR SED, we obtain values of the mean emissivity index $\beta\sim1.42$ and $\beta\sim1.41$ for run~13 and run~25, respectively, both broadly consistent with the value $\beta=1.60 \pm 0.38$ reported by \citet{Casey2012} and derived from observations. We also notice that the consistency with both simulations and observations of $T_{\rm peak}$ represents automatically a good consistency check that the SED of our simulations peaks at the correct wavelength.

%\begin{figure}
%    \centering
    %\includegraphics[width=\columnwidth]{sed_allgal_nuLnu.pdf}
    %caption{SED prediction from the baseline analysis for the six runs analyzed in this work.}
    %\label{fig:sed_allsims}
%\end{figure}

Similarly to \citet{Liang2019}, we notice that, at the redshift we are assuming here ($z \sim 2$), 
the contribution to the radiation field from the cosmic microwave background is negligible \citep[e.g.][]{DaCunha2013}, and hence, we do not include it. Also, we neglect the contribution from the external UV radiation field, whose effect will be investigated in future work based on cosmological boxes with consistent environmental dependencies and cosmic evolution over time.

\begin{figure*}
    \centering
    \includegraphics[width=\textwidth]{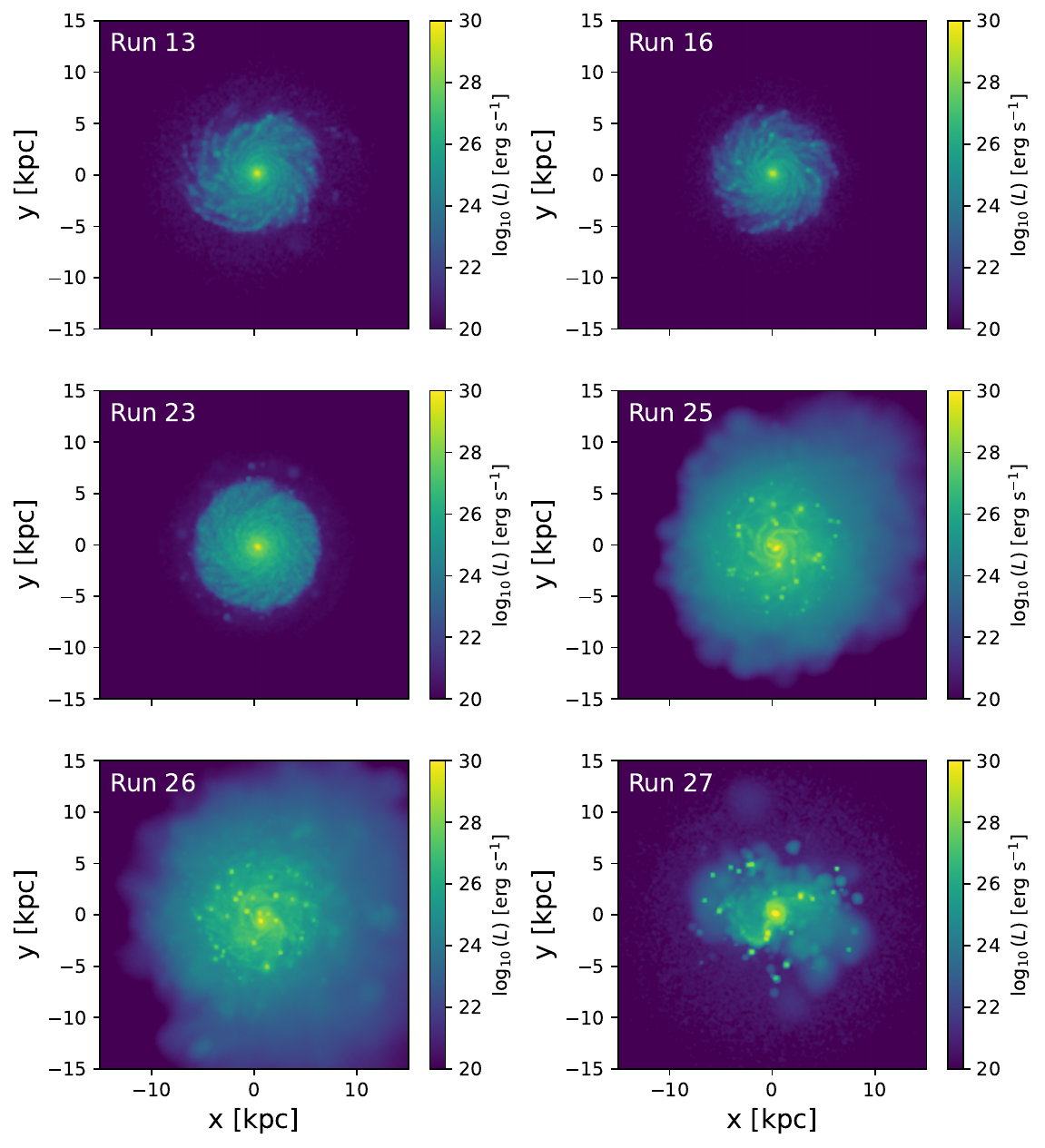}
    \caption{Dust continuum  luminosity images at $\lambda = 100 \, \mu$m, viewed face-on, for the six different runs analyzed in this paper, after $\Delta t=0.2 \, {\rm Gyr}$ from the end of the relaxation phase. The run each image refers to is reported inside the corresponding panel.}
    \label{fig:lum100um_slices}
\end{figure*}

\begin{figure}
    \centering
    \includegraphics[width=\columnwidth]{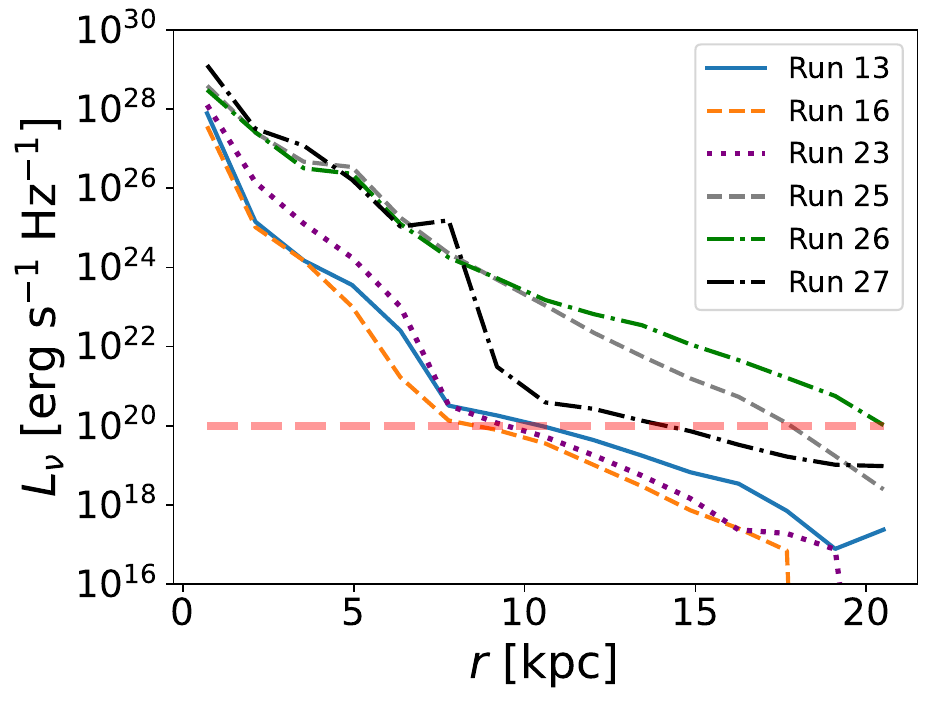}
    \caption{Radial luminosity profiles at $\lambda=100~\mu {\rm m}$ for run~13 (blue solid), run~16 (orange dashed), run~23 (purple dotted), run~25 (gray dashed), run~26 (green dotted-dashed), and run~27 (black dotted-dashed). The red dashed line marks the lower limit of the colorbar in Figure \ref{fig:lum100um_slices}.}
    \label{fig:radial_prof_100um}
\end{figure}

\begin{figure*}
    \centering
    \includegraphics[width=\textwidth]{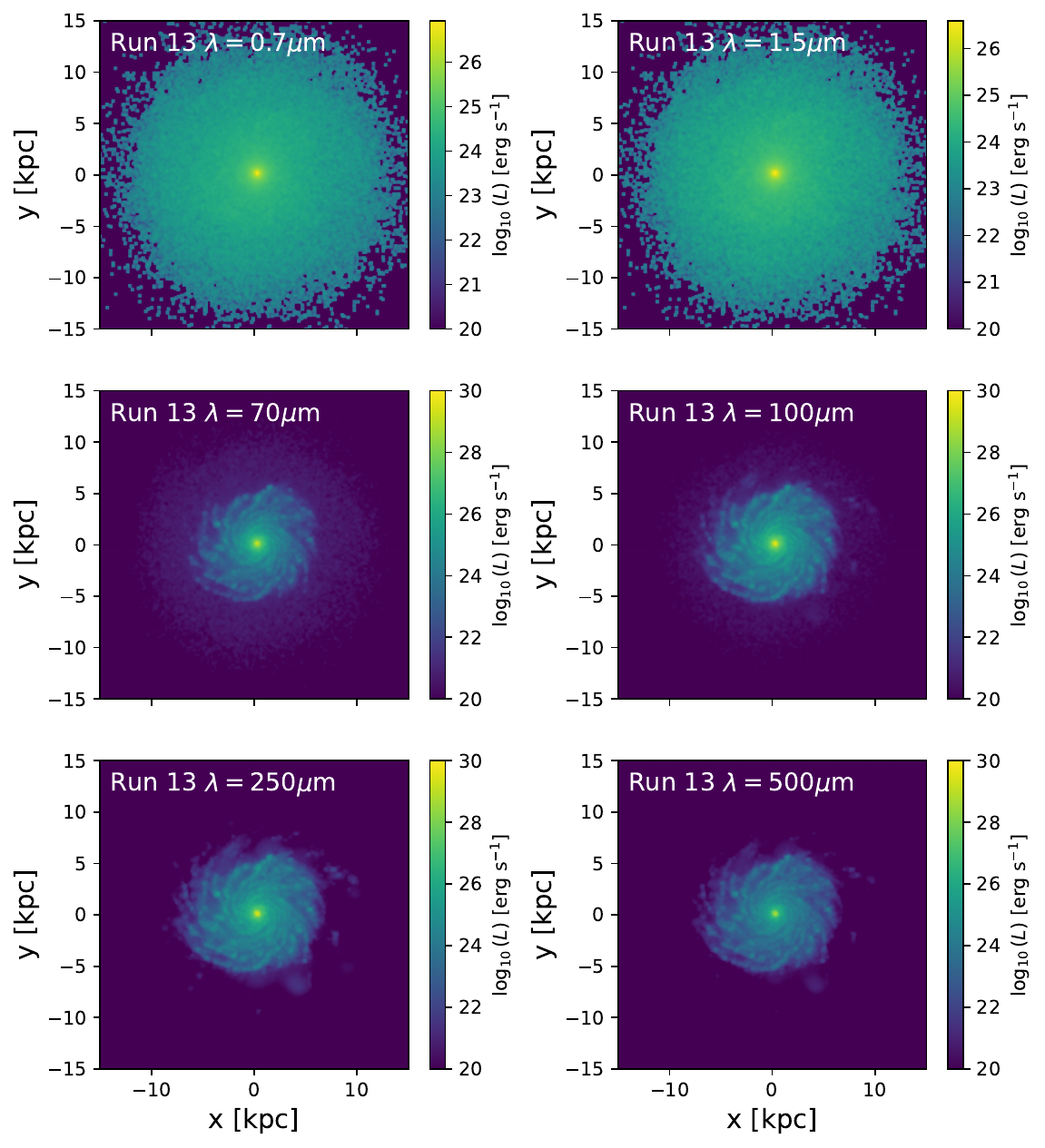}
    \caption{Dust continuum luminosity images at $\lambda = 0.7 \, \mu$m (top left), $\lambda = 1.5 \, \mu$m (top right), $\lambda = 70 \, \mu$m (mid left), $\lambda = 100 \, \mu$m (mid right), $\lambda = 250 \, \mu$m (bottom left), and $\lambda = 500 \, \mu$m (bottom right), viewed face-on, for run~13, after $\Delta t=0.2 \, {\rm Gyr}$ from the end of the relaxation phase.}
    \label{fig:luminosity_run13_lambda}
\end{figure*}

\begin{figure}
    \centering
    \includegraphics[width=\columnwidth]{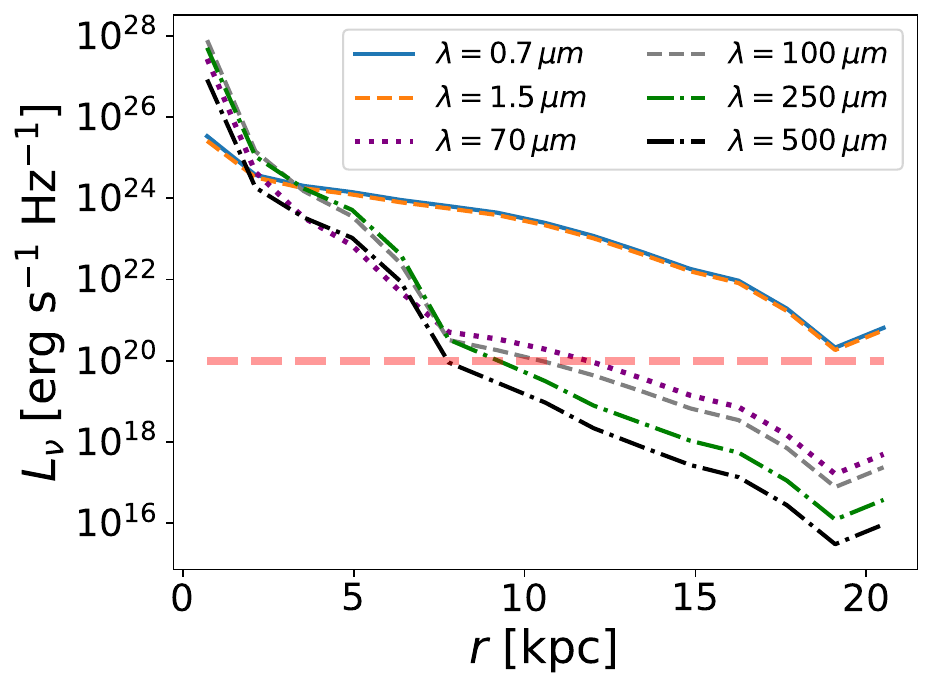}
    \caption{Radial luminosity profiles for run~13 at $\lambda=8~\mu {\rm m}$ (blue solid), $\lambda=24~\mu {\rm m}$ (orange dashed), $\lambda=70~\mu {\rm m}$ (purple dotted), $\lambda=100~\mu {\rm m}$ (gray dashed), $\lambda=250~\mu {\rm m}$ (green dotted-dashed), and $\lambda=500~\mu {\rm m}$ (black dotted-dashed), obtained from the dust continuum luminosity maps shown in Figure~\ref{fig:radial_prof_100um}. The red dashed line marks the lower limit of the colorbar in Figure~\ref{fig:luminosity_run13_lambda}.}
    \label{fig:radial_prof_cont_run13}
\end{figure}

\begin{figure}
    \centering
    \includegraphics[width=\columnwidth]{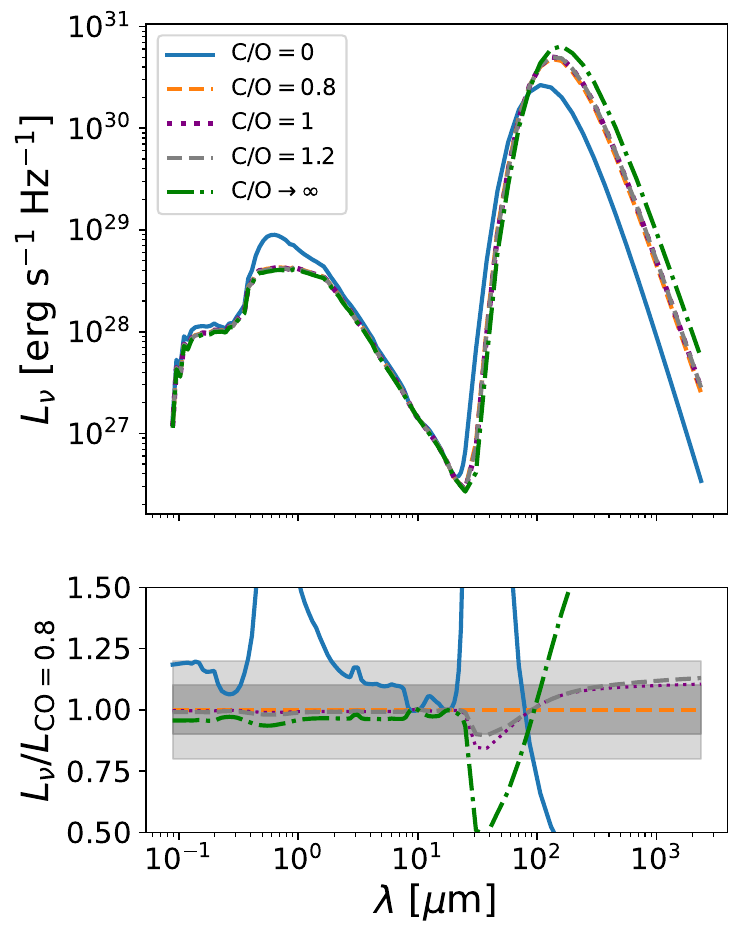}
    \caption{Results from the baseline analysis for run 13 (orange dashed), compared to the results from the other studied dust compositions:  C/O $\sim 0$ (blue solid), C/O $\sim 1$ (purple dotted), C/O $\sim 1.2$ (gray dashed), and C/O $\rightarrow \infty$ (black dotted-dashed). Top: SED predictions as a function of the relative abundance of silicate and carbonaceous grains. Bottom: ratios between the SED of each studied case and the baseline (C/O $\sim 0.8$), with the light and dark gray bands showing a 10$\%$ and 20$\%$ difference, respectively.}
    \label{fig:sed_composition}
\end{figure}

\begin{figure}
    \centering
    \includegraphics[width=\columnwidth]{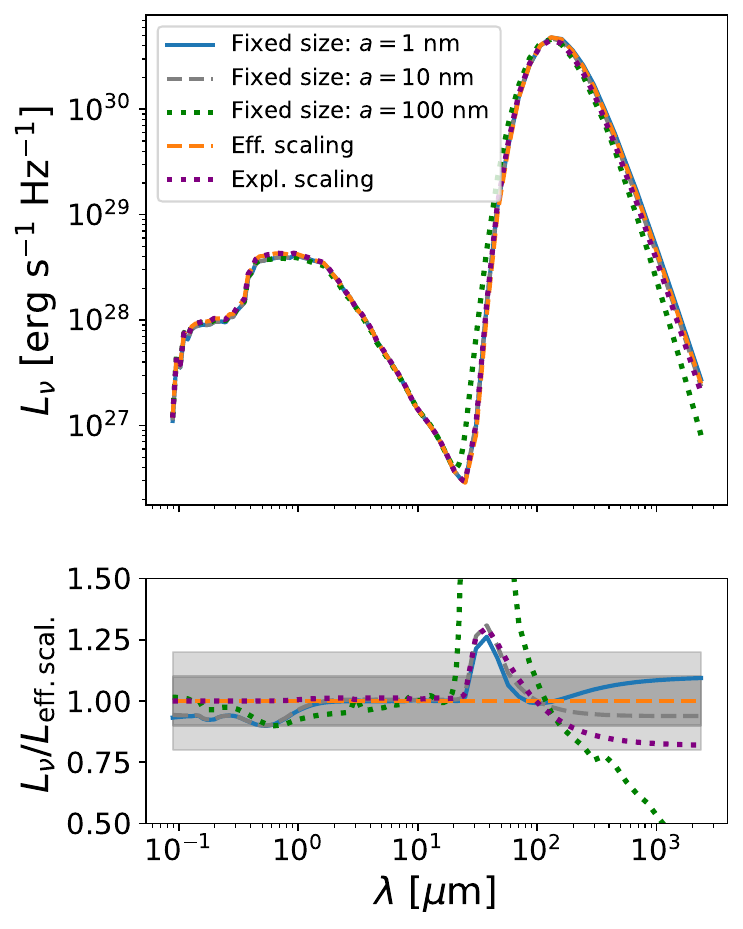}
    \caption{Results from the baseline analysis for run 13 (orange dashed), compared to the results from the other studied grain size distributions: fixed grain size with $a=1$~nm (blue solid), fixed grain size with $a=10$~nm (gray dashed), fixed grain size with $a=100$~nm (green dotted), and explicit modelling of the grain size distribution $n(a)\sim a^{-3.5}$ (purple dotted). Top: SED predictions as a function of the modelling of the grain size distribution. Bottom: ratios between the SED of each studied case and the baseline (effective mixture of grains with distribution $n(a)\sim a^{-3.5}$), with the light and dark gray bands showing a 10$\%$ and 20$\%$ difference, respectively.}
    \label{fig:sed_grainsize}
\end{figure}

We further investigate the results of our radiative transfer computations by visually inspecting the continuum images at different IR wavelengths. In Figure~\ref{fig:lum100um_slices} we show the luminosity images of the different runs at $\lambda = 100 \, \mu$m of the fiducial snapshot after $\Delta t=0.2$~Gyr after the end of the relaxation phase. To ease the comparison, the images share the same colour scale. While the resulting morphology is qualitatively similar to that of the images in Figure~\ref{fig:gas_density} displaying gas mass surface density maps, the small-scale structures are brighter in the continuum images compared to the background than in the density maps.

Furthermore, in some cases (especially runs~25 and 26) there appears to be a diffuse radiation component which extends out to large galactocentric distances. Figure~\ref{fig:radial_prof_100um} displays the radial luminosity profiles at $\lambda=100~\mu {\rm m}$ for the same galaxies shown in Figure~\ref{fig:lum100um_slices}. This quantitatively shows that runs~25, 26, and 27 are indeed generally brighter than the other runs at fixed galactocentric distance. This is mostly due to the fact the these simulations were run with higher gas fraction, which corresponds to a larger dust mass in our model, and hence, to a larger emitted IR luminosity. Figure~\ref{fig:luminosity_run13_lambda} shows instead the continuum images for run~13 at $\lambda=0.7~\mu$m (top left) and $\lambda=1.5 \, \mu$m (top right) to match the median redshift of the {\it JWST NIRCam} F070W and F150W filters \citep{Rieke2005,Rieke2023}, respectively, $\lambda=70~\mu$m (mid left) and  $\lambda=100 \, \mu$m (mid right) to match {\it Herschel} PACS filters \citep{PACS2010}, and $\lambda=250 \, \mu$m (bottom left) and $\lambda=500 \, \mu$m (bottom right) to match {\it Herschel} SPIRE filters \citep{SPIRE2010}. In this case, we do not saturate the mock images to the same colour scale to maximize the visual contrast, since the luminosity images span intrinsically different ranges. While the morphology is reassuringly self-similar in the PACS and SPIRE images, at shorter wavelengths ($\lambda=0.7 \,\mu$m and $\lambda=1.5 \,\mu$m) there appears to be a clear diffuse component extending out to the whole disc, not visible in others, coming from the stellar emission. This can be quantitatively seen in Figure~\ref{fig:radial_prof_cont_run13}, which displays the radial luminosity profiles for run~13 at the same wavelengths shown in Figure~\ref{fig:luminosity_run13_lambda}. In particular, at the shorter wavelengths the luminosity curve is shallower than at longer wavelengths, being fainter towards the centre but brighter towards the outskirts of the galaxy. The different slopes of the short-wavelength and long-wavelength radial luminosity profiles are due to the fact that the former are mainly due to the stellar emission, whereas the latter arise from the dust emission. As such, they mirror the spatial distribution of the stellar and dust density, respectively. Specifically, the dust emission is significantly more luminous in the central part of the galaxy, while the stellar emission has a smoother luminosity decrease with radius.

\subsection{Model of dust composition and grain size}

We address the impact of the model of dust abundance, composition and grain size by looking at the variation in the prediction of the SED with respect to our baseline.

Figure~\ref{fig:sed_composition} shows the comparison between the SEDs obtained assuming the five different relative silicate/carbonaceous compositions outlined in Sect.~\ref{sec:dust_composition_theory}. The top panel shows the resulting SEDs, whereas the bottom panel displays the ratios between the SED of each studied case and that of the baseline (i.e. the case with C/O $=0.8$), with the gray bands marking $10\%$ (darker) and $20\%$ (lighter) deviations, respectively. One easily notices that the three models with ${\rm C/O}=0.8$ (orange dashed), ${\rm C/O}=1$ (purple dotted), and ${\rm C/O}=1.2$ (gray dashed) have well-converged SEDs (within just a few percent deviations) at $\lambda\lesssim 10~\mu$m, and display reasonable deviations at $\sim$10--15$\%$ at $\lambda\gtrsim 100~\mu$m. On the other hand, it is interesting to notice that the cases comprising only silicate grains (${\rm C/O}=0$, blue solid) and only carbonaceous grains (${\rm C/O}\rightarrow \infty$) show larger deviations. In particular, if compared to the other cases, the peaks of the IR region of the SED are clearly displaced towards shorter wavelengths for the ${\rm C/O}=0$ case and towards longer wavelengths for the ${\rm C/O}\rightarrow\infty$ case, consistently with the dust grains being on average hotter and cooler, respectively.

Figure~\ref{fig:sed_grainsize} shows a similar comparison to that in Figure~\ref{fig:sed_composition}, but analyzing different prescriptions for the grain size distribution. The bottom panel is again normalized by the SED predictions from the baseline analysis, which consists this time in assuming an effective scaling of the dust grain size (see Sect. \ref{sec:dust_size_theory}). In particular, one finds that all the curves converge within $<10\%$ deviations with respect to the baseline analysis at $\lambda\lesssim 10~\mu$m, whereas they tend to have a larger dispersion at longer wavelengths -- within $\lesssim 25\%$ deviations -- expect for the run at fixed size $a=100~{\rm nm}$, which diverges significantly more. This indicates that, while assuming a fixed size for the grains is most likely inadequate, adopting an effective scaling may be a reasonable assumption in many cases. Nonetheless, an assessment of the accuracy and the convergence of the models should be performed case by case. 

\subsection{Atomic-to-molecular transition}

\begin{figure}
    \centering
    \includegraphics[width=\columnwidth]{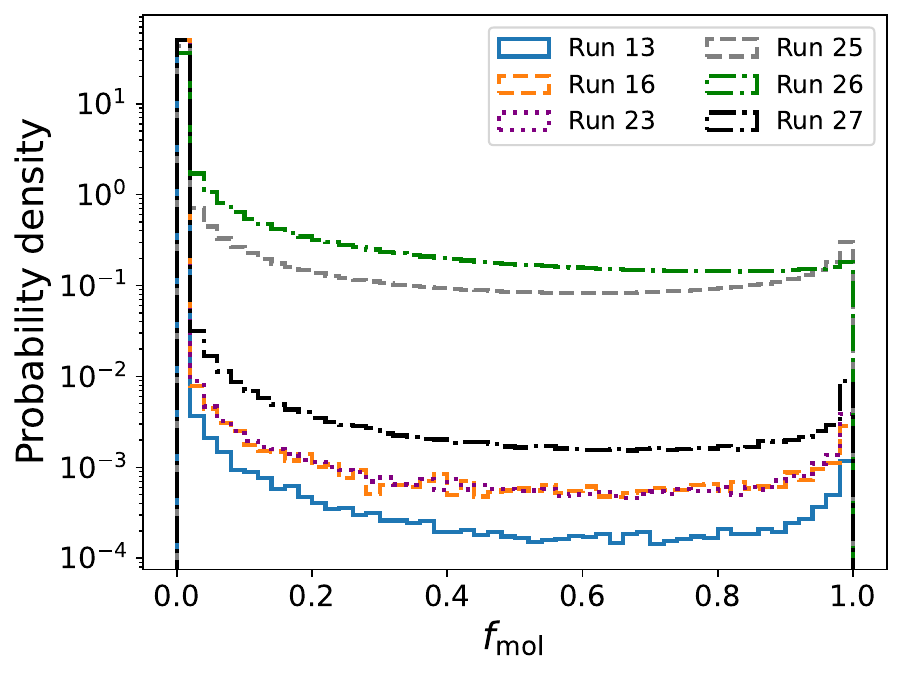}
    \caption{Probability density distribution function for the molecular fraction $f_{\rm mol}$ (see Sect.~\ref{sec:h2hi_modelling}). All the distributions feature a major peak around $f_{\rm mol}\sim 0$ (fully atomic gas) and a less significant peak around $f_{\rm mol}\sim 1$ (fully molecular gas). The displayed curves are normalized so that they subtend an area equal to unity.}
    \label{fig:fmol_hist}
\end{figure}

Figure~\ref{fig:fmol_hist} shows the resulting distribution for the cell-wise molecular fraction $f_{\rm mol}=\rho_{\rm H2}/\rho_{\rm gas,tot}$ for the six different simulations investigated herein. In particular, in this plot $f_{\rm mol}=0$ means fully atomic gas, whereas $f_{\rm mol}=1$ means that the gas is fully molecular. In generating this plot, we have considered only cells with non-zero gas density and discarded all the cells devoid of gas. The displayed curves are normalized so that they subtend an area equal to unity. The resulting distributions are clearly skewed in all the studied cases, and feature a major peak at $f_{\rm mol}=0$ and a much less significant peak at $f_{\rm mol}=1$. This implies that the fully-atomic and fully-molecular gas phases are respectively strongly and just mildly preferred over intermediate cases featuring a mixture of the two. We also notice that this phenomenon is strongly resolution-dependent. In fact, runs~25 and 26, which have higher resolution compared to the other four simulations, display a consistently larger probability of intermediate $f_{\rm mol}$ values by $\gtrsim 2$ order of magnitude with respect to the other cases. This result is consistent with the fact that a higher resolution allows us to better resolve regions with intermediate and high gas density conditions, where the formation of the molecular phase is favoured \citep[see, e.g.][]{Capelo2018}. In contrast, the high-resolution run~13 does not show the same behaviour, due to its lower gas fraction $f_{\rm gas}=0.3$.

We will further investigate the atomic/molecular phase balance under different galaxy formation models in future works, as well as the resulting scaling relations linking them to other galaxy properties and the impact of the gas phase morphology on the star formation activity and history.

\begin{figure*}
    \centering
    \includegraphics[width=\textwidth]{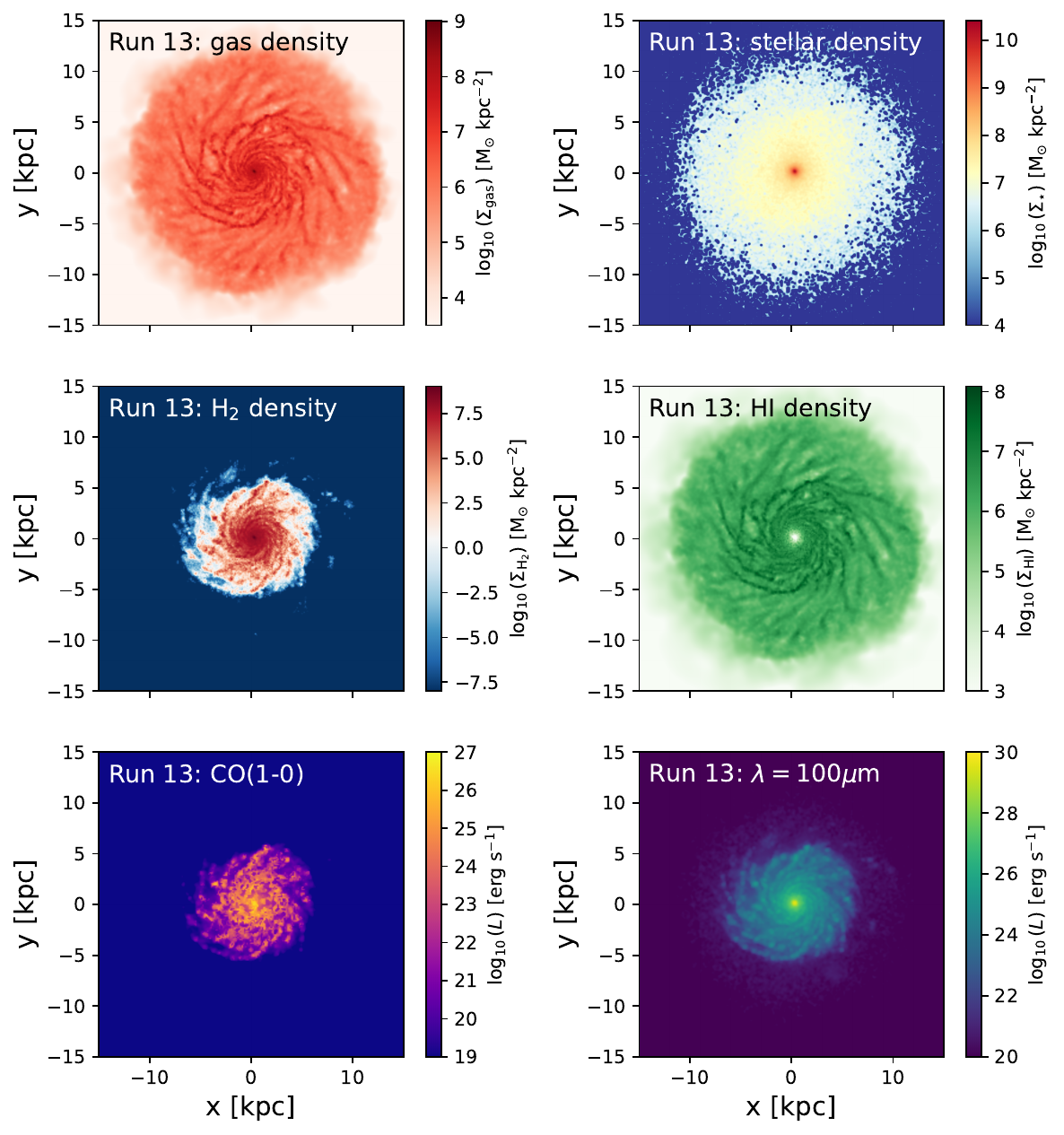}
    \caption{Gas mass surface density (top left), stellar mass surface density (top right), H$_2$ mass surface density (mid left), and \HI{} mass surface density (mid right) maps, continuum luminosity image at $\lambda=100~\mu{\rm m}$ (bottom left), and CO(1-0) luminosity image (bottom right), viewed face-on, for run~13, after $\Delta t=0.2 \, {\rm Gyr}$ from the end of the relaxation phase.}
    \label{fig:comp_run13}
\end{figure*}

\begin{figure*}
    \centering
    \includegraphics[width=\textwidth]{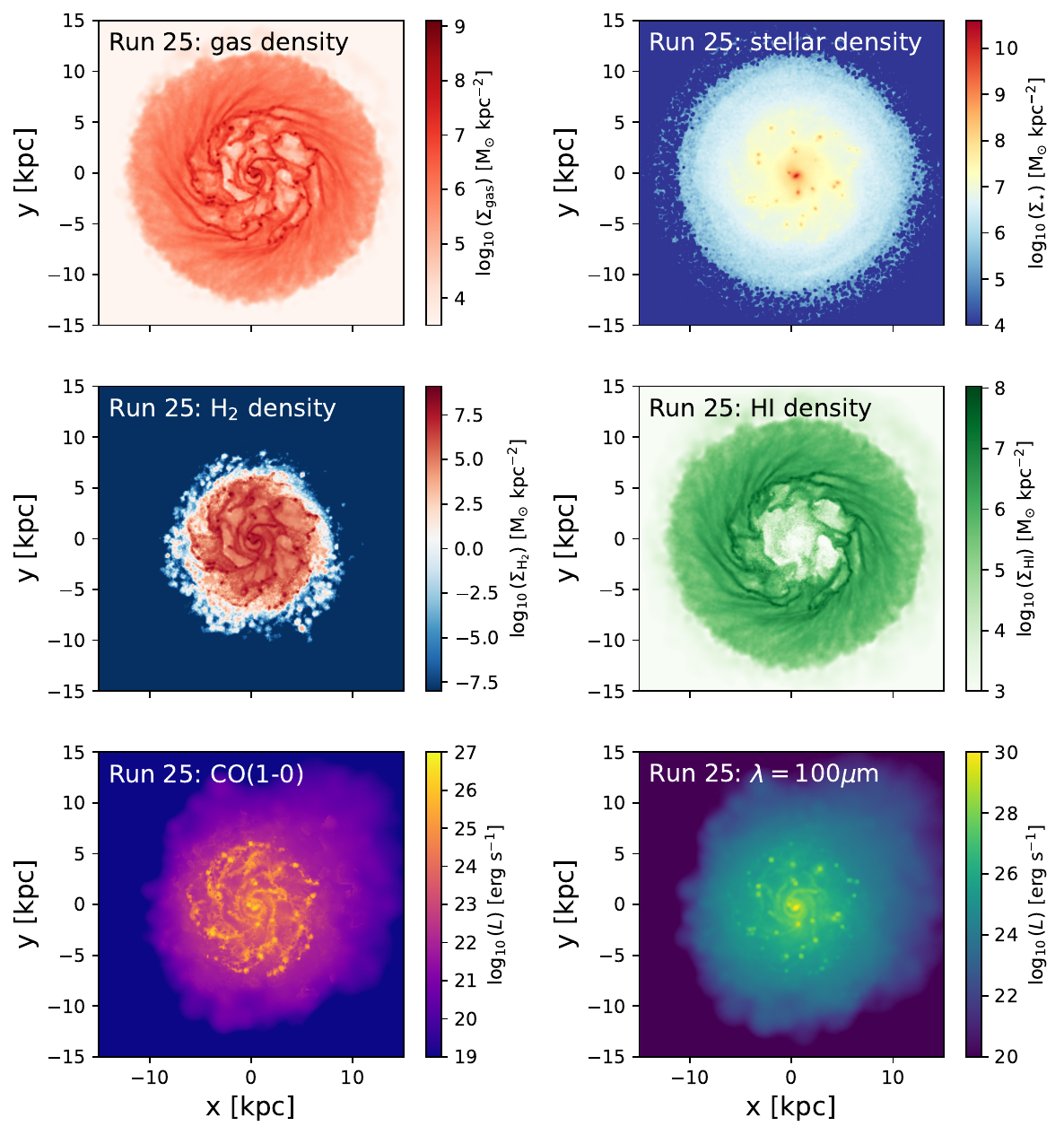}
    \caption{Same as Figure~\ref{fig:comp_run13}, but for run~25.}
    \label{fig:comp_run25}
\end{figure*}

\begin{figure}
    \centering
    \includegraphics[width=\columnwidth]{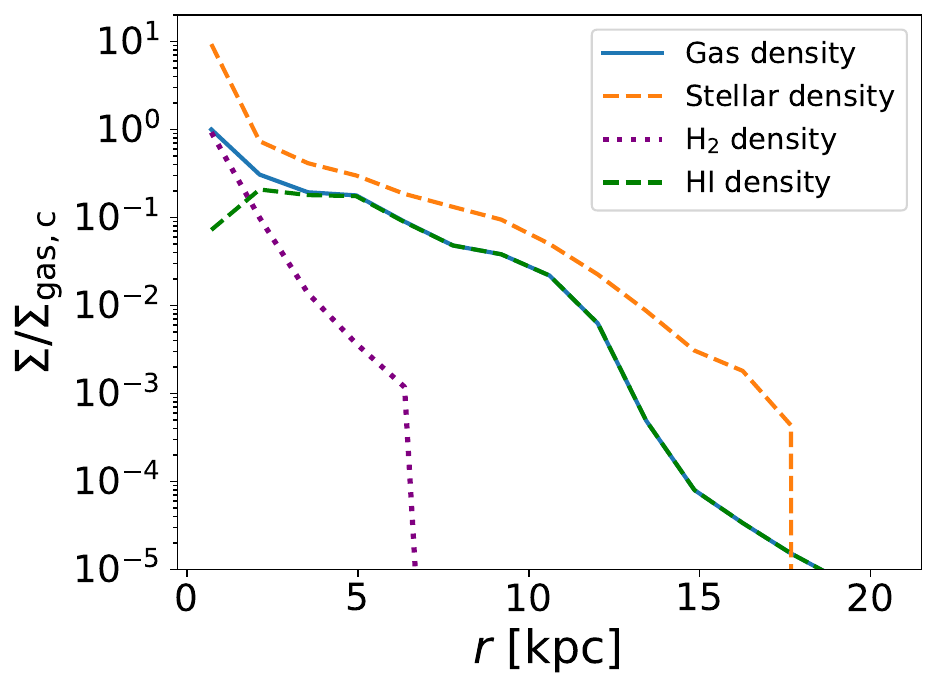}
    \caption{Radial surface density profiles for run~13 of the total gas density (blue solid), stellar density (orange dashed), H$_2$ density (purple dotted), and \HI{} density (green dashed). The curves are normalized to the total gas density in the innermost bin (i.e. at the centre).}
    \label{fig:radial_prof_galprop_run13}
\end{figure}

\begin{figure}
    \centering
    \includegraphics[width=\columnwidth]{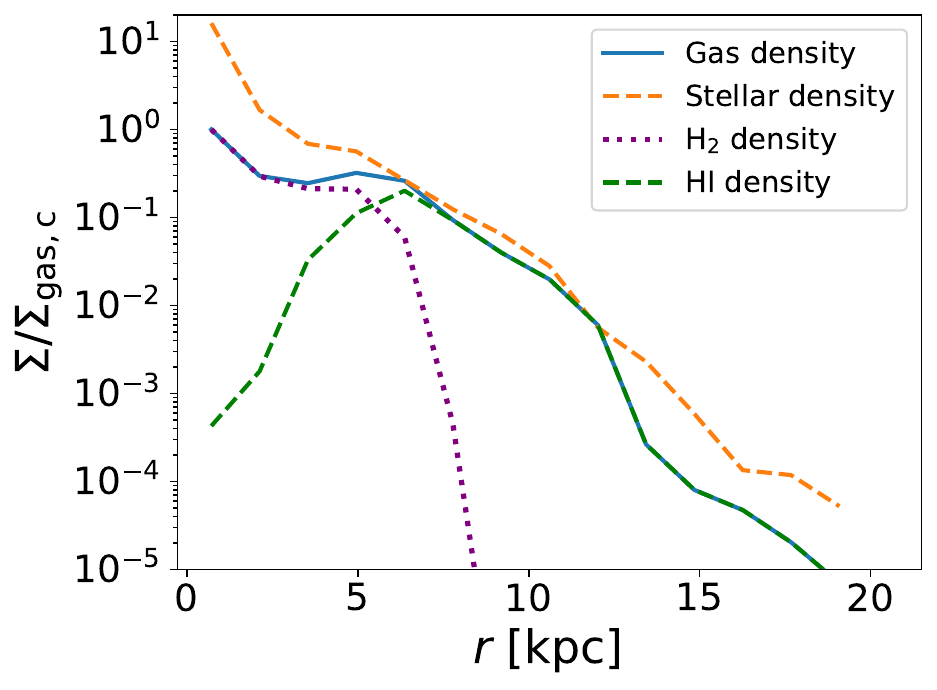}
    \caption{Same as Figure~\ref{fig:radial_prof_galprop_run13}, but for run~25.}
    \label{fig:radial_prof_galprop_run25}
\end{figure}

Figures~\ref{fig:comp_run13} and \ref{fig:comp_run25} show a series of galaxy properties for runs~13 and 25, respectively: the gas density (top left), stellar mass density (top right), H$_2$ mass density (mid left), \HI{} mass density (mid right), CO(1-0) luminosity (bottom left), and continuum luminosity at $\lambda=100~\mu$m (bottom right). By comparing the \HI{} density and H$_2$ density maps, it turns out that the molecular phase is more concentrated towards the centre, while the atomic phase is spread over a more extended disc \citep[as also found in][]{Capelo2018}. This is -- at least partly -- due to the gas in the central region being denser, and hence, more likely to be in the molecular phase. Also, the atomic component is found to be underdense towards the centre of the galaxy with respect to the average, where the molecular phase dominates. This fact can be appreciated more quantitatively in Figures \ref{fig:radial_prof_galprop_run13} and \ref{fig:radial_prof_galprop_run25}, displaying the radial density profiles of the total gas density, stellar density, H$_2$ density, and \HI{} density, for runs~13 and 25, respectively. From the resulting curves, it is clear that the molecular component dominates the total gas density profile at the centre, while the atomic component dominates towards the outer regions of the discs. This fact is consistent with $z=0$ galaxy observations, wherein \HI{} is found to form a disc which extends out to radii $\sim$5--10 times larger than the optical disc \citep[see, e.g.][]{Walter2008}, and where the CO emission often has a much smaller galactocentric extent than the optical disc \citep[see, e.g.][]{Leroy2008,Nagy2022}.

In particular, we notice that -- as anticipated -- the computation of the \HI{} distribution in the studied galaxies within our pipeline enables us to obtain predictions for the 21-cm line, amongst others, which is of paramount importance in the view of the future unprecedented observational data which will be collected by the Square Kilometre Array \citep[SKA,][]{Dewdney2009}.

%\subsection{Continuum transfer: impact of dust modelling}

%\subsection{Continuum transfer: local H$_2$ dissociation field}

\subsection{CO line radiative transfer}

Figures~\ref{fig:comp_run13} and \ref{fig:comp_run25} show also the CO(1-0) luminosity images for runs~13 and 25 (bottom left), respectively. All the runs presented herein are obtained under the local thermal equilibrium (LTE) assumption. We address the impact of LTE versus non-LTE in Appendix~\ref{sec:line_transfer}, but we anticipate that the impact is a deviation of just a few percent, which we neglect in the runs presented in this section due to their tiny contribution. In future studies, we will assess the impact of this assumption on a case-by-case basis, and adopt the non-LTE computation when its deviation from LTE becomes non-negligible.

\begin{figure*}
    \centering
    \includegraphics[width=\textwidth]{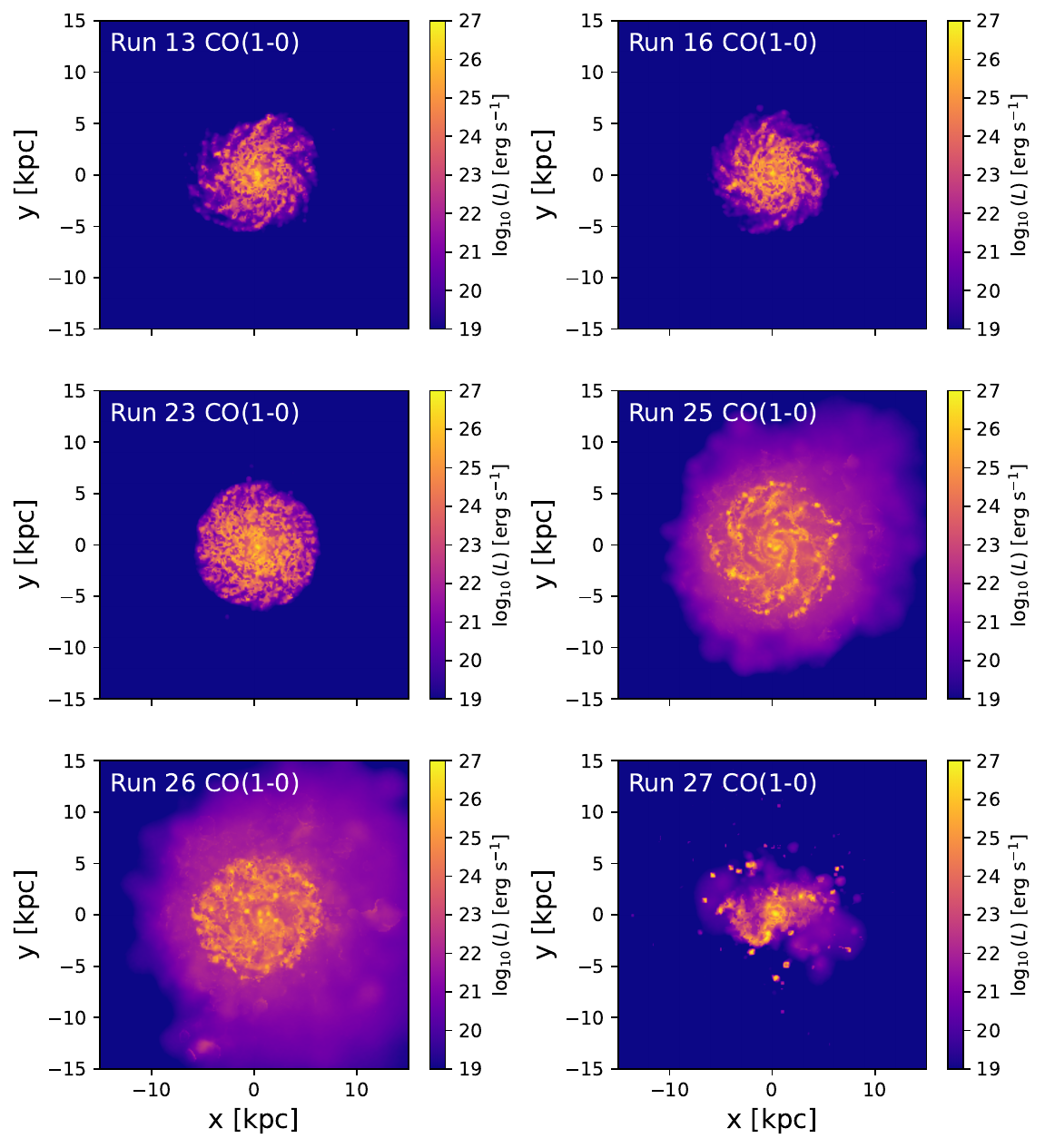}
    \caption{CO(1-0) luminosity images, viewed face-on, for the six different runs analyzed in this paper, after $\Delta t=0.2 \, {\rm Gyr}$ from the end of the relaxation phase. The run each image refers to is reported inside the corresponding panel.}
    \label{fig:co_all}
\end{figure*}

We start by noticing that the CO luminosity highlights the high-density regions of the studied galaxies -- as expected from molecular gas -- and in particular the clumpy structure of runs 25, 26, and 27 (see Figure~\ref{fig:co_all}, wherein we expand this comparison and show the CO(1-0) luminosity image for all the different runs). This is particularly important in the light of using the mock CO images to study the nature of massive clumps in high-redshift galaxies as observed with ALMA \citep[see][]{Dessauges2019,Dessauges2023}, which will be the subject of a forthcoming publication. It is interesting to notice that runs~25 and 26 -- which were run at high resolution and have a gas fraction $f_{\rm gas} = 0.5$ -- display a pronounced diffuse CO radiation in the outskirts of the disc. In contrast, run~27 does not display an analogous diffuse component despite having been run with the same gas fraction ($f_{\rm gas} = 0.5$) as runs~25 and 26. A careful investigation of this fact revealed that the violent fragmentation that the disc undergoes starts from the inner regions of the disc, and that at the time snapshot studied herein ($\Delta t = 0.2$~Gyr from the end of the relaxation phase) the molecular gas has formed preferentially towards the centre. For this reason, we do not observe a diffuse CO component in run~27 at the investigated snapshot. In contrast, we explicitly verified that, at later times, run~27 builds up a molecular gas radial profile extending out to larger galactocentric distances, and shows a consistent CO diffuse component. This is due to the fact that, because run~27 was run at lower halo concentration than runs~25 and 26 ($c=6$ for run~27, $c=10$ for runs~25 and 26), the rotation velocity at the same galactocentric distance is lower  and the orbital time longer in the disc of run~27 than in the discs of runs~25 and 26, which is consistent with a delayed evolution in run~27. We will study this phenomenon more in detail in a forthcoming paper addressing the nature of the giant clumps observed in CO maps.

\begin{figure}
    \centering
    \includegraphics[width=\columnwidth]{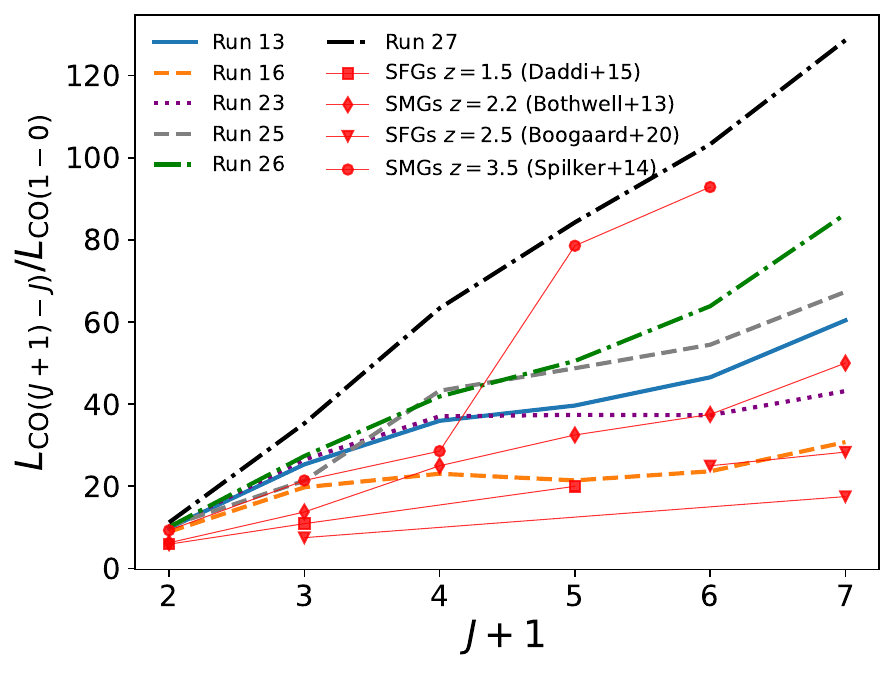}
    \caption{CO SLED for the baseline analysis of the six simulations studied in this work, after $\Delta t=0.2 \, {\rm Gyr}$ from the end of the relaxation phase. We display as thin red lines literature results for different galaxies at cosmic noon.}
    \label{fig:co_sled}
\end{figure}

We also quantify the relative emission of rotational lines at higher quantum rotational number with respect to the CO(1-0) transition. This is shown in Figure~\ref{fig:co_sled}, where we display the luminosity ratio $L_{\rm CO((J+1)-J)}/L_{\rm CO(1-0)}$ for different quantum numbers $J$ -- known as the CO spectral line energy distribution (SLED) -- and for all the runs studied herein. We also report in the same plot the observational results for the CO SLED from cosmic noon galaxies \citep{Bothwell2013,Spilker2014,Daddi2015,Boogaard2020}. We notice that the predicted curves from our model are in reasonable agreement with the observed galaxies, and well within the scatter of the observed CO SLED for galaxies at $1.5<z<3.5$ \citep[see also, e.g.][]{Vallini2018}. The only galaxy with higher CO SLED than the rest of the sample is run 27, consistently with the fact that disc fragmentation, and hence, molecular gas formation, are favored due to the large gas fraction.

%-------------------------------------------------------------------
%-------------------------------------------------------------------
%-------------------------------------------------------------------

\section{Time evolution}\label{sec:time_evolution}

\begin{figure*}
    \centering  
    \includegraphics[width=\textwidth]{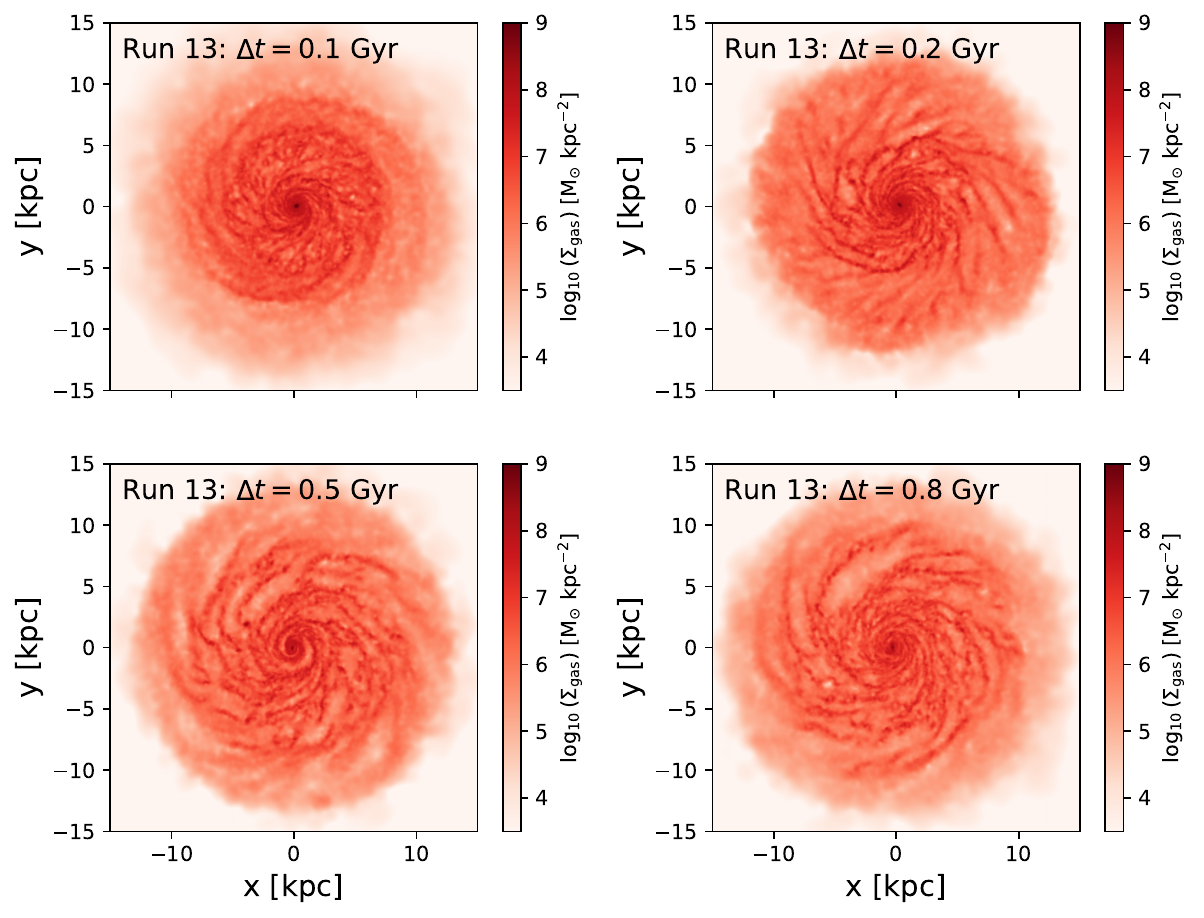}
    \caption{Gas mass surface density maps, viewed face-on, for run~13, after $\Delta t=0.1 \, {\rm Gyr}$ (top left), $\Delta t=0.2 \, {\rm Gyr}$ (top right), $\Delta t=0.5 \, {\rm Gyr}$ (bottom left), and $\Delta t=0.8 \, {\rm Gyr}$ (bottom right) from the end of the relaxation phase, respectively.
    }
    \label{fig:gas_run13_time}
\end{figure*}

\begin{figure*}
    \centering  
    \includegraphics[width=\textwidth]{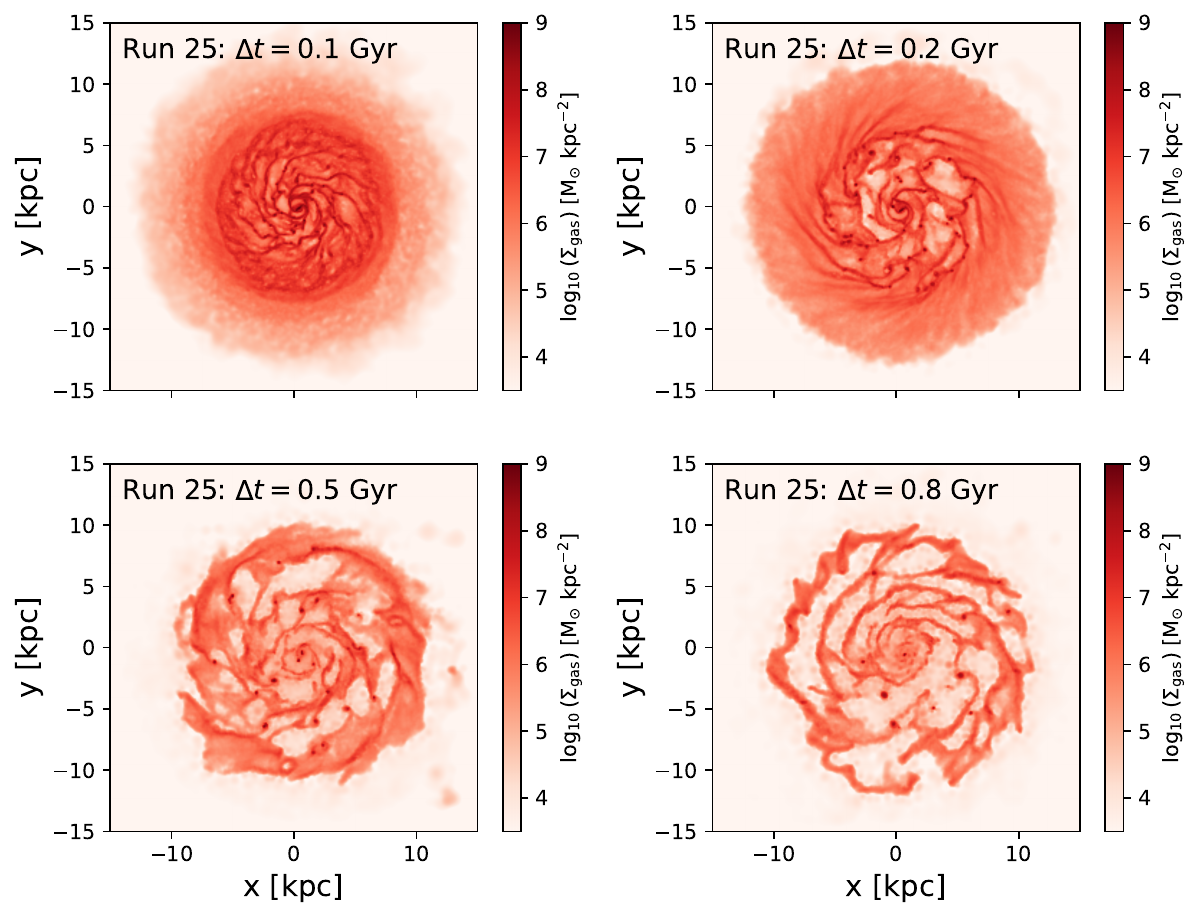}
    \caption{Gas mass surface density maps, viewed face-on, for run~25, after $\Delta t=0.1 \, {\rm Gyr}$ (top left), $\Delta t=0.2 \, {\rm Gyr}$ (top right), $\Delta t=0.5 \, {\rm Gyr}$ (bottom left), and $\Delta t=0.8 \, {\rm Gyr}$ (bottom right) from the end of the relaxation phase, respectively.}
    \label{fig:gas_run25_time}
\end{figure*}

As an additional application of the framework presented in this paper, we study the evolution over cosmic time of two simulations of our sample, namely runs~13 and 25. In particular, we analyze the predicted observables for these two galaxies at four different timesteps: $\Delta t=0.1$, $0.2$, $0.5$, and $0.8$~Gyr after the end of the relaxation phase. Figures~\ref{fig:gas_run13_time} and \ref{fig:gas_run25_time} show the gas surface density maps for runs~13 and 25, respectively, at the four aforementioned cosmic times. By performing a visual inspection, one easily notices how the disc evolves over time in both cases, acquiring a more and more structured morphology the later the time. While run~13 develops a stable spiral structure after $\Delta t=0.8$~Gyr, run~25 undergoes a quicker gas depletion and by the same time the galaxy has largely devoided its gas reservoirs due to an intense star formation activity driven by a larger gas fraction than in run~13 \citep{Tamburello2015}. In addition, the disc in run~25 develops a pronounced clumpy structure, while run~13 does not. This is due to the higher gas fraction in run~25, which favours disc fragmentation \citep[see][]{Tamburello2015}.

\begin{figure}
    \centering
    \includegraphics[width=\columnwidth]{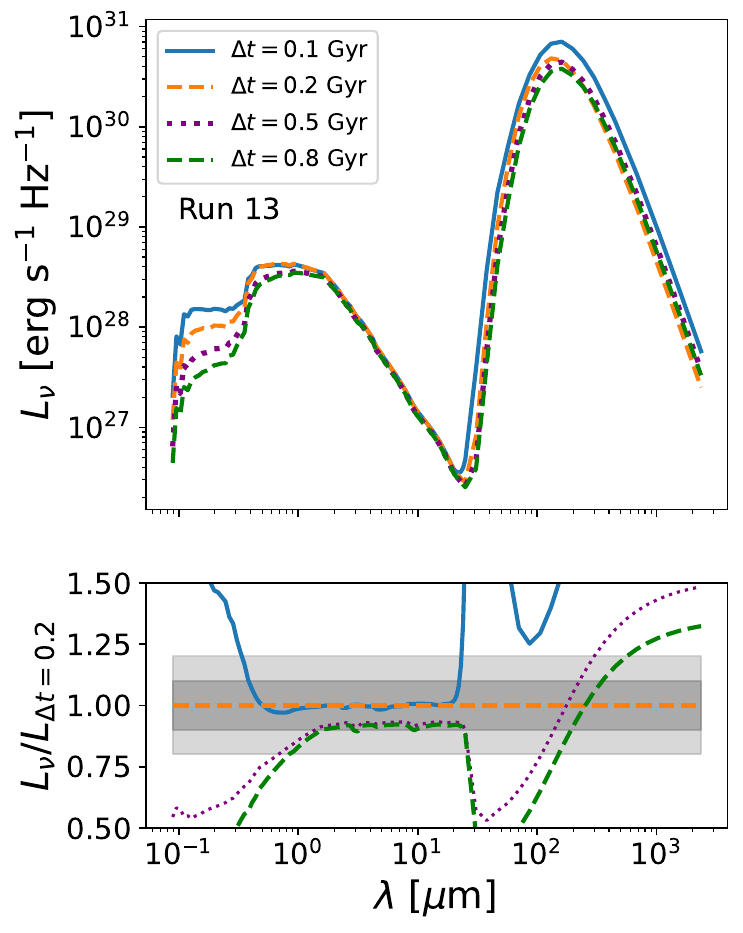}
    \caption{Top: a comparison of the SED predictions for run~13 at different time snapshots, $\Delta t=0.1$ Gyr (blue solid), $\Delta t=0.2$ Gyr (orange dashed), $\Delta t=0.5$ Gyr (purple dotted), and $\Delta t=0.8$ Gyr (green dotted-dashed) after the end of the relaxation phase, respectively. Bottom: ratios between the SED of each studied case and the baseline ($\Delta t=0.2$ Gyr).}
    \label{fig:sed_run13_over_time}
\end{figure}

\begin{figure}
    \centering
    \includegraphics[width=\columnwidth]{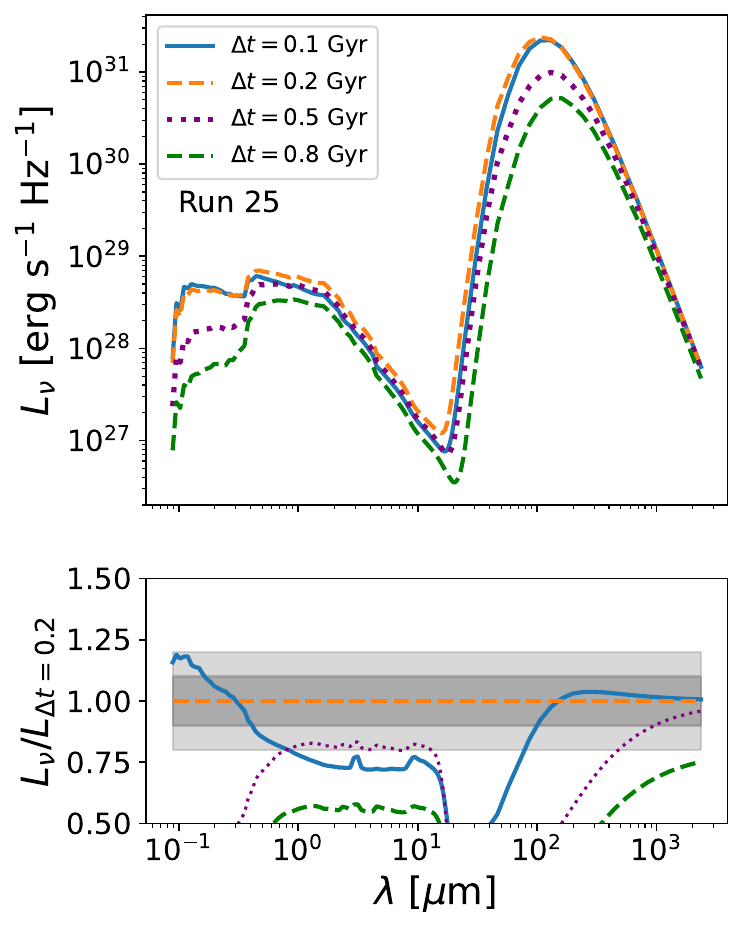}
    \caption{Top: a comparison of the SED predictions for run~25 at different time snapshots, $\Delta t=0.1$ Gyr (blue solid), $\Delta t=0.2$ Gyr (orange dashed), $\Delta t=0.5$ Gyr (purple dotted), and $\Delta t=0.8$ Gyr (green dotted-dashed) after the end of the relaxation phase, respectively. Bottom: ratios between the SED of each studied case and the baseline ($\Delta t=0.2$ Gyr).}
    \label{fig:sed_run25_over_time}
\end{figure}

Figures~\ref{fig:sed_run13_over_time} and \ref{fig:sed_run25_over_time} display the predicted SED for runs~13 and 25, respectively, following a similar structure as that of Figure~\ref{fig:sed_composition}. In particular, the bottom panel is this time normalized to the SED computed at $\Delta t=0.2$ Gyr -- that is the snapshot analyzed throughout this work -- to perform a meaningful comparison with the results presented in the remainder of this paper. Both runs~13 and 25 feature a monotonic temporal evolution of the UV region of the SED, with the both the UV and FIR emission becoming fainter with time. This is consistent with the presence of older stellar population and lower dust density -- the latter due to the fact that the dust density is proportional to the gas density in our model, and that the gas gets depletd by star formation and there is no external gas accretion in our simulations -- respectively. We argue that there may be a significant dependence on the variation of the predicted SED when clumps are formed and disrupted, which we will investigate in a forthcoming paper. We also warn the reader that this result applies in the case of the studied simulations, which are isolated discs, but they may not represent a realistic scenario in cosmological simulations including gas accretion onto galaxies.

\begin{figure}
    \centering
    \includegraphics[width=\columnwidth]{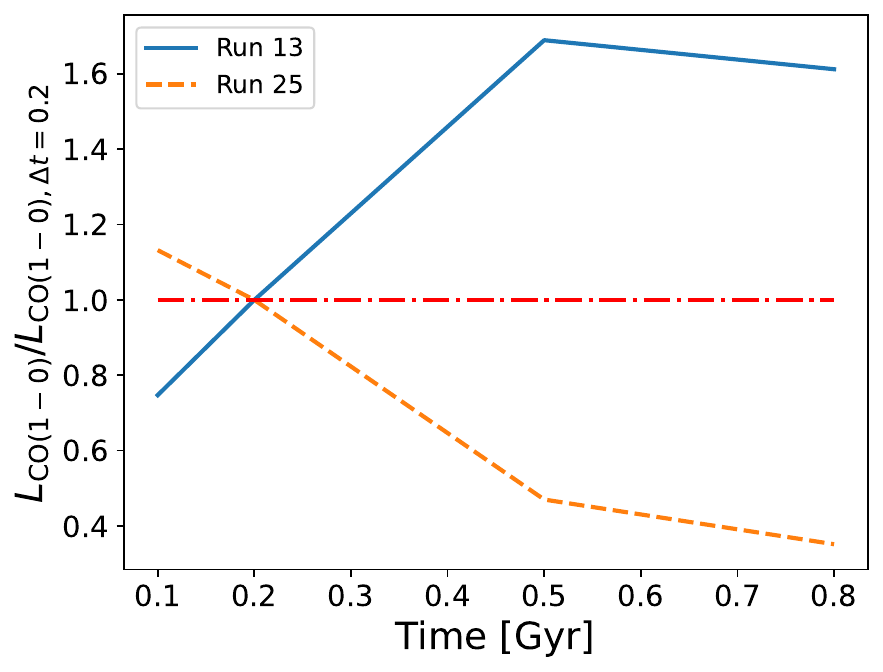}
    \caption{Total CO(1-0) luminosity at different time snapshots normalized to the total CO(1-0) luminosity $\Delta t=0.2$ Gyr after the relaxation phase, for run~13 (blue solid) and run~25 (orange dashed). The red dotted-dashed line marks unity ratios.}
    \label{fig:co_run13_run25_over_time}
\end{figure}

Finally, Figure~\ref{fig:co_run13_run25_over_time} shows the evolution over time of the total CO(1-0) luminosity for runs~13 (blue solid) and run~25 (orange dashed), respectively, normalized to the CO(1-0) luminosity computed after $\Delta t = 0.2$ Gyr (i.e. the baseline analysis). The resulting curves highlight that the CO(1-0) luminosity increases over time in run~13, reaching a maximum at $\Delta t=0.5$ Gyr, and then decreases towards later time. This hints towards the fact that this galaxy undergoes a first phase of molecular cloud formation, which favours the CO emission, while after $\Delta t=0.5$ Gyr gas depletion due to star formation and feedback starts to dominate causing a decrease of CO luminosity. On the contrary, the evolution of run~25 is different, as the CO luminosity decreases monotonically over time. This fact is consistent with what is pointed out earlier, that is the galaxy disc evolved much more quickly and depletes gas more efficiently (possibly because of clump formation)  than what occurs in run~13. 

%-------------------------------------------------------------------
%-------------------------------------------------------------------
%-------------------------------------------------------------------

\section{Summary and conclusions}\label{sec:conclusions}

This work presents the implementation and validation of an end-to-end pipeline -- dubbed \texttt{RTGen} -- based on the \texttt{RADMC-3D} code \citep{Dullemond2012} and aimed at performing a-posteriori continuum and spectral line radiative transfer simulations of hydrodynamic simulations of galaxies. We summarize in what follows the main steps and features of the pipeline, as well as the main results and conclusions.

The implemented framework relies on a Monte Carlo radiative transfer algorithm that first determines the dust temperature and then ray-traces photons at a given wavelength, solving for the continuum or line transfer equation, depending on the requested setup. Given that \texttt{RADMC-3D} is model-agnostic and only runs the radiative transfer computation once the physical problem is fully specified, the pipeline implements a detailed modelling of the gas, dust, and stellar density leveraging state-of-the-art public tools, models, and observational results from the literature. In particular, we first interpolate the gas and star particles onto a grid using an SPH kernel and a CIC scheme, respectively. Afterwards, we model the input stellar spectrum for each of the studied galaxies as sum of the spectra of the different stellar particles, computed by relying on the \texttt{STARBURST99} stellar population synthesis code \citep{Leitherer1999}. Since the dust formation, growth, and destruction are not tracked in the hydrodynamic simulations used in this paper, we predict the dust abundance from the gas density field by applying a scaling relation linking the DGR to the gas metallicity, measured from the \texttt{SIMBA} simulation and found to agree well with state-of-the-art observational results \citep{Li2019}. After running the Monte Carlo dust radiative transfer and the ray tracing to obtain continuum images, we obtain the radiation field self-consistently in our pipeline from radiative transfer at $\lambda = 0.1\, \mu$m and use it to compute the atomic-to-molecular transition presented in \citet{GnedinKravtsov2011} and calibrated onto high-resolution hydrodynamic simulation. This allows us to obtain predictions for the atomic (\HI{}) and molecular (H$_2$) density. Eventually, we use the latter to compute the CO abundance and predict the CO luminosity via line radiative transfer computation. 

Within our dust modelling, we test the assumptions behind dust composition and grain size. In particular, we find that assuming dust composition of approximately similar proportion of silicate and carbonaceous grains yields a reasonable convergence in the predicted SEDs, while one observes significant deviations when the dust is assumed to be composed just by silicate grains. When looking into the results from different models for the dust grain size, we find a good convergence (within $<10\%$ deviations) of the predicted SEDs  at $\lambda\lesssim 10 ~\mu{\rm m}$ for all the studied cases, whereas the SEDs tend to diverge at longer wavelengths. This advocates for a careful study and testing of the adopted model on a case-by-case basis.

The atomic/molecular splitting yields a skewed distribution for the molecular fraction, $f_{\rm mol}$, which is found to have a strong dependence on resolution and on the gas fraction. In addition, the predicted spatial distribution with galactocentric distance of \HI{} and H$_2$ is in good qualitative agreement with well-studied cases in the local Universe, where the latter is found to be confined in a small disc at the centre of the galaxy, whereas the former forms a disc which extends out to much larger radii \citep[e.g.][]{Walter2008,Leroy2008}.

Eventually, we predict the CO luminosity by calibrating the H$_2$-to-CO abundance and forward-modelling the radiative transfer, assuming the Milky Way CO-to-H$_2$ conversion factor $\alpha_{\rm CO}$ \citep[e.g.][]{Bolatto2013}. The predicted CO luminosity images show greatly enhanced morphological structure in the galaxies, and in particular highlight the clumpy structure which will be investigated in detail in a forthcoming paper.

We notice that, while we have thoroughly addressed the impact of the dust modelling, we will perform a systematic investigation of the uncertainties associated with the atomic-to-molecular transition and on the H$_2$-to-CO conversion in future work.

We envision that this pipeline can be used for a variety of studies of galaxy formation and evolution, allowing us to study the impact of several different aspects of galaxy formation and feedback models on typical galaxy observables such as the SEDs and multi-wavelength images. In particular, the mock images produced within the \texttt{RTGen} pipeline can be postprocessed by applying JWST filters to continuum images and ALMA systematics (array configuration and noise, among others) to CO images using \texttt{CASA} \citep{CASA2022}. In a forthcoming publication, we will use these techniques to perform a comparative study of the nature of massive clumps in high-redshift galaxies, as was done by \citet{Tamburello2017} leveraging the H$\alpha$ line. The code implementing the models presented in this work will be made publicly available upon the acceptance of the paper.\footnote{\url{https://github.com/francescosinigaglia/RTGen-public}}

%-------------------------------------------------------------------
%-------------------------------------------------------------------
%-------------------------------------------------------------------

\begin{acknowledgements}
      The authors warmly thank the anonymous referee, whose insightful comments helped improving the quality of the manuscript.
      The authors wish to also acknowledge Cornelis Dullemond -- the author of the \texttt{RADMC-3D} code -- for making the code available and for assistance on the usage of the software. FS acknowledges the support of the Swiss National Science Foundation (SNSF) 200021\_214990/1 grant. FS is also grateful to Dr Gaia Lacedelli, for availability of computing resources. LM acknowledges support from the SNSF under the Grant 200020\_207406. PRC acknowledges support from the SNSF under the Sinergia Grant CRSII5\_213497 (GW-Learn).
\end{acknowledgements}


%\begin{comment}

\break 
\clearpage

%\begin{thebibliography}{}
\bibliographystyle{aa}
\bibliography{lit}

\begin{thebibliography}{141}
\expandafter\ifx\csname natexlab\endcsname\relax\def\natexlab#1{#1}\fi

\bibitem[{{Allen} {et~al.}(????){Allen}, {Groves}, {Dopita}, {Sutherland}, \& {Kewley}}]{Groves2008}
{Allen}, M.~G., {Groves}, B.~A., {Dopita}, M.~A., {Sutherland}, R.~S., \& {Kewley}, L.~J. ????, \apjs, 178, 20

\bibitem[{{Aoyama} {et~al.}(2017){Aoyama}, {Hou}, {Shimizu}, {Hirashita}, {Todoroki}, {Choi}, \& {Nagamine}}]{Aoyama2017}
{Aoyama}, S., {Hou}, K.-C., {Shimizu}, I., {et~al.} 2017, \mnras, 466, 105

\bibitem[{{Baes} \& {Camps}(2015)}]{Baes2015}
{Baes}, M. \& {Camps}, P. 2015, Astronomy and Computing, 12, 33

\bibitem[{{Bekki}(2015)}]{Bekki2015}
{Bekki}, K. 2015, \mnras, 449, 1625

\bibitem[{{Bialy} {et~al.}(2017){Bialy}, {Burkhart}, \& {Sternberg}}]{Bialy2017}
{Bialy}, S., {Burkhart}, B., \& {Sternberg}, A. 2017, \apj, 843, 92

\bibitem[{{Bjorkman} \& {Wood}(2001)}]{BjorkmanWood2001}
{Bjorkman}, J.~E. \& {Wood}, K. 2001, \apj, 554, 615

\bibitem[{{Blitz} \& {Rosolowsky}(2004)}]{Blitz2004}
{Blitz}, L. \& {Rosolowsky}, E. 2004, \apjl, 612, L29

\bibitem[{{Blitz} \& {Rosolowsky}(2006)}]{Blitz2006}
{Blitz}, L. \& {Rosolowsky}, E. 2006, \apj, 650, 933

\bibitem[{{Bohren} \& {Huffman}(1983)}]{BohrenHuffman1983}
{Bohren}, C.~F. \& {Huffman}, D.~R. 1983, {Absorption and scattering of light by small particles}, Research supported by the University of Arizona and Institute of Occupational and Environmental Health. New York, Wiley-Interscience, 1983, 541 p.

\bibitem[{{Bolatto} {et~al.}(2013){Bolatto}, {Wolfire}, \& {Leroy}}]{Bolatto2013}
{Bolatto}, A.~D., {Wolfire}, M., \& {Leroy}, A.~K. 2013, \araa, 51, 207

\bibitem[{{Boogaard} {et~al.}(2020){Boogaard}, {van der Werf}, {Weiss}, {Popping}, {Decarli}, {Walter}, {Aravena}, {Bouwens}, {Riechers}, {Gonz{\'a}lez-L{\'o}pez}, {Smail}, {Carilli}, {Kaasinen}, {Daddi}, {Cox}, {D{\'\i}az-Santos}, {Inami}, {Cortes}, \& {Wagg}}]{Boogaard2020}
{Boogaard}, L.~A., {van der Werf}, P., {Weiss}, A., {et~al.} 2020, \apj, 902, 109

\bibitem[{{Bothwell} {et~al.}(2013){Bothwell}, {Smail}, {Chapman}, {Genzel}, {Ivison}, {Tacconi}, {Alaghband-Zadeh}, {Bertoldi}, {Blain}, {Casey}, {Cox}, {Greve}, {Lutz}, {Neri}, {Omont}, \& {Swinbank}}]{Bothwell2013}
{Bothwell}, M.~S., {Smail}, I., {Chapman}, S.~C., {et~al.} 2013, \mnras, 429, 3047

\bibitem[{{Bressan} {et~al.}(1993){Bressan}, {Fagotto}, {Bertelli}, \& {Chiosi}}]{Bressan1993}
{Bressan}, A., {Fagotto}, F., {Bertelli}, G., \& {Chiosi}, C. 1993, \aaps, 100, 647

\bibitem[{{Bressan} {et~al.}(2012){Bressan}, {Marigo}, {Girardi}, {Salasnich}, {Dal Cero}, {Rubele}, \& {Nanni}}]{Bressan2012}
{Bressan}, A., {Marigo}, P., {Girardi}, L., {et~al.} 2012, \mnras, 427, 127

\bibitem[{{Bullock} {et~al.}(2001){Bullock}, {Kolatt}, {Sigad}, {Somerville}, {Kravtsov}, {Klypin}, {Primack}, \& {Dekel}}]{Bullock2001}
{Bullock}, J.~S., {Kolatt}, T.~S., {Sigad}, Y., {et~al.} 2001, \mnras, 321, 559

\bibitem[{{Camps} \& {Baes}(2015)}]{Camps2015}
{Camps}, P. \& {Baes}, M. 2015, Astronomy and Computing, 9, 20

\bibitem[{{Camps} \& {Baes}(2020)}]{Camps2020}
{Camps}, P. \& {Baes}, M. 2020, Astronomy and Computing, 31, 100381

\bibitem[{{Capelo} {et~al.}(2018){Capelo}, {Bovino}, {Lupi}, {Schleicher}, \& {Grassi}}]{Capelo2018}
{Capelo}, P.~R., {Bovino}, S., {Lupi}, A., {Schleicher}, D. R.~G., \& {Grassi}, T. 2018, \mnras, 475, 3283

\bibitem[{{CASA Team} {et~al.}(2022){CASA Team}, {Bean}, {Bhatnagar}, {Castro}, {Donovan Meyer}, {Emonts}, {Garcia}, {Garwood}, {Golap}, {Gonzalez Villalba}, {Harris}, {Hayashi}, {Hoskins}, {Hsieh}, {Jagannathan}, {Kawasaki}, {Keimpema}, {Kettenis}, {Lopez}, {Marvil}, {Masters}, {McNichols}, {Mehringer}, {Miel}, {Moellenbrock}, {Montesino}, {Nakazato}, {Ott}, {Petry}, {Pokorny}, {Raba}, {Rau}, {Schiebel}, {Schweighart}, {Sekhar}, {Shimada}, {Small}, {Steeb}, {Sugimoto}, {Suoranta}, {Tsutsumi}, {van Bemmel}, {Verkouter}, {Wells}, {Xiong}, {Szomoru}, {Griffith}, {Glendenning}, \& {Kern}}]{CASA2022}
{CASA Team}, {Bean}, B., {Bhatnagar}, S., {et~al.} 2022, \pasp, 134, 114501

\bibitem[{{Casey}(2012)}]{Casey2012}
{Casey}, C.~M. 2012, \mnras, 425, 3094

\bibitem[{{Casey} {et~al.}(2014){Casey}, {Narayanan}, \& {Cooray}}]{Casey2014}
{Casey}, C.~M., {Narayanan}, D., \& {Cooray}, A. 2014, \physrep, 541, 45

\bibitem[{{Crain} \& {van de Voort}(2023)}]{Crain2023}
{Crain}, R.~A. \& {van de Voort}, F. 2023, \araa, 61, 473

\bibitem[{{da Cunha} {et~al.}(2013){da Cunha}, {Groves}, {Walter}, {Decarli}, {Weiss}, {Bertoldi}, {Carilli}, {Daddi}, {Elbaz}, {Ivison}, {Maiolino}, {Riechers}, {Rix}, {Sargent}, \& {Smail}}]{DaCunha2013}
{da Cunha}, E., {Groves}, B., {Walter}, F., {et~al.} 2013, \apj, 766, 13

\bibitem[{{Daddi} {et~al.}(2015){Daddi}, {Dannerbauer}, {Liu}, {Aravena}, {Bournaud}, {Walter}, {Riechers}, {Magdis}, {Sargent}, {B{\'e}thermin}, {Carilli}, {Cibinel}, {Dickinson}, {Elbaz}, {Gao}, {Gobat}, {Hodge}, \& {Krips}}]{Daddi2015}
{Daddi}, E., {Dannerbauer}, H., {Liu}, D., {et~al.} 2015, \aap, 577, A46

\bibitem[{{Dav{\'e}} {et~al.}(2019){Dav{\'e}}, {Angl{\'e}s-Alc{\'a}zar}, {Narayanan}, {Li}, {Rafieferantsoa}, \& {Appleby}}]{Dave2019}
{Dav{\'e}}, R., {Angl{\'e}s-Alc{\'a}zar}, D., {Narayanan}, D., {et~al.} 2019, \mnras, 486, 2827

\bibitem[{{Dav{\'e}} {et~al.}(2016){Dav{\'e}}, {Thompson}, \& {Hopkins}}]{Dave2016}
{Dav{\'e}}, R., {Thompson}, R., \& {Hopkins}, P.~F. 2016, \mnras, 462, 3265

\bibitem[{{De Vis} {et~al.}(2019){De Vis}, {Jones}, {Viaene}, {Casasola}, {Clark}, {Baes}, {Bianchi}, {Cassara}, {Davies}, {De Looze}, {Galametz}, {Galliano}, {Lianou}, {Madden}, {Manilla-Robles}, {Mosenkov}, {Nersesian}, {Roychowdhury}, {Xilouris}, \& {Ysard}}]{DeVis2019}
{De Vis}, P., {Jones}, A., {Viaene}, S., {et~al.} 2019, \aap, 623, A5

\bibitem[{{Dell'Agli} {et~al.}(2023){Dell'Agli}, {Tosi}, {Kamath}, {Stanghellini}, {Bianchi}, {Ventura}, {Marini}, \& {Garc{\'\i}a-Hern{\'a}ndez}}]{DellAgli2023}
{Dell'Agli}, F., {Tosi}, S., {Kamath}, D., {et~al.} 2023, \mnras, 526, 5386

\bibitem[{{den Brok} {et~al.}(2023){den Brok}, {Bigiel}, {Chastenet}, {Sandstrom}, {Leroy}, {Usero}, {Schinnerer}, {Rosolowsky}, {Koch}, {Chiang}, {Barnes}, {Puschnig}, {Saito}, {Be{\v{s}}li{\'c}}, {Chevance}, {Dale}, {Eibensteiner}, {Glover}, {Jim{\'e}nez-Donaire}, {Teng}, \& {Williams}}]{DenBrok2023}
{den Brok}, J.~S., {Bigiel}, F., {Chastenet}, J., {et~al.} 2023, \aap, 676, A93

\bibitem[{{Dessauges-Zavadsky} {et~al.}(2023){Dessauges-Zavadsky}, {Richard}, {Combes}, {Messa}, {Nagy}, {Mayer}, {Schaerer}, {Egami}, \& {Adamo}}]{Dessauges2023}
{Dessauges-Zavadsky}, M., {Richard}, J., {Combes}, F., {et~al.} 2023, \mnras, 519, 6222

\bibitem[{{Dessauges-Zavadsky} {et~al.}(2019){Dessauges-Zavadsky}, {Richard}, {Combes}, {Schaerer}, {Rujopakarn}, {Mayer}, {Cava}, {Boone}, {Egami}, {Kneib}, {P{\'e}rez-Gonz{\'a}lez}, {Pfenniger}, {Rawle}, {Teyssier}, \& {van der Werf}}]{Dessauges2019}
{Dessauges-Zavadsky}, M., {Richard}, J., {Combes}, F., {et~al.} 2019, Nature Astronomy, 3, 1115

\bibitem[{{Dessauges-Zavadsky} {et~al.}(2015){Dessauges-Zavadsky}, {Zamojski}, {Schaerer}, {Combes}, {Egami}, {Swinbank}, {Richard}, {Sklias}, {Rawle}, {Rex}, {Kneib}, {Boone}, \& {Blain}}]{Dessauges2015}
{Dessauges-Zavadsky}, M., {Zamojski}, M., {Schaerer}, D., {et~al.} 2015, \aap, 577, A50

\bibitem[{{Dewdney} {et~al.}(2009){Dewdney}, {Hall}, {Schilizzi}, \& {Lazio}}]{Dewdney2009}
{Dewdney}, P.~E., {Hall}, P.~J., {Schilizzi}, R.~T., \& {Lazio}, T.~J.~L.~W. 2009, IEEE Proceedings, 97, 1482

\bibitem[{{Diemer} {et~al.}(2018){Diemer}, {Stevens}, {Forbes}, {Marinacci}, {Hernquist}, {Lagos}, {Sternberg}, {Pillepich}, {Nelson}, {Popping}, {Villaescusa-Navarro}, {Torrey}, \& {Vogelsberger}}]{Diemer2018}
{Diemer}, B., {Stevens}, A. R.~H., {Forbes}, J.~C., {et~al.} 2018, \apjs, 238, 33

\bibitem[{{Draine}(1978)}]{Draine1978}
{Draine}, B.~T. 1978, \apjs, 36, 595

\bibitem[{{Draine}(2003)}]{Draine2003}
{Draine}, B.~T. 2003, \araa, 41, 241

\bibitem[{{Draine} \& {Bertoldi}(1996)}]{Draine1996}
{Draine}, B.~T. \& {Bertoldi}, F. 1996, \apj, 468, 269

\bibitem[{{Draine} \& {Lee}(1984)}]{DraineLee1984}
{Draine}, B.~T. \& {Lee}, H.~M. 1984, \apj, 285, 89

\bibitem[{{Dubois} {et~al.}(2024){Dubois}, {Rodr{\'\i}guez Montero}, {Guerra}, {Trebitsch}, {Han}, {Beckmann}, {Yi}, {Lewis}, \& {Jang}}]{Dubois2024}
{Dubois}, Y., {Rodr{\'\i}guez Montero}, F., {Guerra}, C., {et~al.} 2024, \aap, 687, A240

\bibitem[{{Dullemond} {et~al.}(2012){Dullemond}, {Juhasz}, {Pohl}, {Sereshti}, {Shetty}, {Peters}, {Commercon}, \& {Flock}}]{Dullemond2012}
{Dullemond}, C.~P., {Juhasz}, A., {Pohl}, A., {et~al.} 2012, {RADMC-3D: A multi-purpose radiative transfer tool}, Astrophysics Source Code Library, record ascl:1202.015

\bibitem[{{Esmerian} \& {Gnedin}(2022)}]{Esmerian2022}
{Esmerian}, C.~J. \& {Gnedin}, N.~Y. 2022, \apj, 940, 74

\bibitem[{{Feldmann} \& {Mayer}(2015)}]{Feldmann2015}
{Feldmann}, R. \& {Mayer}, L. 2015, \mnras, 446, 1939

\bibitem[{{Ferland} {et~al.}(2017){Ferland}, {Chatzikos}, {Guzm{\'a}n}, {Lykins}, {van Hoof}, {Williams}, {Abel}, {Badnell}, {Keenan}, {Porter}, \& {Stancil}}]{Ferland2017}
{Ferland}, G.~J., {Chatzikos}, M., {Guzm{\'a}n}, F., {et~al.} 2017, \rmxaa, 53, 385

\bibitem[{{Ferland} {et~al.}(1998){Ferland}, {Korista}, {Verner}, {Ferguson}, {Kingdon}, \& {Verner}}]{Ferland1998}
{Ferland}, G.~J., {Korista}, K.~T., {Verner}, D.~A., {et~al.} 1998, \pasp, 110, 761

\bibitem[{{Ferrarotti} \& {Gail}(2006)}]{Ferrarotti2006}
{Ferrarotti}, A.~S. \& {Gail}, H.~P. 2006, \aap, 447, 553

\bibitem[{{Fiacconi} {et~al.}(2015){Fiacconi}, {Feldmann}, \& {Mayer}}]{Fiacconi2015}
{Fiacconi}, D., {Feldmann}, R., \& {Mayer}, L. 2015, \mnras, 446, 1957

\bibitem[{{Gall} \& {Hjorth}(2018)}]{Gall2018}
{Gall}, C. \& {Hjorth}, J. 2018, \apj, 868, 62

\bibitem[{{Gardner} {et~al.}(2006){Gardner}, {Mather}, {Clampin}, {Doyon}, {Greenhouse}, {Hammel}, {Hutchings}, {Jakobsen}, {Lilly}, {Long}, {Lunine}, {McCaughrean}, {Mountain}, {Nella}, {Rieke}, {Rieke}, {Rix}, {Smith}, {Sonneborn}, {Stiavelli}, {Stockman}, {Windhorst}, \& {Wright}}]{Gardner2006}
{Gardner}, J.~P., {Mather}, J.~C., {Clampin}, M., {et~al.} 2006, \ssr, 123, 485

\bibitem[{{Gebek} {et~al.}(2023){Gebek}, {Baes}, {Diemer}, {de Blok}, {Nelson}, {Kapoor}, {Camps}, {Rabyang}, \& {Leeuw}}]{Gebek2023}
{Gebek}, A., {Baes}, M., {Diemer}, B., {et~al.} 2023, \mnras, 521, 5645

\bibitem[{{Gnedin} \& {Abel}(2001)}]{Gnedin2001}
{Gnedin}, N.~Y. \& {Abel}, T. 2001, \na, 6, 437

\bibitem[{{Gnedin} \& {Draine}(2014)}]{GnedinDraine2014}
{Gnedin}, N.~Y. \& {Draine}, B.~T. 2014, \apj, 795, 37

\bibitem[{{Gnedin} \& {Kravtsov}(2010)}]{Gnedin2010}
{Gnedin}, N.~Y. \& {Kravtsov}, A.~V. 2010, \apj, 714, 287

\bibitem[{{Gnedin} \& {Kravtsov}(2011)}]{GnedinKravtsov2011}
{Gnedin}, N.~Y. \& {Kravtsov}, A.~V. 2011, \apj, 728, 88

\bibitem[{{Gnedin} {et~al.}(2008){Gnedin}, {Kravtsov}, \& {Chen}}]{Gnedin2008}
{Gnedin}, N.~Y., {Kravtsov}, A.~V., \& {Chen}, H.-W. 2008, \apj, 672, 765

\bibitem[{{Gnedin} {et~al.}(2009){Gnedin}, {Tassis}, \& {Kravtsov}}]{Gnedin2009}
{Gnedin}, N.~Y., {Tassis}, K., \& {Kravtsov}, A.~V. 2009, \apj, 697, 55

\bibitem[{{Goldman} {et~al.}(2022){Goldman}, {Boyer}, {Dalcanton}, {McDonald}, {Girardi}, {Williams}, {Srinivasan}, \& {Gordon}}]{Goldman2022}
{Goldman}, S.~R., {Boyer}, M.~L., {Dalcanton}, J., {et~al.} 2022, \apjs, 259, 41

\bibitem[{{Griffin} {et~al.}(2010){Griffin}, {Abergel}, {Abreu}, {Ade}, {Andr{\'e}}, {Augueres}, {Babbedge}, {Bae}, {Baillie}, {Baluteau}, {Barlow}, {Bendo}, {Benielli}, {Bock}, {Bonhomme}, {Brisbin}, {Brockley-Blatt}, {Caldwell}, {Cara}, {Castro-Rodriguez}, {Cerulli}, {Chanial}, {Chen}, {Clark}, {Clements}, {Clerc}, {Coker}, {Communal}, {Conversi}, {Cox}, {Crumb}, {Cunningham}, {Daly}, {Davis}, {de Antoni}, {Delderfield}, {Devin}, {di Giorgio}, {Didschuns}, {Dohlen}, {Donati}, {Dowell}, {Dowell}, {Duband}, {Dumaye}, {Emery}, {Ferlet}, {Ferrand}, {Fontignie}, {Fox}, {Franceschini}, {Frerking}, {Fulton}, {Garcia}, {Gastaud}, {Gear}, {Glenn}, {Goizel}, {Griffin}, {Grundy}, {Guest}, {Guillemet}, {Hargrave}, {Harwit}, {Hastings}, {Hatziminaoglou}, {Herman}, {Hinde}, {Hristov}, {Huang}, {Imhof}, {Isaak}, {Israelsson}, {Ivison}, {Jennings}, {Kiernan}, {King}, {Lange}, {Latter}, {Laurent}, {Laurent}, {Leeks}, {Lellouch}, {Levenson}, {Li}, {Li}, {Lilienthal}, {Lim}, {Liu}, {Lu}, {Madden}, {Mainetti}, {Marliani},
  {McKay}, {Mercier}, {Molinari}, {Morris}, {Moseley}, {Mulder}, {Mur}, {Naylor}, {Nguyen}, {O'Halloran}, {Oliver}, {Olofsson}, {Olofsson}, {Orfei}, {Page}, {Pain}, {Panuzzo}, {Papageorgiou}, {Parks}, {Parr-Burman}, {Pearce}, {Pearson}, {P{\'e}rez-Fournon}, {Pinsard}, {Pisano}, {Podosek}, {Pohlen}, {Polehampton}, {Pouliquen}, {Rigopoulou}, {Rizzo}, {Roseboom}, {Roussel}, {Rowan-Robinson}, {Rownd}, {Saraceno}, {Sauvage}, {Savage}, {Savini}, {Sawyer}, {Scharmberg}, {Schmitt}, {Schneider}, {Schulz}, {Schwartz}, {Shafer}, {Shupe}, {Sibthorpe}, {Sidher}, {Smith}, {Smith}, {Smith}, {Spencer}, {Stobie}, {Sudiwala}, {Sukhatme}, {Surace}, {Stevens}, {Swinyard}, {Trichas}, {Tourette}, {Triou}, {Tseng}, {Tucker}, {Turner}, {Vaccari}, {Valtchanov}, {Vigroux}, {Virique}, {Voellmer}, {Walker}, {Ward}, {Waskett}, {Weilert}, {Wesson}, {White}, {Whitehouse}, {Wilson}, {Winter}, {Woodcraft}, {Wright}, {Xu}, {Zavagno}, {Zemcov}, {Zhang}, \& {Zonca}}]{SPIRE2010}
{Griffin}, M.~J., {Abergel}, A., {Abreu}, A., {et~al.} 2010, \aap, 518, L3

\bibitem[{{Henyey} \& {Greenstein}(1941)}]{HenveyGreestein1941}
{Henyey}, L.~G. \& {Greenstein}, J.~L. 1941, \apj, 93, 70

\bibitem[{{Hernquist}(1993)}]{Hernquist1993}
{Hernquist}, L. 1993, \apjs, 86, 389

\bibitem[{{Hirashita}(2015)}]{Hirashita2015}
{Hirashita}, H. 2015, \mnras, 447, 2937

\bibitem[{{Hirashita} \& {Kobayashi}(2013)}]{Hirashita2013}
{Hirashita}, H. \& {Kobayashi}, H. 2013, Earth, Planets and Space, 65, 1083

\bibitem[{{Hopkins} \& {Lee}(2016)}]{Hopkins2016}
{Hopkins}, P.~F. \& {Lee}, H. 2016, \mnras, 456, 4174

\bibitem[{{Jones} {et~al.}(2017){Jones}, {K{\"o}hler}, {Ysard}, {Bocchio}, \& {Verstraete}}]{Jones2017}
{Jones}, A.~P., {K{\"o}hler}, M., {Ysard}, N., {Bocchio}, M., \& {Verstraete}, L. 2017, \aap, 602, A46

\bibitem[{{Jonsson} {et~al.}(2010){Jonsson}, {Groves}, \& {Cox}}]{Jonsson2010}
{Jonsson}, P., {Groves}, B.~A., \& {Cox}, T.~J. 2010, \mnras, 403, 17

\bibitem[{{Kannan} {et~al.}(2022){Kannan}, {Garaldi}, {Smith}, {Pakmor}, {Springel}, {Vogelsberger}, \& {Hernquist}}]{Kannan2022}
{Kannan}, R., {Garaldi}, E., {Smith}, A., {et~al.} 2022, \mnras, 511, 4005

\bibitem[{{Kapoor} {et~al.}(2023){Kapoor}, {Baes}, {van der Wel}, {Gebek}, {Camps}, {Nersesian}, {Meidt}, {Smith}, {Vicens}, {D'Eugenio}, {Martorano}, {Barrientos}, \& {Sartorio}}]{Kapoor2023}
{Kapoor}, A.~U., {Baes}, M., {van der Wel}, A., {et~al.} 2023, \mnras, 526, 3871

\bibitem[{{Khatri} {et~al.}(2024){Khatri}, {Porciani}, {Romano-D{\'\i}az}, {Seifried}, \& {Sch{\"a}be}}]{Khatri2024}
{Khatri}, P., {Porciani}, C., {Romano-D{\'\i}az}, E., {Seifried}, D., \& {Sch{\"a}be}, A. 2024, \aap, 688, A194

\bibitem[{{Kroupa}(2001)}]{Kroupa2001}
{Kroupa}, P. 2001, \mnras, 322, 231

\bibitem[{{Krumholz}(2013)}]{Krumholz2013}
{Krumholz}, M.~R. 2013, \mnras, 436, 2747

\bibitem[{{Krumholz}(2014)}]{Krumholz2014}
{Krumholz}, M.~R. 2014, \mnras, 437, 1662

\bibitem[{{Leitherer} {et~al.}(1999){Leitherer}, {Schaerer}, {Goldader}, {Delgado}, {Robert}, {Kune}, {de Mello}, {Devost}, \& {Heckman}}]{Leitherer1999}
{Leitherer}, C., {Schaerer}, D., {Goldader}, J.~D., {et~al.} 1999, \apjs, 123, 3

\bibitem[{{Leroy} {et~al.}(2008){Leroy}, {Walter}, {Brinks}, {Bigiel}, {de Blok}, {Madore}, \& {Thornley}}]{Leroy2008}
{Leroy}, A.~K., {Walter}, F., {Brinks}, E., {et~al.} 2008, \aj, 136, 2782

\bibitem[{{Li} {et~al.}(2019){Li}, {Narayanan}, \& {Dav{\'e}}}]{Li2019}
{Li}, Q., {Narayanan}, D., \& {Dav{\'e}}, R. 2019, \mnras, 490, 1425

\bibitem[{{Li} {et~al.}(2020){Li}, {Gu}, {Yajima}, {Zhu}, \& {Maji}}]{Li2020}
{Li}, Y., {Gu}, M.~F., {Yajima}, H., {Zhu}, Q., \& {Maji}, M. 2020, \mnras, 494, 1919

\bibitem[{{Liang} {et~al.}(2019){Liang}, {Feldmann}, {Kere{\v{s}}}, {Scoville}, {Hayward}, {Faucher-Gigu{\`e}re}, {Schreiber}, {Ma}, {Hopkins}, \& {Quataert}}]{Liang2019}
{Liang}, L., {Feldmann}, R., {Kere{\v{s}}}, D., {et~al.} 2019, \mnras, 489, 1397

\bibitem[{{Lucy}(1999)}]{Lucy1999}
{Lucy}, L.~B. 1999, \aap, 344, 282

\bibitem[{{Ludlow} {et~al.}(2014){Ludlow}, {Navarro}, {Angulo}, {Boylan-Kolchin}, {Springel}, {Frenk}, \& {White}}]{Ludlow2014}
{Ludlow}, A.~D., {Navarro}, J.~F., {Angulo}, R.~E., {et~al.} 2014, \mnras, 441, 378

\bibitem[{{Lupi} {et~al.}(2018){Lupi}, {Bovino}, {Capelo}, {Volonteri}, \& {Silk}}]{Lupi2018}
{Lupi}, A., {Bovino}, S., {Capelo}, P.~R., {Volonteri}, M., \& {Silk}, J. 2018, \mnras, 474, 2884

\bibitem[{{Ma} {et~al.}(2019){Ma}, {Hayward}, {Casey}, {Hopkins}, {Quataert}, {Liang}, {Faucher-Gigu{\`e}re}, {Feldmann}, \& {Kere{\v{s}}}}]{Ma2019}
{Ma}, X., {Hayward}, C.~C., {Casey}, C.~M., {et~al.} 2019, \mnras, 487, 1844

\bibitem[{{Magnelli} {et~al.}(2014){Magnelli}, {Lutz}, {Saintonge}, {Berta}, {Santini}, {Symeonidis}, {Altieri}, {Andreani}, {Aussel}, {B{\'e}thermin}, {Bock}, {Bongiovanni}, {Cepa}, {Cimatti}, {Conley}, {Daddi}, {Elbaz}, {F{\"o}rster Schreiber}, {Genzel}, {Ivison}, {Le Floc'h}, {Magdis}, {Maiolino}, {Nordon}, {Oliver}, {Page}, {P{\'e}rez Garc{\'\i}a}, {Poglitsch}, {Popesso}, {Pozzi}, {Riguccini}, {Rodighiero}, {Rosario}, {Roseboom}, {Sanchez-Portal}, {Scott}, {Sturm}, {Tacconi}, {Valtchanov}, {Wang}, \& {Wuyts}}]{Magnelli2014}
{Magnelli}, B., {Lutz}, D., {Saintonge}, A., {et~al.} 2014, \aap, 561, A86

\bibitem[{{Marigo}(2001)}]{Marigo2001}
{Marigo}, P. 2001, \aap, 370, 194

\bibitem[{{Mathis} {et~al.}(1977){Mathis}, {Rumpl}, \& {Nordsieck}}]{Mathis1977}
{Mathis}, J.~S., {Rumpl}, W., \& {Nordsieck}, K.~H. 1977, \apj, 217, 425

\bibitem[{{Matsumoto} {et~al.}(2023){Matsumoto}, {Camps}, {Baes}, {De Ceuster}, {Wada}, {Nakagawa}, \& {Nagamine}}]{Matsumoto2023}
{Matsumoto}, K., {Camps}, P., {Baes}, M., {et~al.} 2023, \aap, 678, A175

\bibitem[{{Mayer} {et~al.}(2001){Mayer}, {Governato}, {Colpi}, {Moore}, {Quinn}, {Wadsley}, {Stadel}, \& {Lake}}]{Mayer2001b}
{Mayer}, L., {Governato}, F., {Colpi}, M., {et~al.} 2001, \apjl, 547, L123

\bibitem[{{Mayer} {et~al.}(2002){Mayer}, {Moore}, {Quinn}, {Governato}, \& {Stadel}}]{Mayer2002}
{Mayer}, L., {Moore}, B., {Quinn}, T., {Governato}, F., \& {Stadel}, J. 2002, \mnras, 336, 119

\bibitem[{{Mie}(1908)}]{Mie1908}
{Mie}, G. 1908, Annalen der Physik, 330, 377

\bibitem[{{Nagy} {et~al.}(2022){Nagy}, {Dessauges-Zavadsky}, {Richard}, {Schaerer}, {Combes}, {Messa}, \& {Chisholm}}]{Nagy2022}
{Nagy}, D., {Dessauges-Zavadsky}, M., {Richard}, J., {et~al.} 2022, \aap, 657, A25

\bibitem[{{Narayanan} {et~al.}(2021){Narayanan}, {Turk}, {Robitaille}, {Kelly}, {McClellan}, {Sharma}, {Garg}, {Abruzzo}, {Choi}, {Conroy}, {Johnson}, {Kimock}, {Li}, {Lovell}, {Lower}, {Privon}, {Roberts}, {Sethuram}, {Snyder}, {Thompson}, \& {Wise}}]{Narayanan2021}
{Narayanan}, D., {Turk}, M.~J., {Robitaille}, T., {et~al.} 2021, \apjs, 252, 12

\bibitem[{{Navarro} {et~al.}(1997){Navarro}, {Frenk}, \& {White}}]{NFW1997}
{Navarro}, J.~F., {Frenk}, C.~S., \& {White}, S. D.~M. 1997, \apj, 490, 493

\bibitem[{{Noebauer} \& {Sim}(2019)}]{Noebauer2019}
{Noebauer}, U.~M. \& {Sim}, S.~A. 2019, Living Reviews in Computational Astrophysics, 5, 1

\bibitem[{{Nozawa} {et~al.}(2007){Nozawa}, {Kozasa}, {Habe}, {Dwek}, {Umeda}, {Tominaga}, {Maeda}, \& {Nomoto}}]{Nozawa2007}
{Nozawa}, T., {Kozasa}, T., {Habe}, A., {et~al.} 2007, \apj, 666, 955

\bibitem[{{Nozawa} {et~al.}(2011){Nozawa}, {Maeda}, {Kozasa}, {Tanaka}, {Nomoto}, \& {Umeda}}]{Nozawa2011}
{Nozawa}, T., {Maeda}, K., {Kozasa}, T., {et~al.} 2011, \apj, 736, 45

\bibitem[{{O'Donnell} \& {Mathis}(1997)}]{ODonnel1997}
{O'Donnell}, J.~E. \& {Mathis}, J.~S. 1997, \apj, 479, 806

\bibitem[{{Ossenkopf}(1997)}]{Ossenkopf1997}
{Ossenkopf}, V. 1997, \na, 2, 365

\bibitem[{{Ouchi} {et~al.}(2008){Ouchi}, {Shimasaku}, {Akiyama}, {Simpson}, {Saito}, {Ueda}, {Furusawa}, {Sekiguchi}, {Yamada}, {Kodama}, {Kashikawa}, {Okamura}, {Iye}, {Takata}, {Yoshida}, \& {Yoshida}}]{Ouchi2008}
{Ouchi}, M., {Shimasaku}, K., {Akiyama}, M., {et~al.} 2008, \apjs, 176, 301

\bibitem[{{Pawlik} \& {Schaye}(2008)}]{Pawlik2008}
{Pawlik}, A.~H. \& {Schaye}, J. 2008, \mnras, 389, 651

\bibitem[{{P{\'e}roux} \& {Howk}(2020)}]{Peroux2020}
{P{\'e}roux}, C. \& {Howk}, J.~C. 2020, \araa, 58, 363

\bibitem[{{Poglitsch} {et~al.}(2010){Poglitsch}, {Waelkens}, {Geis}, {Feuchtgruber}, {Vandenbussche}, {Rodriguez}, {Krause}, {Renotte}, {van Hoof}, {Saraceno}, {Cepa}, {Kerschbaum}, {Agn{\`e}se}, {Ali}, {Altieri}, {Andreani}, {Augueres}, {Balog}, {Barl}, {Bauer}, {Belbachir}, {Benedettini}, {Billot}, {Boulade}, {Bischof}, {Blommaert}, {Callut}, {Cara}, {Cerulli}, {Cesarsky}, {Contursi}, {Creten}, {De Meester}, {Doublier}, {Doumayrou}, {Duband}, {Exter}, {Genzel}, {Gillis}, {Gr{\"o}zinger}, {Henning}, {Herreros}, {Huygen}, {Inguscio}, {Jakob}, {Jamar}, {Jean}, {de Jong}, {Katterloher}, {Kiss}, {Klaas}, {Lemke}, {Lutz}, {Madden}, {Marquet}, {Martignac}, {Mazy}, {Merken}, {Montfort}, {Morbidelli}, {M{\"u}ller}, {Nielbock}, {Okumura}, {Orfei}, {Ottensamer}, {Pezzuto}, {Popesso}, {Putzeys}, {Regibo}, {Reveret}, {Royer}, {Sauvage}, {Schreiber}, {Stegmaier}, {Schmitt}, {Schubert}, {Sturm}, {Thiel}, {Tofani}, {Vavrek}, {Wetzstein}, {Wieprecht}, \& {Wiezorrek}}]{PACS2010}
{Poglitsch}, A., {Waelkens}, C., {Geis}, N., {et~al.} 2010, \aap, 518, L2

\bibitem[{{Polzin} {et~al.}(2024){Polzin}, {Kravtsov}, {Semenov}, \& {Gnedin}}]{Polzin2024}
{Polzin}, A., {Kravtsov}, A.~V., {Semenov}, V.~A., \& {Gnedin}, N.~Y. 2024, arXiv e-prints, arXiv:2407.11125

\bibitem[{{Popping} \& {P{\'e}roux}(2022)}]{Popping2022}
{Popping}, G. \& {P{\'e}roux}, C. 2022, \mnras, 513, 1531

\bibitem[{{R{\'e}my-Ruyer} {et~al.}(2014){R{\'e}my-Ruyer}, {Madden}, {Galliano}, {Galametz}, {Takeuchi}, {Asano}, {Zhukovska}, {Lebouteiller}, {Cormier}, {Jones}, {Bocchio}, {Baes}, {Bendo}, {Boquien}, {Boselli}, {DeLooze}, {Doublier-Pritchard}, {Hughes}, {Karczewski}, \& {Spinoglio}}]{RemyRuyer2014}
{R{\'e}my-Ruyer}, A., {Madden}, S.~C., {Galliano}, F., {et~al.} 2014, \aap, 563, A31

\bibitem[{{Rieke} {et~al.}(2005){Rieke}, {Kelly}, \& {Horner}}]{Rieke2005}
{Rieke}, M.~J., {Kelly}, D., \& {Horner}, S. 2005, in Society of Photo-Optical Instrumentation Engineers (SPIE) Conference Series, Vol. 5904, Cryogenic Optical Systems and Instruments XI, ed. J.~B. {Heaney} \& L.~G. {Burriesci}, 1--8

\bibitem[{{Rieke} {et~al.}(2023){Rieke}, {Kelly}, {Misselt}, {Stansberry}, {Boyer}, {Beatty}, {Egami}, {Florian}, {Greene}, {Hainline}, {Leisenring}, {Roellig}, {Schlawin}, {Sun}, {Tinnin}, {Williams}, {Willmer}, {Wilson}, {Clark}, {Rohrbach}, {Brooks}, {Canipe}, {Correnti}, {DiFelice}, {Gennaro}, {Girard}, {Hartig}, {Hilbert}, {Koekemoer}, {Nikolov}, {Pirzkal}, {Rest}, {Robberto}, {Sunnquist}, {Telfer}, {Wu}, {Ferry}, {Lewis}, {Baum}, {Beichman}, {Doyon}, {Dressler}, {Eisenstein}, {Ferrarese}, {Hodapp}, {Horner}, {Jaffe}, {Johnstone}, {Krist}, {Martin}, {McCarthy}, {Meyer}, {Rieke}, {Trauger}, \& {Young}}]{Rieke2023}
{Rieke}, M.~J., {Kelly}, D.~M., {Misselt}, K., {et~al.} 2023, \pasp, 135, 028001

\bibitem[{{Robertson} {et~al.}(2024){Robertson}, {Johnson}, {Tacchella}, {Eisenstein}, {Hainline}, {Arribas}, {Baker}, {Bunker}, {Carniani}, {Cargile}, {Carreira}, {Charlot}, {Chevallard}, {Curti}, {Curtis-Lake}, {D'Eugenio}, {Egami}, {Hausen}, {Helton}, {Jakobsen}, {Ji}, {Jones}, {Maiolino}, {Maseda}, {Nelson}, {P{\'e}rez-Gonz{\'a}lez}, {Pusk{\'a}s}, {Rieke}, {Smit}, {Sun}, {{\"U}bler}, {Whitler}, {Williams}, {Willmer}, {Willott}, \& {Witstok}}]{Robertson2024}
{Robertson}, B., {Johnson}, B.~D., {Tacchella}, S., {et~al.} 2024, \apj, 970, 31

\bibitem[{{Robertson} \& {Kravtsov}(2008)}]{Robertson2008}
{Robertson}, B.~E. \& {Kravtsov}, A.~V. 2008, \apj, 680, 1083

\bibitem[{{Rosdahl} {et~al.}(2018){Rosdahl}, {Katz}, {Blaizot}, {Kimm}, {Michel-Dansac}, {Garel}, {Haehnelt}, {Ocvirk}, \& {Teyssier}}]{Rosdahl2018}
{Rosdahl}, J., {Katz}, H., {Blaizot}, J., {et~al.} 2018, \mnras, 479, 994

\bibitem[{{Sarangi} \& {Cherchneff}(2015)}]{Sarangi2015}
{Sarangi}, A. \& {Cherchneff}, I. 2015, \aap, 575, A95

\bibitem[{{Schneider} {et~al.}(2014){Schneider}, {Valiante}, {Ventura}, {dell'Agli}, {Di Criscienzo}, {Hirashita}, \& {Kemper}}]{Schneider2014}
{Schneider}, R., {Valiante}, R., {Ventura}, P., {et~al.} 2014, \mnras, 442, 1440

\bibitem[{{Schreiber} {et~al.}(2018){Schreiber}, {Elbaz}, {Pannella}, {Ciesla}, {Wang}, \& {Franco}}]{Schreiber2018}
{Schreiber}, C., {Elbaz}, D., {Pannella}, M., {et~al.} 2018, \aap, 609, A30

\bibitem[{{Shen} {et~al.}(2013){Shen}, {Madau}, {Guedes}, {Mayer}, {Prochaska}, \& {Wadsley}}]{Shen2013}
{Shen}, S., {Madau}, P., {Guedes}, J., {et~al.} 2013, \apj, 765, 89

\bibitem[{{Shen} {et~al.}(2010){Shen}, {Wadsley}, \& {Stinson}}]{Shen2010}
{Shen}, S., {Wadsley}, J., \& {Stinson}, G. 2010, \mnras, 407, 1581

\bibitem[{{Shetty} {et~al.}(2011){Shetty}, {Glover}, {Dullemond}, \& {Klessen}}]{Shetty2011}
{Shetty}, R., {Glover}, S.~C., {Dullemond}, C.~P., \& {Klessen}, R.~S. 2011, \mnras, 412, 1686

\bibitem[{{Simpson} {et~al.}(2017){Simpson}, {Smail}, {Swinbank}, {Ivison}, {Dunlop}, {Geach}, {Almaini}, {Arumugam}, {Bremer}, {Chen}, {Conselice}, {Coppin}, {Farrah}, {Ibar}, {Hartley}, {Ma}, {Micha{\l}owski}, {Scott}, {Spaans}, {Thomson}, \& {van der Werf}}]{Simpson2017}
{Simpson}, J.~M., {Smail}, I., {Swinbank}, A.~M., {et~al.} 2017, \apj, 839, 58

\bibitem[{{Spilker} {et~al.}(2014){Spilker}, {Marrone}, {Aguirre}, {Aravena}, {Ashby}, {B{\'e}thermin}, {Bradford}, {Bothwell}, {Brodwin}, {Carlstrom}, {Chapman}, {Crawford}, {de Breuck}, {Fassnacht}, {Gonzalez}, {Greve}, {Gullberg}, {Hezaveh}, {Holzapfel}, {Husband}, {Ma}, {Malkan}, {Murphy}, {Reichardt}, {Rotermund}, {Stalder}, {Stark}, {Strandet}, {Vieira}, {Wei{\ss}}, \& {Welikala}}]{Spilker2014}
{Spilker}, J.~S., {Marrone}, D.~P., {Aguirre}, J.~E., {et~al.} 2014, \apj, 785, 149

\bibitem[{{Spitzer} \& {Zweibel}(1974)}]{SpitzerZweibel1974}
{Spitzer}, Lyman, J. \& {Zweibel}, E.~G. 1974, \apjl, 191, L127

\bibitem[{{Steinacker} {et~al.}(2013){Steinacker}, {Baes}, \& {Gordon}}]{Steinacker2013}
{Steinacker}, J., {Baes}, M., \& {Gordon}, K.~D. 2013, \araa, 51, 63

\bibitem[{{Sternberg}(1988)}]{Sternberg1988}
{Sternberg}, A. 1988, \apj, 332, 400

\bibitem[{{Sternberg} {et~al.}(2014){Sternberg}, {Le Petit}, {Roueff}, \& {Le Bourlot}}]{Sternberg2014}
{Sternberg}, A., {Le Petit}, F., {Roueff}, E., \& {Le Bourlot}, J. 2014, \apj, 790, 10

\bibitem[{{Stinson} {et~al.}(2006){Stinson}, {Seth}, {Katz}, {Wadsley}, {Governato}, \& {Quinn}}]{Stinson2006}
{Stinson}, G., {Seth}, A., {Katz}, N., {et~al.} 2006, \mnras, 373, 1074

\bibitem[{{Tacconi} {et~al.}(2020){Tacconi}, {Genzel}, \& {Sternberg}}]{Tacconi2020}
{Tacconi}, L.~J., {Genzel}, R., \& {Sternberg}, A. 2020, \araa, 58, 157

\bibitem[{{Tamburello} {et~al.}(2015){Tamburello}, {Mayer}, {Shen}, \& {Wadsley}}]{Tamburello2015}
{Tamburello}, V., {Mayer}, L., {Shen}, S., \& {Wadsley}, J. 2015, \mnras, 453, 2490

\bibitem[{{Tamburello} {et~al.}(2017){Tamburello}, {Rahmati}, {Mayer}, {Cava}, {Dessauges-Zavadsky}, \& {Schaerer}}]{Tamburello2017}
{Tamburello}, V., {Rahmati}, A., {Mayer}, L., {et~al.} 2017, \mnras, 468, 4792

\bibitem[{{Teng} {et~al.}(2023){Teng}, {Sandstrom}, {Sun}, {Gong}, {Bolatto}, {Chiang}, {Leroy}, {Usero}, {Glover}, {Klessen}, {Liu}, {Querejeta}, {Schinnerer}, {Bigiel}, {Cao}, {Chevance}, {Eibensteiner}, {Grasha}, {Israel}, {Murphy}, {Neumann}, {Pan}, {Pinna}, {Sormani}, {Smith}, {Walter}, \& {Williams}}]{Teng2023}
{Teng}, Y.-H., {Sandstrom}, K.~M., {Sun}, J., {et~al.} 2023, \apj, 950, 119

\bibitem[{{Thompson} {et~al.}(2014){Thompson}, {Nagamine}, {Jaacks}, \& {Choi}}]{Thompson2014}
{Thompson}, R., {Nagamine}, K., {Jaacks}, J., \& {Choi}, J.-H. 2014, \apj, 780, 145

\bibitem[{{Thomson} {et~al.}(2017){Thomson}, {Simpson}, {Smail}, {Swinbank}, {Best}, {Sobral}, {Geach}, {Ibar}, \& {Johnson}}]{Thomson2017}
{Thomson}, A.~P., {Simpson}, J.~M., {Smail}, I., {et~al.} 2017, \apj, 838, 119

\bibitem[{{Todini} \& {Ferrara}(2001)}]{Todini2001}
{Todini}, P. \& {Ferrara}, A. 2001, \mnras, 325, 726

\bibitem[{{Toomre}(1964)}]{Toomre1964}
{Toomre}, A. 1964, \apj, 139, 1217

\bibitem[{{Toublanc}(1996)}]{Toublanc1996}
{Toublanc}, D. 1996, \ao, 35, 3270

\bibitem[{{Valentino} {et~al.}(2024){Valentino}, {Fujimoto}, {Gim{\'e}nez-Arteaga}, {Brammer}, {Kohno}, {Sun}, {Kokorev}, {Bauer}, {Di Cesare}, {Espada}, {Lee}, {Dessauges-Zavadsky}, {Ao}, {Koekemoer}, {Ouchi}, {Wu}, {Egami}, {Jolly}, {Lagos}, {Magdis}, {Schaerer}, {Shimasaku}, {Umehata}, \& {Wang}}]{Valentino2024}
{Valentino}, F., {Fujimoto}, S., {Gim{\'e}nez-Arteaga}, C., {et~al.} 2024, arXiv e-prints, arXiv:2402.17845

\bibitem[{{Vallini} {et~al.}(2018){Vallini}, {Pallottini}, {Ferrara}, {Gallerani}, {Sobacchi}, \& {Behrens}}]{Vallini2018}
{Vallini}, L., {Pallottini}, A., {Ferrara}, A., {et~al.} 2018, \mnras, 473, 271

\bibitem[{{van de Hulst}(1958)}]{VanDerHulst1958}
{van de Hulst}, H.~C. 1958, Quarterly Journal of the Royal Meteorological Society, 84, 198

\bibitem[{{van der Tak} {et~al.}(2007){van der Tak}, {Black}, {Sch{\"o}ier}, {Jansen}, \& {van Dishoeck}}]{VanDerTak2007}
{van der Tak}, F.~F.~S., {Black}, J.~H., {Sch{\"o}ier}, F.~L., {Jansen}, D.~J., \& {van Dishoeck}, E.~F. 2007, \aap, 468, 627

\bibitem[{{Vogelsberger} {et~al.}(2020){Vogelsberger}, {Marinacci}, {Torrey}, \& {Puchwein}}]{Vogelsberger2020}
{Vogelsberger}, M., {Marinacci}, F., {Torrey}, P., \& {Puchwein}, E. 2020, Nature Reviews Physics, 2, 42

\bibitem[{{Wadsley} {et~al.}(2017){Wadsley}, {Keller}, \& {Quinn}}]{Wadsley2017}
{Wadsley}, J.~W., {Keller}, B.~W., \& {Quinn}, T.~R. 2017, \mnras, 471, 2357

\bibitem[{{Wadsley} {et~al.}(2004){Wadsley}, {Stadel}, \& {Quinn}}]{Wadsley2004}
{Wadsley}, J.~W., {Stadel}, J., \& {Quinn}, T. 2004, \na, 9, 137

\bibitem[{{Walter} {et~al.}(2008){Walter}, {Brinks}, {de Blok}, {Bigiel}, {Kennicutt}, {Thornley}, \& {Leroy}}]{Walter2008}
{Walter}, F., {Brinks}, E., {de Blok}, W.~J.~G., {et~al.} 2008, \aj, 136, 2563

\bibitem[{Wendland(1995)}]{Wendland1995}
Wendland, H. 1995, Advances in Computational Mathematics, 4, 389

\bibitem[{{Wootten} \& {Thompson}(2009)}]{Wootten2009}
{Wootten}, A. \& {Thompson}, A.~R. 2009, IEEE Proceedings, 97, 1463

\bibitem[{{Ysard} {et~al.}(2024){Ysard}, {Jones}, {Guillet}, {Demyk}, {Decleir}, {Verstraete}, {Choubani}, {Miville-Desch{\^e}nes}, \& {Fanciullo}}]{Ysard2024}
{Ysard}, N., {Jones}, A.~P., {Guillet}, V., {et~al.} 2024, \aap, 684, A34

\bibitem[{{Zanella} {et~al.}(2018){Zanella}, {Daddi}, {Magdis}, {Diaz Santos}, {Cormier}, {Liu}, {Cibinel}, {Gobat}, {Dickinson}, {Sargent}, {Popping}, {Madden}, {Bethermin}, {Hughes}, {Valentino}, {Rujopakarn}, {Pannella}, {Bournaud}, {Walter}, {Wang}, {Elbaz}, \& {Coogan}}]{Zanella2018}
{Zanella}, A., {Daddi}, E., {Magdis}, G., {et~al.} 2018, \mnras, 481, 1976

\bibitem[{{Zavala} {et~al.}(2018){Zavala}, {Aretxaga}, {Dunlop}, {Micha{\l}owski}, {Hughes}, {Bourne}, {Chapin}, {Cowley}, {Farrah}, {Lacey}, {Targett}, \& {van der Werf}}]{Zavala2018}
{Zavala}, J.~A., {Aretxaga}, I., {Dunlop}, J.~S., {et~al.} 2018, \mnras, 475, 5585

\end{thebibliography}

\begin{appendix}

%-------------------------------------------------------------------
%-------------------------------------------------------------------
%-------------------------------------------------------------------

\section{Monte Carlo radiative transfer in RADMC-3D}\label{sec:appendix_A}

In this appendix, we summarize the main working principle behind the \texttt{RADMC-3D} code, as well as that of the Monte Carlo radiative transfer implemented therein.

\subsection*{Dust temperature computation} \label{sec:dust_transfer}

\texttt{RADMC-3D} implements the radiative transfer Monte Carlo method described in \citet{BjorkmanWood2001}, including the continuous absorption method \citep{Lucy1999}. 

The whole method is based on enforcing radiative equilibrium, for which the absorbed energy in a volume element (in our case, a cell of the grid) must be re-radiated. 

We can express the energy emitted over a time interval $\Delta t$ as
%\begin{multline}
%\frac{dE_{\rm em}}{dt} = 4\pi \int dV \int j_\nu \, d\nu = \\ = 4\pi \int k_p(T) \, B(T) \, \rho dV =  \frac{dE_{\rm abs}}{dt}   
%end{multline}
\begin{equation} \label{eq:radeq}
E_{\rm em} = 4\pi \Delta t \int dV \int j_\nu \, d\nu = 4\pi \Delta t \int k_p(T) \, B_{\nu}(T) \, \rho dV \, ,
\end{equation}
where $j_\nu = \rho  k_\nu  B_\nu(T)$ is the dust thermal emissivity, $\rho$ is the dust density, $k_\nu$ is the dust absorption 
opacity, $B_\nu(T)$ is the Planck function at temperature $T$, $k_p(T)=\int k_\nu B_\nu dv/B$ is the Planck mean opacity, and $B=\sigma T^4/\pi$ is the Planck function integrated over frequency, which is the result of the Stefan-Boltzmann law with $\sigma$ the Stefan-Boltzmann constant.

Briefly, \texttt{RADMC-3D} injects in the grid a given number of photon packages $N_\gamma$ (which is an input parameter), each package with the same initial energy $E_\gamma=E/N_\gamma$, where $E$ is the total energy in the model. In our framework, because the source of luminosity in our radiative transfer simulations is a stellar density grid (in contrast to single stars, as the code allows in other configurations), the energy of each photon package is computed using the total luminosity in a cell, instead of that of a single star. 
 
After this step, each photon package in \texttt{RADMC-3D} is assigned a random frequency sampled from the chosen SED. In this way, the number of photons constituting the photon package will be automatically determined by energy conservation. We describe the modelling of stellar sources and the associated SED in Sect.~\ref{sec:star_model}.

Once a photon package has been initialized, its motion is started in a random direction. At this point, a random number $z$ is obtained via a uniform random sampling from the interval $\mathcal{I}=[0,1)$. In this formalism, the mean free path of a photon package between two events is assumed to be $x=-\ln(z)$, which corresponds to a physical displacement $l=x/[\rho(k_\nu+\sigma_\nu)]$, where $\sigma_\nu$ is the dust scattering opacity 
\citep[][]{Lucy1999}. After displacing each photon package by $l$ along the initial direction, if the package goes beyond the boundary of the initial cell, then one computes $l$ again. If instead a photon package remains inside the same cell, it gets scattered by a dust grain if $z<\sigma_\nu/(k_\nu+\sigma_\nu)$, or it gets absorbed otherwise.

When a photon package gets scattered, the frequency is kept unchanged, while only the direction is reassigned, depending on the scattering approximation used. \texttt{RADMC-3D} implements isotropic scattering, as well anisotropic scattering in different degrees of complexity and realism. The different scattering mechanisms are described in Sect.~\ref{sec:opac_comp}.

When a photon package gets absorbed, it must be re-radiated instead. In this case, both the direction and the frequency must be reassigned. Assuming LTE, the emitted energy after an absorption event is given by the radiative equilibrium condition expressed in Eq.~\ref{eq:radeq}.

Each time a photon package enters a cell, the energy associated with the cell increases and the dust temperature must be recomputed. This treatment is different from that in \citet{BjorkmanWood2001}, in which the temperature is recomputed only if an absorption event occurs. Hence, in \texttt{RADMC-3D} the energy increase in each cell is given by the energy $E_\gamma$ associated with the photon package. Enforcing the radiative equilibrium condition and setting $E_{\rm em}=E_\gamma$ gives an implicit equation for the dust temperature, which must be solved iteratively at each step.

In this way, a photon package travels throughout the probed volume until it escapes, at which point a new package is launched and the simulation continues until all photons have been injected into the box and they have escaped it. We notice from now that the calculation will be subject to the number of photon packages used in the calculation, given the intrinsic degree of stochasticity. In general, one expects that the higher the number of photons, the less prone the results will be to stochastic fluctuations, but the more expensive a Monte Carlo will become. In Sect.~\ref{sec:convergence}, we show that the computation converges to sub-percent deviations accuracy if a sufficiently large number of photon packages is chosen, and we will adopt the minimum number of packages which ensures convergence throughout this work. 

We refer the reader to \citet{Lucy1999} and \citet{BjorkmanWood2001} for a more detailed treatment of the Monte Carlo radiative transfer method used herein.

\subsection{Continuum transfer} 

Once the dust temperature has been determined using the Monte Carlo method described in Sect.~\ref{sec:dust_transfer}, \texttt{RADMC-3D} performs the ray-tracing of individual photons at an arbitrary wavelength defined by the user. In particular, given a photon of frequency $\nu$, the code solves the radiative transfer equation, 
\begin{equation} \label{eq:radtransf}
    \frac{dI_\nu}{d\tau_\nu} = j_\nu - \alpha_{\nu} I_\nu \,,
\end{equation}
where $\tau_\nu$ denotes the optical depth at frequency $\nu$, $I_\nu$ is the intensity of the local radiation field, 
and $\alpha_\nu$  is the extinction coefficient. In the present case, the emissivity has both a thermal emission and a scattering term: $j_\nu=j_\nu^{\rm th} + j_\nu^{\rm sc}$. The thermal part, $j_\nu^{\rm th}=\alpha_\nu^{\rm abs}B_\nu(T)$, is given by product of the absorption extinction $\alpha_\nu^{\rm abs}=\rho k_\nu$ and the Planck function. The scattering term can instead be expressed as
\begin{equation} \label{eq:scat_term}
    j_\nu^{\rm sc}= \alpha_\nu^{\rm sc} S_\nu^{\rm sc} \, ,
\end{equation}
where, $\alpha_\nu^{\rm sc}=\rho \sigma_\nu$ is the scattering extinction and $S_\nu^{\rm sc}$ is the scattering source function, which will be discussed more in detail in Sect.~\ref{sec:opac_comp}. 

Similarly, also the extinction term of Eq. \ref{eq:radtransf} can be expressed as the sum of two components, namely a scattering and an absorption part. In particular, one can write $\alpha_{\nu}=\alpha_\nu^{\rm sc}+\alpha_\nu^{\rm abs}$, where the two addends are defined as above. 

The equation is fully specified if the dust temperature $T$, the dust 
density $\rho$, the opacities $k_\nu$, and the scattering phase functions are known. In this case, we notice that $T$ is the outcome of the Monte Carlo thermal run described in Sect.~\ref{sec:dust_transfer}, $\rho$ is an input of the model, the scattering phase  function is either known analytically or is pre-tabulated and the $k_\nu$ coefficients are also pre-computed using Mie theory (see Sect.~\ref{sec:opac_comp}). Hence, the equation can be readily solved at this point.

\subsection{Spectral line transfer}
\label{sec:line_transfer}

After employing the Monte Carlo run to obtain the dust temperature, \texttt{RADMC-3D} can also solve the radiative transfer equation (Eq.~\ref{eq:radtransf}) for spectral lines. Given a spectral line arising from a transition between two energy levels $i$ and $j$, Eq. \ref{eq:radtransf} reads
\begin{equation}
    \frac{dI_\nu}{d\tau_\nu}=j_{\nu}^{\, ij} - \alpha_{\nu}^{ij}I_\nu \, ,  
\end{equation}
with
\begin{equation}
    j_{\nu}^{\, ij}=\frac{h\nu}{4\pi}N\, n_i\, A^{ij}\varphi_{\nu}^{ij} \, , \quad  
    \alpha_\nu^{ij}=\frac{h\nu}{4\pi}N(n^jB^{ji} - n^iB^{ij})\varphi_\nu^{ij} \, ,
\end{equation}
where $N$ is the total number density of the studied atom/molecule, $n_i$ is the relative abundance of the atom/molecule at level $i$, $A^{ij}$ is the Einstein coefficient for the emission from level $i$ and $j$, $B^{ij}$ and $B^{ji}$ are the Einstein B-coefficients, and $\varphi_\nu^{ij}$ is the line profile function. Here, $N$, $A^{ij}$, and $b^{ij}$ are supplied by the user as inputs, while $n_i$ and $\varphi_\nu^{ij}$ are computed internally within the code. We refer the reader to the documentation of \texttt{RADMC-3D} for a complete description of the implementation, which goes beyond the scope of this paper.  

\texttt{RADMC-3D} implements the solution of the radiative transfer equation in LTE conditions, as well as the approximated solution in non-LTE conditions with different methods. We notice that a full non-LTE treatment for line transfer is not available in \texttt{RADMC-3D} \citep[see, instead, e.g.][for a self-consistent treatment of dust and non-LTE CO radiative transfer]{Matsumoto2023}, but for most of the cases it would probably become too computationally expensive and probably not necessary. %The two approximations tested in this work are:
%\begin{itemize}
In this work, we test the Large Velocity Gradient (hereafter LVG, also known as Sobolov approximation) + Escape Probability (hereafter EscProb) non-LTE method \citep{Ossenkopf1997,Shetty2011}. In this method, the local mean intensity $J=(1-\beta)S$ is integrated over the line, where $S$ is the line intensity and $\beta$ is the escape fraction parametrized as $\beta=(1-\exp(-\tau))/\tau$, where $\tau=\min(\tau_{\rm LVG},\tau_{\rm EscProb})$: in this formalism, one seeks to determine the minimum optical depth which causes a photon to escape due to either the presence of large velocity gradients ($\tau_{\rm LVG}$) or due to the finite width of the line ($\tau_{\rm EscProb}$). In the LVG method, $\tau_{\rm LVG}\propto 1/|\nabla\vec{v}|$ \citep[see, e.g.][]{VanDerTak2007,Shetty2011}, where $\nabla\vec{v}$ is the gradient of the local velocity field, such that the escape probability is larger than the absolute value of the velocity gradient. In the escape probability method, $\tau_{\rm EscProb}\propto L/w_{\rm line}$, where $L$ is a free scale parameter (to be supplied by the user) and $w_{\rm line }$ is the width of the spectral line, where again the smaller is $L$ the more non-local is the method becomes and for $L\rightarrow\infty$ the case is equivalent to pure LVG.

\begin{figure}
    \centering
    \includegraphics[width=\columnwidth]{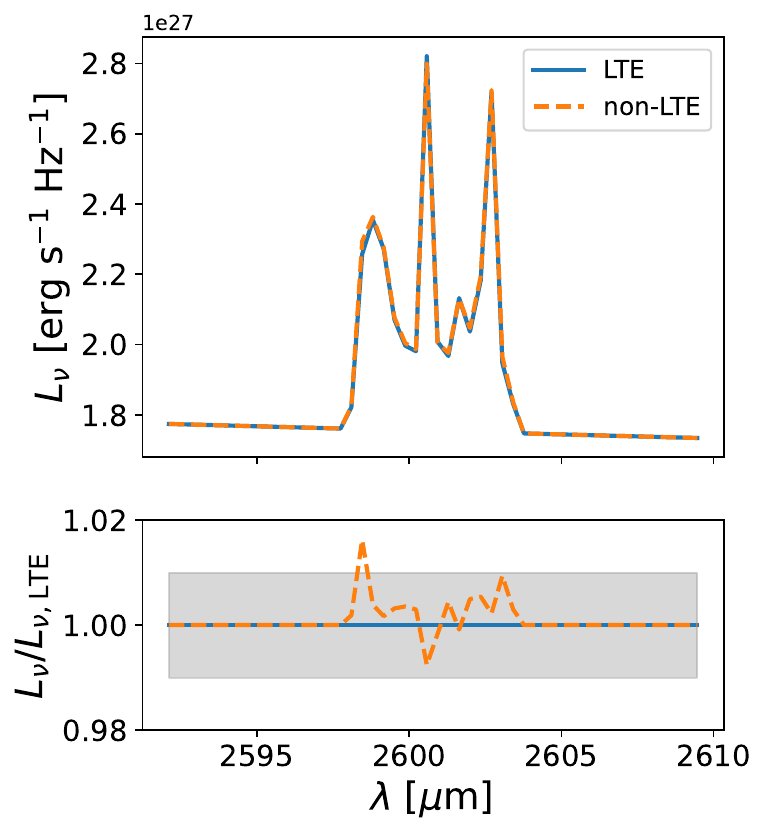}
    \caption{CO(1-0) line spectra for run~13, obtained projecting the galaxy edge-on. The top panel shows the comparison between the spectrum obtained in LTE approximation (blue solid) and the one obtained adopting the 
    non-LTE method (orange dashed). The bottom panel displays the wavelength-wise ratio between the spectra, highlighting that the difference is $\lesssim 1\%$ in almost all bins.}
    \label{fig:co_spec}
\end{figure}

To address the impact of assuming a simpler LTE formalism over performing the non-LTE computation, we study the CO(1-0) resulting spectra from such two methods. The results are shown in Figure~\ref{fig:co_spec}, where we show the spectra in the top panel, and the ratio between the non-LTE and the LTE spectra in the bottom panel. The ratios clearly show that the difference between the LTE and non-LTE computation is $\lesssim 1\%$ across the whole spectrum. This means that, in this specific problem and under the studied configuration, the adoption of a non-LTE method is not crucial, and goes at the expense of a significantly longer computing time.

\subsection{Opacity computation and scattering treatment} \label{sec:opac_comp}

The scattering and absorption opacity must be computed given the features of the dust grains in the model. While the opacity is not directly computed inside \texttt{RADMC-3D}, we have included in the \texttt{RTGen} pipeline an external package to compute the opacity as a function of wavelength.\footnote{The package is a Python rewriting by C.P. Dullemond of the original Fortran77 code by B. Draine available at \url{https://www.astro.princeton.edu/~draine/scattering.html}.}

Briefly, the opacity computation relies on Mie scattering theory \citep{Mie1908,VanDerHulst1958,BohrenHuffman1983}. Following \citet{BohrenHuffman1983} and expressing the dust absorption opacity as $k_\nu=s_\nu n/\rho$, where $s_\nu$ is the interaction cross section, and $n$ and $\rho$ are the number and mass density of the dust respectively, one can express the total and the scattering cross sections $\sigma_{\rm tot}$ and $\sigma_{\rm sc}$ in terms of vector spherical harmonics expansion as
\begin{equation}
    \sigma_{{\rm tot},\nu} = \frac{2\pi}{k_\nu^2} \sum_{i=1}^{\infty} (2i+1) \, {\rm Re}\{a_i+b_i\} \, ,
\end{equation}
\begin{equation}
    \sigma_{{\rm sc},\nu} = \frac{2\pi}{k_\nu^2} \sum_{i=1}^{\infty} (2i+1) \, (|a_i|^2+|b_i|^2) \, ,
\end{equation}
where $a_i$ and $b_i$ are the scattering coefficients, defined as
\begin{equation}   a_i=\frac{m\psi_i(mx)\psi'_i(x)- \psi_i(x)\psi'_i(mx)}{m\psi_i(mx)\xi'_i(x) - \xi_i(x)\psi'_i{mx)}} \, ,
\end{equation}
\begin{equation}   
b_i=\frac{\psi_i(mx)\psi'_i(x)- m\psi_i(x)\psi'_i(mx)}{\psi_i(mx)\xi'_i(x) - m\xi_i(x)\psi'_i(mx)} \, ,
\end{equation}
where $\psi_i$ and $\xi_i$ are the Riccati-Bessel functions, $m$ is the complex refractive index of the considered medium, and $x=2\pi a/\lambda$ is the ratio between the size of the particle $a$ and the wavelength $\lambda$.

In this way, the used piece of software allows us to obtain the scattering and absorption opacity. Also, while an exhaustive treatment of this topic goes beyond the scope of the paper \citep[see][and references therein, for a detailed explanation]{BohrenHuffman1983}, the same code also returns the scattering Müller matrix $\mathcal{Z}$ transforming an input Stokes vector $\vec{S}_0=(I_0,Q_0,U_0,V_0)$ into the output Stokes vector $\vec{S}=(I,Q,U,V)$ via the relation $\vec{S}=\mathcal{Z}\vec{S}_0$. For randomly-oriented particles, $\mathcal{Z}$ is a symmetric matrix, for which only $6$ elements (e.g. the upper triangular part including the diagonal) are independent. Also, the elements of $\mathcal{Z}$ will depend on the angle between the incoming and outgoing scattering directions $\theta$ and on the wavelength $\lambda$. Therefore, it is necessary to compute that Müller matrix for a grid on $\theta$ and for all the $\lambda$ included in the model.

With these ingredients, \texttt{RADMC-3D} implements six treatments of scattering, of increasing realism and complexity:

\begin{itemize}

    \item no scattering: scattering is not included. This option is clearly not realistic, but it is still instructive to test it and compare it with other ways of treating scattering;
    
    \item isotropic scattering: in this case, the scattering source function $S_\nu^{\rm sc}$ in Eq. \ref{eq:scat_term} is the integral over the solid angle of the intensity $I_\nu$:
    \begin{equation}
        S_\nu^{\rm sc}=\frac{1}{4\pi}\int I_\nu \, d\Omega \, ,
    \end{equation}
    which is independent of the direction;
    
    \item anisotropic scattering with the Henyey-Greenstein phase fuction: in the anisotropic scattering case, the scattering source function can be obtained by updating the previous expression with the introduction of the scattering phase function $\Phi(\vec{n}_{\rm in}, \vec{n}_{\rm out})$, which depends on incoming and outgoing direction, parametrized through the unit vectors $\vec{n}_{\rm in}$ and $\vec{n}_{\rm out}$. Defining $\theta$ as the angle subtended by $\vec{n}_{\rm in}$ and $\vec{n}_{\rm out}$ and  $\cos\theta=\mu=\vec{n}_{\rm in}\cdot \vec{n}_{\rm out}$, we can express $\Phi(\vec{n}_{\rm in}, \vec{n}_{\rm out})=\Phi(\mu)$ and write the scattering source function as
    \begin{equation}
    S_\nu^{\rm sc}(\mu)= \frac{1}{4\pi} \int I_\nu \, \Phi(\mu) \, d\Omega \, .
    \end{equation}
    In this option, the scattering phase function is assumed to be the Henyey-Greenstein function \citep[e.g.][]{HenveyGreestein1941,Toublanc1996}:
    \begin{equation}
        \Phi(\mu)=\frac{1-g^2}{(1+g^2-2g\mu)^{3/2}} \, ,
    \end{equation}
    where $g$ is the anisotropy parameter and is computed numerically from Mie theory.
    
    \item anisotropic scattering with the phase function tabulated by the user: in this case, one has to specify the $\mathcal{Z}_{11}$ element of the Müller matrix, which is then linked to the phase function as $\Phi(\mu)=4\pi \mathcal{Z}_{11}/\sigma$;
    
    \item anisotropic scattering using the full scattering matrix for polarization and randomly-oriented particles: in this case, the full Müller matrix is used to compute the phase function, but it is still a symmetric matrix with just $6$ independent elements;
    
    \item anisotropic scattering with full polarization treatment: in this case, the potential alignment of non-spherical dust grains is taken into account and the full Müller matrix (with $16$ independent elements) is used in the computation.

\end{itemize}

We notice that, since we are not interested in studying polarization at this stage, we do not investigate the three last options, which are also the most computationally expensive. We leave such a study for future work.

\begin{figure}
    \centering
    \includegraphics[width=\columnwidth]{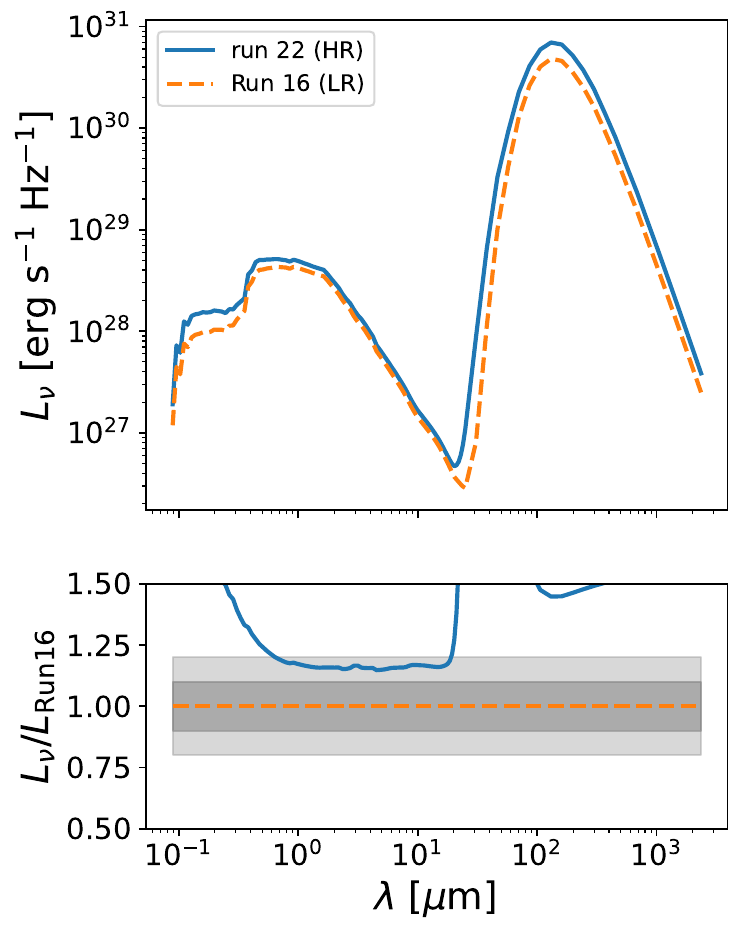}
    \caption{Results from the baseline analysis for run 16 (orange dashed) and run 22 (blue solid). Top: SED predictions as a function of the simulation's resolution (number of SPH particles). Bottom: ratios between the SED of each studied case and the baseline, with the light and dark gray bands showing a 10$\%$ and 20$\%$ difference, respectively.}
    \label{fig:sed_resolution}
\end{figure}

\begin{figure}
    \centering
    \includegraphics[width=\columnwidth]{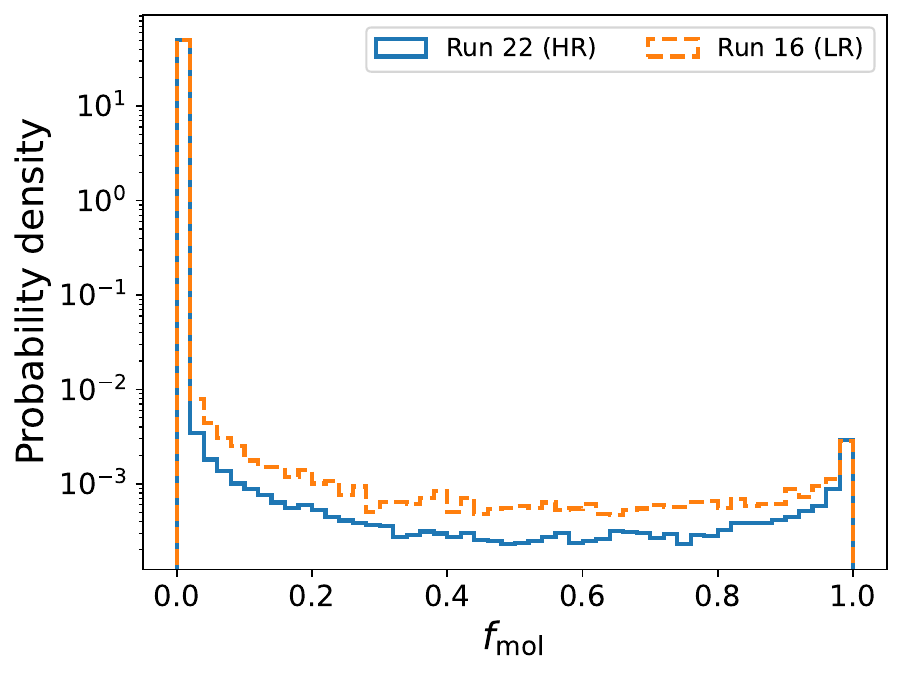}
    \caption{Same as Figure~\ref{fig:fmol_hist}, but for runs~22 and 16, i.e. as a function of the simulations resolution (number of SPH particles).}
    \label{fig:fmol_hist_resolution}
\end{figure}

\begin{figure}
    \centering  \includegraphics[width=\columnwidth]{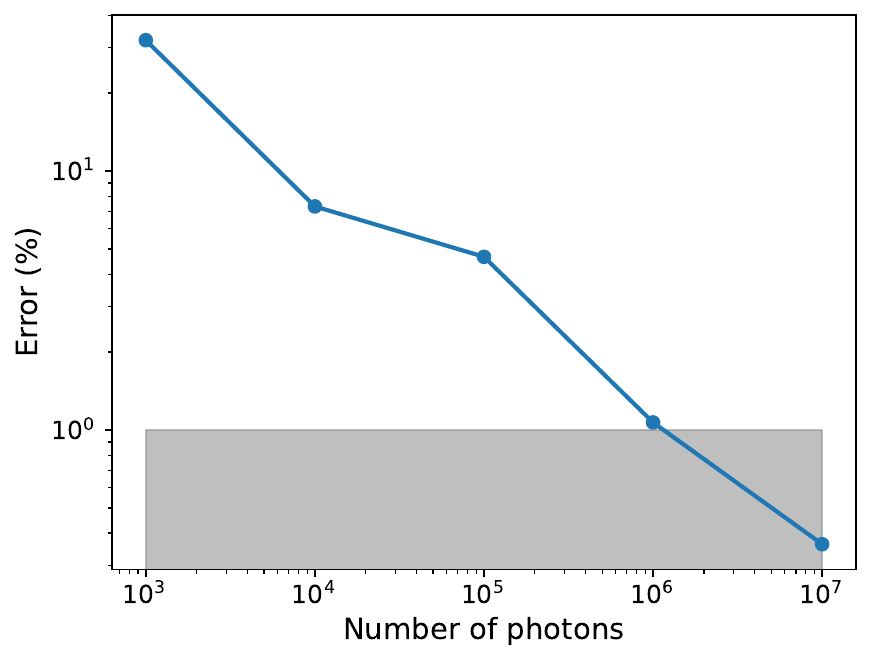}
    \caption{Maximum percentage uncertainty on the SED as a function of the number of photons used in the Monte Carlo dust radiative transfer simulations. Sub-percent convergence is reached for $n_{\rm phot}=10^7$.}
    \label{fig:convergence}
\end{figure}

\subsection{Impact of resolution}\label{sec:test_resolution}

In this section, we study the impact of the simulation's resolution on the predicted observables. In particular, we consider the high-resolution simulation counterpart of run~16 -- i.e. run~22 from \citet{Tamburello2015} -- and run our pipeline over it. Figure~\ref{fig:sed_resolution} shows the comparison between the SEDs obtained from run~16 (low resolution, orange dashed) and run~22 (high resolution, blue solid). The top panel shows the resulting SEDs, whereas the bottom panel displays the ratios between the SED of each studied case and that of run 16, with the gray bands marking $10\%$ (darker) and $20\%$ (lighter) deviations, respectively. One clearly sees that the high-resolution case predicts a brighter SED across all wavelengths. Figure~\ref{fig:fmol_hist_resolution} displays the resulting distribution for the cell-wise molecular fraction $f_{\rm mol}=\rho_{\rm H2}/\rho_{\rm gas,tot}$, similarly to Figure~\ref{fig:fmol_hist}, but for runs~16 and 22. This plot shows that run~22 has generally a larger molecular fraction than that of run~16. These results can be easily understood by noticing that a higher resolution implies the formation of higher-density regions in the simulation, thereby favouring the molecular cloud formation and consequently star formation. This translates into having in such regions, on average, younger stellar populations, as well as larger dust density, given that the latter is proportional to the gas density in our framework. These two facts motivate the SED of run~22 being brighter than that of run~16 in the UV/optical regions and in the FIR region, respectively.

\subsection{Convergence tests against number of photons}\label{sec:convergence}

To study the impact of stochastic fluctuations due to the intrinsic Monte Carlo nature of our simulations and the consequent robustness of our results, we perform convergence tests studying the variation in the SED as a function of the number of photons used in our runs. In particular, the goal is to establish the minimum number of photons for which the simulations are well converged, to strike a balance between the computational cost and the stability of the results. To do so, we vary the number of photons in the range $n_{\rm phot}=10^3$--$10^7$ (one case per order of magnitude, i.e. $5$ cases in total) and run $50$ dust continuum transfer simulations for each case. Afterwards, we compute the average $\bar{S}(\lambda)$ and SED for all the simulations of the same case and its standard deviation $\sigma(\lambda)$, and quantify the percentage error as $\sigma_{\rm perc}(\lambda)=\sigma(\lambda)/\bar{S}(\lambda\times 100)$. We then compute the maximum error over the wavelength range, to have a conservative estimate of the stochastic uncertainty. Figure~\ref{fig:convergence} shows the results of this test. Our findings evidence that the maximum error in the SED is $\sim$$30\%$ for $n_{\rm phot}=10^3$, it approaches $\sim$$1\%$ for $n_{\rm phot}=10^6$, and it is $\sim$$0.1\%$ for $n_{\rm phot}=10^7$. Therefore, for all the simulations discussed in the remainder of the paper, we set $n_{\rm phot}=10^7$, to be sure to have a sub-percent convergence even in extreme cases (a 5$\sigma$ deviation from the mean would correspond to a $\sim$$0.5\%$ error). 

\end{appendix}
% *************************************

%\section*{Appendix A: LTE versus non-LTE line transfer}

%\section*{Appendix B: opacity computation and scattering treatment}

%\section*{Appendix C: Molecular/atomic gas splitting}

%\section*{Appendix D: Dust composition and grain size}

%  \bibitem[Baker(1966)]{baker} Baker, N. 1966,
%      in Stellar Evolution,
%      ed.\ R. F. Stein,\& A. G. W. Cameron
%      (Plenum, New York) 333

%   \bibitem[Balluch(1988)]{balluch} Balluch, M. 1988,
%      A\&A, 200, 58

%   \bibitem[Cox(1980)]{cox} Cox, J. P. 1980,
%      Theory of Stellar Pulsation
%      (Princeton University Press, Princeton) 165

%   \bibitem[Cox(1969)]{cox69} Cox, A. N.,\& Stewart, J. N. 1969,
%      Academia Nauk, Scientific Information 15, 1

%   \bibitem[Mizuno(1980)]{mizuno} Mizuno H. 1980,
%      Prog. Theor. Phys., 64, 544
   
%   \bibitem[Tscharnuter(1987)]{tscharnuter} Tscharnuter W. M. 1987,
%      A\&A, 188, 55
  
%   \bibitem[Terlevich(1992)]{terlevich} Terlevich, R. 1992, in ASP Conf. Ser. 31, 
%      Relationships between Active Galactic Nuclei and Starburst Galaxies, 
%      ed. A. V. Filippenko, 13

%   \bibitem[Yorke(1980a)]{yorke80a} Yorke, H. W. 1980a,
%      A\&A, 86, 286

%   \bibitem[Zheng(1997)]{zheng} Zheng, W., Davidsen, A. F., Tytler, D. \& Kriss, G. A.
%     1997, preprint
%\end{thebibliography}
%\end{comment}

\end{document}